\documentclass[a4paper,11pt]{article}

\usepackage[colorlinks=true,breaklinks=true,bookmarks=true,urlcolor=blue,
citecolor=blue,linkcolor=blue,bookmarksopen=false,draft=false]{hyperref}

\usepackage{amsmath}
\usepackage{amssymb}
\usepackage{amsthm}
\usepackage{tikz}
\usepackage{graphicx}
\usepackage{color}
\usepackage{calc}
\usepackage{dsfont}
\usepackage{makeidx}
\usepackage[square]{natbib}
\usepackage{subcaption}
\usepackage{bbm}
\usepackage{geometry}
\usepackage{setspace}

\usetikzlibrary{arrows,positioning}

\usepackage{bm}  
\usepackage{verbatim}

\numberwithin{equation}{section}

\newtheorem{theorem}{Theorem}
\newtheorem{definition}{Definition}[section]
\newtheorem{example}[theorem]{Example}
\newtheorem{corollary}{Corollary}[section]

\newtheorem{lemma}[theorem]{Lemma}

\newtheorem{remark}{Remark}[section]

\newtheorem{open}{Open Question}[section]

\renewcommand{\S}{\mathcal{S}}
\newcommand{\N}{\mathbb{N}}
\newcommand{\K}{\mathcal{K}}

\newcommand{\R}{\mathbb{R}}

\newcommand{\C}{\mathcal{C}}
\newcommand{\G}{\mathcal{G}}
\newcommand{\Gamespace}{\G(G,\K,\S)}
\newcommand{\GGamespace}{\overline{\G}(G,\K,\S)}
\newcommand{\ga}{\Gamma}
\newcommand{\gga}{\bar{\Gamma}}
\newcommand{\norm}[1]{|\!|#1|\!|_{\infty}}
\newcommand{\gnorm}[1]{|\!|#1|\!|}
\newcommand{\mb}{\nu}
\newcommand{\h}{\varkappa}

\newcommand{\rhm}[1]{{#1}}

\newcommand{\wu}[1]{#1}

\newcommand{\wusec}[1]{#1}

\newcommand{\WU}[1]{#1}
\newcommand{\WUFinal}[1]{#1}
\newcommand{\wuu}[1]{{#1}}
\newcommand{\wulast}[1]{{#1}}

\title{A sensitivity analysis of the PoA in non-atomic congestion games}

\author{Zijun Wu\\
	Institute for Applied Optimization\\
	School of Artificial Intelligence and Big Data\\
	Hefei University\\
	Jinxiu 99, 230601, Jingkai District, Hefei, Anhui, China\\
	 \texttt{wuzj@hfuu.edu.cn}\\[2ex]
	 Rolf H. M{\"o}hring\\
	 Kombinatorische Optimierung und Graphenalgorithmen (COGA)\\
	 Fakult{\"a}t II-Mathematik und Naturwissenschaften\\
	 Institut f{\"u}r Mathematik, Sekr. MA 5-1\\
	 Technische Universit{\"a}t Berlin\\
	  Stra{\ss}e des 17. Juni 136, Berlin, Germany, 10623\\
	  \texttt{rolf.moehring@tu-berlin.de}}

\begin{document}

\maketitle

\begin{abstract}
	\wuu{The price of anarchy(PoA) is a standard measure to quantify the inefﬁciency of equilibria in non-atomic congestion 
		games. Most publications have focused on worst-case bounds for the PoA, only few have analyzed the sensitivity of the PoA against changes of the demands or cost functions, although that is crucial for empirical computation of the PoA.
		We analyze the sensitivity of the PoA w.r.t. simultaneous changes of demands and cost 
		functions. The key to this analysis is a metric for the distance between two games that deﬁnes a topological metric space consisting of all games with the same combinatorial structure.
		The PoA is then a locally pointwise H{\"o}lder continuous function of the demands and cost functions, and 
		we analyze the H{\"o}lder exponent for different classes of cost functions. We also apply our approach to the convergence analysis of the PoA when the total demand tends to zero or inﬁnity.
		Our results \WUFinal{further develop the} recent \WUFinal{seminal work} by Englert etal., Takalloo and Kwon, and Cominetti et al., who \WUFinal{have considered} the sensitivity of the PoA w.r.t. changes of the demands under special conditions.}
\end{abstract}

    \newpage
    
	\section{Introduction}
	\label{sec:Introduction}

    The \emph{Price of Anarchy} (PoA) (\citet{Papadimitriou1999} and \citet{Papadimitriou2001Algorithms}) is 
    a classic
    notion in algorithmic game theory and
    has been intensively studied in the
    last two decades, see, e.g., the book by \citet{Nisan2007} for an overview.
    Much of this work has been devoted
    to worst-case upper bounds of the PoA
    in congestion games for
    cost  functions of different types, starting with
    the pioneering paper of \citet{Roughgarden2000How}. 
    
    Much less attention has been paid to the evolution of the PoA 
    as a ``function'' of the demands and the cost functions, although
    this is quite important for \emph{traffic networks}.
    \wuu{In fact,} the real
    demands are usually hard to  measure accurately since they may
    fluctuate within a certain range. Also
    the actual latency on a street in a traffic network is almost impossible to
    obtain, and usually modeled by an \emph{idealized} flow-dependent cost function that is estimated empirically from real traffic data \WUFinal{and may also include tolls}. Thus modeling errors will inevitably occur.
	This raises the natural question if and how much  \wuu{such modeling errors} \wuu{will} influence the PoA,
	in particular, if the PoA \wuu{may differ} largely \wuu{under} small modeling
	\wuu{or measuring} errors \wuu{of}  the demands and cost functions. This is \wuu{crucial} for \wuu{applications}
	and asks for a sensitivity analysis of the
	PoA  in a traffic network w.r.t. small changes
\wuu{of} the demands and the cost functions of the network.
	

First results in this direction have been obtained by 
 \citet{EnglertFraOlb2010}, \citet{TakallooKwon2020}, and  \citet{Cominetti2019}.


\citet{EnglertFraOlb2010}  considered a  traffic network
with a \emph{single} \emph{origin-destination} (O/D) pair and \emph{polynomial} cost functions of degree at most
$\beta.$ They view the PoA in the
network as a function of the demand of that
O/D pair \emph{when the cost functions \wusec{stay} unchanged.}
They showed that the increase of this PoA function is bounded by a factor of $(1 + \epsilon)^\beta$ from above when the demand  increases by a factor of $1 + \epsilon$.
 
\citet{TakallooKwon2020} have generalized
this result to traffic networks
with \emph{multiple} O/D pairs and polynomial cost functions
of degree at most $\beta$. They \wusec{obtained} the same upper bound 
for the increase of the PoA when the demands of all O/D pairs
increase \emph{synchronously} by the {\em same} factor  $1 + \epsilon$ and \emph{the cost functions stay unchanged.}
\WU{Moreover, they showed that}
the increase of the PoA is bounded \wusec{from below} by 
a factor $\frac{1}{(1+\epsilon)^{\beta}}$ under the same conditions.

Similar \wusec{to} \citet{EnglertFraOlb2010},
\citet{Cominetti2019} considered also a traffic network with 
a \emph{single} O/D pair, and viewed the PoA as a function
of the demand of that O/D pair \emph{when the cost functions
are not varying.}
For cost functions \wusec{with}  \emph{strictly positive}
derivatives, they showed that the resulting
PoA function is differentiable at each demand
level \wu{of the O/D pair} where all the optimal paths carry a strictly positive
flow. For affine \wuu{linear} cost functions, they showed further
that the \emph{equilibrium cost} is \wusec{piece-wise} linear and
differentiable except at so-called \emph{$\mathcal{E}$-breakpoints},
whose number is finite, though \wusec{possibly} exponentially large in the size
of the network.
They then showed that, in any interval between two consecutive $\mathcal{E}$-breakpoints, the PoA function
is differentiable, and either monotone or unimodal
with a unique minimum on the interior of \wusec{that interval.}
So the maximum of the PoA function is
attained at an $\mathcal{E}$-breakpoint.
They also presented several examples showing how these properties fail for general cost functions.

We will come back to their results in more detail in Section~\ref{sec:Theory}.

These results  undoubtedly \wusec{indicate} that the change of the PoA
in a traffic network
\wusec{is} within an acceptable and predictable range w.r.t. small changes
of the demands
when the cost functions are \wu{fixed} and have
certain good properties, and the changes of the demands
fulfill certain regularity conditions.
They are thus seminal first results for a sensitivity
analysis of the PoA in traffic networks, but \wusec{are still restricted to special cases.}
In particular, a sensitivity analysis of the PoA w.r.t. \wusec{simultaneous} changes of 
\wusec{demands and} cost functions
is still missing.




\subsection{Our contribution}
	\label{subsec:OurResults}

This paper continues the studies of \citet{EnglertFraOlb2010},
\citet{TakallooKwon2020} and \citet{Cominetti2019}.
We consider a general traffic network \wuu{that have}
\emph{multiple} O/D pairs and \wuu{general cost functions that may include tolls.}
\wuu{We} view the PoA
as a real-valued map of the demands and the cost functions of that network,
and then analyze the sensitivity of the resulting map 
\emph{when both the demands 
	and the cost functions may change.}

To that purpose, we need first a well defined
measure for simultaneous changes of the demands and the
cost functions in a traffic network with multiple O/D pairs.

We thus fix in our analysis an \emph{arbitrary} directed network $G=(V,A)$ and an \emph{arbitrary}
finite set $\K$ of O/D pairs \WU{in} that network, and consider \emph{only} the
cost functions $\tau_a(\cdot)$ of arcs $a\in A$ and the demands $d_k$ of O/D pairs
$k\in\K$ as ``variables".
Each pair $\ga=(\tau,d)$ consisting of a cost function vector
$\tau=(\tau_a)_{a\in A}$
and a demand vector $d=(d_k)_{k\in\K}$
then represents a ``state'' of the network and
corresponds to an instance of the \emph{traffic game} (\citet{Roughgarden2000How})
defined on the network.
We then view
the collection of all these states as a \emph{game space}, and
the PoA in the network as a real-valued map $\rho(\cdot)$ from this space
to the unbounded interval $[1,\infty).$ Then the value
$\rho(\ga)$ at a point $\ga=(\tau,d)$
\wusec{of} the game space \wu{is} the resulting PoA when the network 
is at state $\ga,$ i.e., the resulting PoA when the network has the cost function vector \wuu{$\tau$ and 
the demand vector $d.$}
	
By adapting the	\emph{$L^\infty$-norms} of functions
and vectors, we define a binary operator
$\text{Dist}(\cdot,\cdot),$ \wusec{see}  \eqref{eq:GeneralizedNorm},
	on the game space by putting
	\begin{displaymath}
		\text{Dist}(\ga,\ga')\!=\!
		\max\left\{\norm{d\!-\!d'},\ \max_{a\in A,x\in [0,\min\{T(d),T(d')\}]}
		|\tau_a(x)\!-\!\sigma_{a}(x)|,\ \norm{\tau(T(d))\!-\!\sigma(T(d'))}\right\}
	\end{displaymath}
	for \wusec{two} arbitrary points $\ga=(\tau,d)$
	and $\ga'=(\sigma,d')$ \wusec{of} the game space,
	where $T(d)=\sum_{k\in \K}d_k$ and $T(d')=\sum_{k\in \K}d'_k$
	are the \emph{total demands} of the demand vectors
	$d$ and $d'$ \wusec{respectively,} $\tau(T(d))=(\tau_a(T(d)))_{a\in A}$
	and $\sigma(T(d'))=(\sigma_a(T(d')))_{a\in A}.$
	
	This binary operator \wusec{$\text{Dist}(\cdot,\cdot)$} is actually  a \emph{metric}
	on the game space w.r.t.\ an equivalence relation \wuu{defined in} \eqref{eq:GameEquivalence},
	see Lemma~\ref{lemma:Metric}.
	The game space then becomes a \emph{metric space}, and a simultaneous change of the cost functions and the demands
	\WU{is then quantified by the ``distance"} \wusec{$\gnorm{\ga-\ga'}:=\text{Dist}(\ga,\ga')$}
	between two points $\ga$ and $\ga'$ \wusec{of} the game space.
	
	With this metric, \wusec{our} sensitivity analysis of the PoA in the network \wusec{then transforms}
	to \wusec{an analysis of the} \emph{pointwise H{\"o}lder continuity}
	of the map $\rho(\cdot)$ on the game space, aiming at
	finding for each point $\ga=(\tau,d)$ a \emph{H{\"o}lder exponent}
	$\gamma_{\ga}>0$ s.t. $|\rho(\ga)-\rho(\ga')|<\h_{\ga}\cdot \gnorm{\ga-\ga'}^{\gamma_{\ga}}$
	when $\gnorm{\ga-\ga'}<\epsilon_\ga$
	for two positive constants $\h_\ga,\epsilon_{\ga}>0$
	depending only on $\ga.$
	Trivially, the larger the H{\"o}lder exponent
	$\gamma_{\ga}$ at a point $\ga=(\tau,d)$, the less sensitive is the PoA
	at the network state $\ga$ w.r.t. 
	a small change $\gnorm{\ga-\ga'}$ of the state $\ga$.

    As our first result, we show that both the equilibrium cost
    and the socially optimal cost are continuous \wusec{maps} (w.r.t. the metric) on the whole game space.
    This implies directly that
    the map $\rho(\cdot)$ is continuous on the whole game space, see Theorem~\ref{thm:PoA_Continuity}. 
    Hence, the PoA \wusec{changes} only slightly when the changes
    \wusec{of} the cost functions and the demands are very small
    in \WU{terms} of the metric.

We then 
show that $\rho(\cdot)$ is not \emph{uniformly H{\"o}lder  continuous} on the
whole game space, \wusec{i.e.,} there are no 
constants $\gamma,\h>0$ \wusec{such that}
\begin{displaymath}
    |\rho(\ga)-\rho(\ga')|\wuu{\le}\h\cdot \gnorm{\ga-\ga'}^{\gamma}\quad \forall 
    \ga,\ga'\in\Gamespace,
\end{displaymath}
see Theorem~\ref{thm:NotUniformlyHoelder}a).
Moreover, we show that no point $\ga$ \wusec{of} the game space
has the whole game space as its \emph{H{\"o}lder neighborhood},
see Theorem~\ref{thm:NotUniformlyHoelder}b).
Here, a H{\"o}lder neighborhood 
of a game $\ga$ refers to an open subset 
$U_\ga$ of the game space
with $\ga\in U_{\ga},$ for which there are 
constants $\gamma_{\ga},\h_{\ga}>0$ depending
only on $\ga$ s.t. $|\rho(\ga)-\rho(\ga')|
\wuu{\le}\h_{\ga}\cdot \gnorm{\ga-\ga'}^{\gamma_{\ga}}$
\wu{for each} $\ga'\in U_{\ga}.$

Hence, \wusec{$\rho(\cdot)$ may only be}
\emph{pointwise} and \emph{locally} \wusec{H{\"o}lder continuous} on the game space.
In particular, Theorem~\ref{thm:NotUniformlyHoelder}a)
implies that \emph{$\rho(\cdot)$ is not Lipschitz
continuous on the whole game space.}


Along with Theorem~\ref{thm:NotUniformlyHoelder}, our first H{\"o}lder continuity result shows that  
$\rho(\cdot)$ is \emph{pointwise} H{\"o}lder 
continuous with H{\"o}lder exponent 
$\frac{1}{2}$ at each point $\ga=(\tau,d)$ \wu{where} the
cost functions $\tau_a(\cdot)$ are \emph{Lipschitz continuous}
on the interval $[0,T(d)],$ see Theorem~\ref{thm:HalfHoelderContinuity}. Hence,
the change of the PoA is bounded from above by $\h_{\ga}\cdot \sqrt{\gnorm{\ga-\ga'}}$
for a \wu{H{\"o}lder} constant $\h_\ga>0$ 
when the network undergoes only a \emph{small} change $\gnorm{\ga-\ga'}$
at \wuu{a state $\ga=(\tau,d)$ with Lipschitz continuous cost functions
$\tau_a(\cdot)$ on 
$[0,T(d)].$} Since the interval $[0,T(d)]$ is compact, this result
applies obviously to all continuously differentiable cost functions.
Because of Theorem~\ref{thm:NotUniformlyHoelder}b), \wu{however, it may} not apply
when the network undergoes a large change \wusec{of} the cost functions and/or the demands in \WU{terms} of the metric.

To obtain this result, we show first that $|\rho(\ga)-\rho(\ga')|\wuu{\le}\h_{1,\ga}\cdot 
\sqrt{\gnorm{\ga-\ga'}}$ for a constant $\h_{1,\ga}>0$ when 
$\ga=(\tau,d)$ and $\ga'=(\sigma,d)$ have \emph{the same demand vector} $d,$
the \wusec{change} $\gnorm{\ga-\ga'}$ is small
and the cost functions $\tau_a(\cdot)$ of $\ga$ are Lipschitz continuous
on $[0,T(d)],$
see Lemma~\ref{lemma:PoA_Lipschiz}.
We then show that $|\rho(\ga)-\rho(\ga')|\wuu{\le}\h_{2,\ga}\cdot 
\sqrt{\gnorm{\ga-\ga'}}$ for a constant $\h_{2,\ga}>0$
when $\ga=(\tau,d)$ and $\ga'=(\tau,d')$ have \emph{the same cost functions} $\tau_a(\cdot)$, the
\wusec{change} $\gnorm{\ga-\ga'}$ is small and the cost functions
$\tau_a(\cdot)$ are Lipschitz continuous on $[0,T(d)],$ 
see Lemma~\ref{lemma:HalfHoelderCostSlice}.

Lemma~\ref{lemma:PoA_Lipschiz} \wusec{and} Lemma~\ref{lemma:HalfHoelderCostSlice}
then imply Theorem~\ref{thm:HalfHoelderContinuity}.
In particular,
Lemma~\ref{lemma:HalfHoelderCostSlice} generalizes
the sensitivity results of \citet{EnglertFraOlb2010} and \citet{TakallooKwon2020}
by considering more general cost functions and removing the requirement that 
the demands \wu{of different O/D pairs} increase synchronously by the same factor. \wusec{However,
this comes at the cost of}
a \wu{weaker} H{\"o}lder exponent $\frac{1}{2}$
than that of \citet{EnglertFraOlb2010} and \citet{TakallooKwon2020}, 
\wusec{who obtain a \wuu{H{\"o}lder} exponent of $1$}.

\wuu{The above H{\"o}lder continuity results build essentially upon Lemma~\ref{lem:EpsilonNE}c)
	in Section~\ref{subsec:Potential_epsilon_NE}, which shows 
	that the total cost of an $\epsilon$-approximate \emph{Wardrop equilibrium}
	(\citet{Wardrop1952ROAD}) deviates by at most $O(\sqrt{\epsilon})$ from that of a precise Wardrop equilibrium.
This is already a tight upper bound on the cost deviation for arbitrary Lipschitz continuous
cost functions on $[0,T(d)],$ see Example~\ref{example:Tightness_Of_epsilon_WE},
and
we are thus presently unable to improve the H{\"o}lder exponent $\frac{1}{2}$
in Theorem~\ref{thm:HalfHoelderContinuity}, Lemma~\ref{lemma:PoA_Lipschiz} and
Lemma~\ref{lemma:HalfHoelderCostSlice}, see Remark~\ref{remark:Exponent-LischpitzCost-Cannot-Improve}.
Nevertheless, a stronger H{\"o}lder exponent is still possible when the
cost functions have special properties similar to those in the work of
\citet{EnglertFraOlb2010}, \citet{Cominetti2019}, and
\citet{TakallooKwon2020}.}

\wusec{When the cost functions $\tau_a(\cdot)$
	of a point $\ga=(\tau,d)$  are constants
	or have strictly positive derivatives
	on the interval $[0,T(d)],$ we obtain the stronger result that}
$\rho(\cdot)$
is pointwise H{\"o}lder continuous with \wu{H{\"o}lder}
exponent $1$ at the point $\ga,$
 see Theorem~\ref{thm:Finer_PoA_Approx_Particular}.
Hence, the resulting change of the PoA is bounded from above by 
$\h_{\ga}\cdot \gnorm{\ga-\ga'}$ for 
a constant $\h_\ga>0$ when
the network undergoes a small change $\gnorm{\ga-\ga'}$ 
at \wuu{such a state $\ga$.} 
Again by Theorem~\ref{thm:NotUniformlyHoelder}, this \wu{may} not hold when the network undergoes a large
change \wusec{of} the cost functions and/or demands in \wu{terms} of
the metric.

    Finally, we demonstrate that \wusec{our} H{\"o}lder continuity
    results \wusec{also} help \wusec{to analyze the}  \emph{convergence \wusec{rate}
    of the PoA} in traffic networks, which is an
emerging research topic
    initiated recently by 
    \citet{Colini2017WINE,Colini2020OR}, see
    Section~\ref{subsec:RelatedWorks}
    or Section~\ref{subsec:ConvergenceAnalysis-Reference-Summary} for
    an overview of the related work on this new topic.
    
    \wusec{For $T(d)\to 0$,
    \citet{Colini2017WINE,Colini2020OR} have \wu{obtained} the first convergence result
    \wu{stating}  that the PoA converges to $1$ at a rate of 
   $O(T(d))$ when the cost functions are \wuu{of the form
   	$\tau_a(x)=\sum_{n\in\N}\xi_{a,n}\cdot x^n$}
   for each \WU{$a\in A$ and each $x\in [0,\infty)$,} and the demands of
all O/D pairs follow a \emph{specific} 
decreasing pattern, i.e., all of them  decrease to $0$ at the same rate of $\Theta(T(d)),$
see Section~\ref{subsec:RelatedWorks} or Section~\ref{subsec:ConvergenceAnalysis-Reference-Summary}.}
    With Theorem~\ref{thm:Finer_PoA_Approx_Particular}a),
    we show \wusec{a stronger result}
     that the PoA converges to $1$
    at a rate of $O(T(d))$ as $T(d)\to 0$ \emph{regardless
    of the decreasing pattern of the demands}
    when the cost functions are strictly positive and Lipschitz continuous
    within a small neighborhood
    around the origin $0$, see Corollary~\ref{corollary:ConvergencePoA_Light}.
    
    When $T(d)\to \infty,$ it has been shown by \wuu{\citet{Colini2017WINE,Colini2020OR} and} \citet{Wu2019}
    	that 
    	the PoA converges to $1$ as $T(d)\to\infty$ for arbitrary
    	regularly varying cost functions.
    For the special case of arbitrary polynomials 
as cost functions, \citet{Colini2017WINE,Colini2020OR} obtained 
the first convergence rate of $O(\frac{1}{T(d)})$
for the PoA when 
the demands of all O/D pairs grow to $\infty$  \wu{at}
the same rate of \wu{$\Theta(T(d))$.} For the more specific case of BPR cost functions (\citet{BPR})
with degree $\beta\ge 0$, \citet{Wu2019} showed a \wu{stronger} convergence rate of 
$O(\frac{1}{T(d)^{\beta}})$
for the PoA when the total demand $T(d)$ grows to $\infty.$
\wuu{See  Section~\ref{sec:Applications}}
for an overview of related results on the convergence \wu{rates}
of the PoA.
    With Theorem~\ref{thm:HalfHoelderContinuity}, we show that 
    the PoA converges to $1$ at 
    a rate of $O(\sqrt{1/\ln (T(d)+1)})$ as $T(d)\to \infty$
    for  \emph{regularly varying} (\citet{Bingham1987Regular})
    cost functions of the form $\zeta_a\cdot x^{\beta}\cdot \ln ^{\alpha}(x+1),$ $\beta>0,\ \alpha\ge 0,$ see Corollary~\ref{corollary:TailConvergence}.
    \wusec{This is} the first explicit convergence rate
    of the PoA in traffic networks with regularly varying cost functions
    \wusec{that are not polynomials}
    for the case $T(d)\to \infty$.

    Altogether, we have considerably enhanced the sensitivity results of \citet{EnglertFraOlb2010}, \citet{TakallooKwon2020}
    and \citet{Cominetti2019} by considering
    general traffic networks and  simultaneous changes
    of demands and cost functions.
    \wuu{Our} results 
    establish 
    the first sensitivity analysis of the PoA  for \emph{simultaneous changes} of demands and cost functions in traffic networks with multiple
    O/D pairs. 
    
    These sensitivity results \wuu{give also new insights into congestion pricing with tolls. Tolls
    change the cost functions, and so---due to our sensitivity results---tolls need to be considerable
in order to reduce the PoA significantly. This has, e.g., been observed by \citet{Harks2015Computing}.
They  consider tolls on a limited number of streets and use steepest descent methods to reduce the PoA. Their empirical calculations stabilize quickly with decreasing changes of the tolls, as justified in hindsight by our results. Furthermore, our results} 
    help to analyze the convergence rates
    of the PoA when the total demand $T(d)$ 
    tends to zero or infinity.
    
    Although we use the terminology
    of traffic networks in this paper, our analysis and results 
    do not depend on this view and carry over naturally to arbitrary non-atomic congestion games.
	
	\subsection{Related work}
	\label{subsec:RelatedWorks}
	
	
	
	\citet{Papadimitriou1999} proposed to quantify the inefficiency of equilibria in arbitrary congestion games from a worst-case perspective. This then resulted in the concept of the {\em price of
	anarchy} (PoA) that is usually defined as the ratio of the worst case equilibrium cost over
	the socially optimal cost, see  \citet{Papadimitriou2001Algorithms}.
	
    \subsubsection{Early development}
	
	A wave of research has been started with the pioneering paper of \citet{Roughgarden2000How}
	on the PoA in traffic networks with affine linear cost functions. Examples are \citet{Roughgarden2001Designing,Roughgarden2002The,Roughgarden2004Bounding,Roughgarden2005Selfish,Roughgarden2015Intrinsic,CorreaSchSti04,Correa2005On}. They
	investigated the worst-case upper bound of the PoA for different types of cost functions $\tau_a(\cdot)$, and
	analyzed the influence of the network topology on this bound. In particular, they showed that this bound 
	is $\tfrac{4}{3}$ when
	all $\tau_a(\cdot)$ are affine linear (\citet{Roughgarden2000How}), and
	\wu{$\Theta(\frac{\beta}{\ln \beta})$} when all $\tau_a(\cdot)$ are polynomials with maximum degree \wu{$\beta >0$} (\citet{Roughgarden2004Bounding} and \citet{Roughgarden2015Intrinsic}).
	Moreover, they proved that this bound
	is independent of the network topology, see,
	e.g., 
	\citet{Roughgarden2002The}.
	They also developed a $(\lambda,\mu)$-smooth
	method by which one can obtain 
	a {\em tight} and {\em robust} worst-case upper bound 
	of the PoA for a large class of
	cost functions, \wu{see, e.g.,} \citet{Roughgarden2002The}, \citet{Roughgarden2004Bounding} and \citet{Roughgarden2015Intrinsic}.
	This method was reproved by \citet{Correa2005On} from
	a geometric perspective. Moreover, \citet{Perakis2007} considered worst-case upper bounds for non-separable cost functions.
	See
	\citet{Roughgarden2007Introduction} for a comprehensive overview of the early development of that research.
	
	\subsubsection{Convergence analysis \wu{of} the PoA in traffic networks}Recent papers have empirically studied the PoA in traffic networks with \emph{BPR cost functions}
	(\citet{BPR}) and real \wusec{traffic} data.   
	\citet{Youn2008Price}
	observed that the empirical PoA
	in a traffic network depends crucially on the total demand. Starting from~1, it grows with some oscillations, and
	ultimately becomes 1 again as the total demand increases. A similar observation was made by \citet{O2016Mechanisms}.
	They even conjectured that the PoA converges to $1$ in the power law $1+O\big(T(d)^{-2\cdot\beta}\big)$ when the total
	demand $T(d)$ becomes large. \citet{Monnot2017} showed that traffic choices of commuting students in Singapore are near-optimal and that the empirical PoA is much smaller than known worst-case upper bounds. Similar observations have been reported by \citet{Jahn2004System}.


These empirical observations have been \wu{recently} confirmed  by \citet{Colini2016On,Colini2017WINE,Colini2020OR} and \citet{Wu2019}.
\citet{Colini2016On,Colini2017WINE,Colini2020OR} were
the first to theoretically analyze the convergence of the PoA in traffic networks with growing or
decreasing total
demand.

As a first step, \citet{Colini2016On} showed
the convergence of the PoA to $1$ as the total demand 
$T(d)\to\infty$ for traffic networks with a \emph{single} O/D pair
and regularly varying cost functions.
This convergence result was then substantially extended \wu{by} \citet{Colini2017WINE}
to traffic networks with multiple O/D pairs
for both the case $T(d)\to 0$ and the case $T(d)\to\infty,$
when the ratio of the demand of each O/D pair
over the total demand $T(d)$ remains positive as $T(d)\to 0$ or $\infty$.
Finally, \citet{Colini2020OR} 
extended these results to the cases
where the demands and the network together fulfill certain \emph{tightness and salience conditions}
that allow the ratios of demands to vary in a certain \wusec{pattern}
as $T(d)\to 0$ or $\infty.$ Moreover, they illustrated
by an example that the PoA need not converge to $1$ as $T(d)\to \infty$
when the cost functions are not regularly varying.
In particular, they obtained the first convergence rates of the PoA 
in traffic network with polynomial cost functions when the ratio
of the demand of each O/D pair over the total demand $T(d)$
\wusec{stays positive}
as $T(d)\to 0$ or $\infty.$
We \wu{will} come back to these rates in Section~\ref{sec:Applications}.

\citet{Wu2019} extended the \wu{work} of \citet{Colini2016On,Colini2017WINE,Colini2020OR}
\wu{for growing total demand} by a new framework. They showed for traffic networks
	with {\em arbitrary} regularly varying functions that the PoA converges to $1$
as the total demand tends to $\infty$ regardless of the growth pattern of the demands.
In particular, they proved a convergence rate of $O(T(d)^{-\beta})$ for BPR cost functions
\wu{of degree $\beta$}
and illustrated by examples that the conjecture proposed by \citet{O2016Mechanisms} need not hold.

\subsubsection{Sensitivity analysis of the PoA in traffic networks}



\wuu{First results on the sensitivity of the PoA in traffic networks
have been obtained recently by
%
\citet{EnglertFraOlb2010}, \citet{TakallooKwon2020}, and  \citet{Cominetti2019}.}

\citet{EnglertFraOlb2010}  considered traffic \wu{networks}
with a \emph{single}  \emph{origin-destination} (O/D) pair and \emph{polynomial} cost functions of degree at most
$\beta.$ They viewed the PoA as a function of the demand of \wu{that}
O/D pair \emph{when the polynomial cost functions \wusec{do not change.}}
They showed that the increase of this PoA function is bounded by a factor of $(1 + \epsilon)^\beta$ from above when the demand \wusec{increases} by a factor of $1 + \epsilon$.

\citet{TakallooKwon2020} generalized
this result to traffic networks
with \emph{multiple} O/D pairs and polynomial cost functions
of degree at most $\beta$. They \wu{proved} the same upper bound 
\wusec{on} the increase of the PoA when the demands of all O/D pairs
increase \emph{synchronously} by the {\em same} factor  $1 + \epsilon$ and \emph{the polynomial cost functions \wusec{do not change.}}
\wu{They also} showed that
the increase of the PoA is bounded by 
a factor $\frac{1}{(1+\epsilon)^{\beta}}$ from below under the same conditions.

Similar to \citet{EnglertFraOlb2010},
\citet{Cominetti2019} considered also traffic \wu{networks} with 
a \emph{single} O/D pair, and viewed the PoA as a function
of the demand of that O/D pair for \emph{fixed cost functions.}
For cost functions with  \emph{strictly positive}
derivatives, they showed that the
PoA function is differentiable at each demand
level where all the optimal paths carry a strictly positive
flow. For affine cost functions, they showed further
that the \emph{equilibrium cost} is piece-wise linear and
differentiable except at so-called \emph{$\mathcal{E}$-breakpoints}
whose number is finite though possibly exponentially large \wu{in the size of the network.}
In the interval between
	any two  consecutive $\mathcal{E}$-breakpoints, the PoA function
is differentiable, and either monotone or unimodal
with a unique minimum on the interior of \wusec{that} interval.
So the maximum of the PoA function is
attained at an $\mathcal{E}$-breakpoint.
They also presented several examples showing how these properties fail for general cost functions.


\subsubsection{\wuu{Sensitivity analysis of equilibria in traffic networks}}

\wuu{Related results on the sensitivity of  Wardrop equilibria have been obtained by \citet{Hall1978}, \citet{Patriksson2004},
\citet{Patriksson2007}, \citet{Lu2017},   \citet{Klimm2021} and others.
\citet{Hall1978} proved that the Wardrop equilibrium path cost of an O/D pair is continuous, and even montonically non-decreasing  with the growth
of its demand when both the demands of other
O/D pairs and the cost functions are fixed. Consequently, the total cost of Wardrop equilibria is a continuous function of the demands. Theorem~\ref{thm:PoA_Continuity} generalizes this continuity to the whole topological game space.}

\wuu{\citet{Patriksson2004}, \citet{Patriksson2007} and 
\citet{Lu2017} considered 
the sensitivity of Wardrop equilibria w.r.t. changes of parameters of 
the demands and the cost functions when both the demands and the cost functions 
are parametric and contain parameters defined on finite dimensional Euclidean spaces.
In this case, the  non-atomic congestion game is characterized completely by
 a parameter vector $\mu\in \R^n$ for a fixed integer $n>0$, and the Wardrop equilibrium arc flows and 
 cost are then functions of the parameter vector $\mu\in\R^n$ that map $\mu$ to ``points'' on 
 the Euclidean space $\R^A$.
 When the cost functions are differentiable,
\citet{Patriksson2004} characterized the existence of a directional derivative of the Wardrop equilibrium
solution (arc flow and arc cost) in an arbitrary direction of $\mu$. \citet{Patriksson2007} further improved
 \citet{Patriksson2004}, and showed that 
the Wardrop equilibrium arc cost is always directionally differentiable w.r.t. $\mu$, while the Wardrop equilibrium 
arc flow may not. Moreover,
\citet{Lu2017} derived
methods to calculate the semiderivatives of the Wardrop equilibrium arc flow
 w.r.t. $\mu$ under
general conditions, and the derivatives of the Wardrop equilibrium arc flow w.r.t. $\mu$ under 
more restrictive conditions.
Here, the \emph{semiderivative} of a function $h: \R^n\to \R^A$ at a point 
$\mu_0\in \R^n$ refers to a continuous and positively homogeneous function $\delta_{\mu_0}: \R^n\to \R^m$ s.t.
$h(\mu)=h(\mu_0)+\delta_{\mu_0}(\mu-\mu_0)+o(\|\mu-\mu_0\|)$ for each $\mu\in \R^n$
with sufficiently small $\|\mu-\mu_0\|.$ The vector $\delta_{\mu_0}(\mu-\mu_0)$
is then the \emph{directional derivative} of $h$ at $\mu_0$ w.r.t. direction $\mu-\mu_0.$
In particular, the \emph{derivative} of $h$ at $\mu_0$ exists and equals $\delta_{\mu_0}(\cdot)$ when 
the semiderivative $\delta_{\mu_0}(\cdot)$ at $\mu_0$ is a linear function.}

\wuu{The recent seminal work by \citet{Klimm2021}, see also the conference version  \citet{Klimm2019}, developed 
	a Katzenelson's homotopy method to compute all Wardrop equilibria for a non-atomic congestion game with piece-wise linear cost when the demand vector has the form $d=\lambda\cdot d^{(0)}$ 
	for a fixed direction $d^{(0)}\in (0,\infty)^{\K}$ and a variable parameter $\lambda\in (0.\infty).$
	They viewed the Wardrop equilibrium arc flow  as a function of the parameter $\lambda$, and proved  that this function is actually piece-wise linear in $\lambda$, and, in particular, may have exponentially many breakpoints when the cost functions are affine linear.}

\wuu{While these \wulast{parametric} sensitivity results of Wardrop equilibira are very interesting, they do not apply
	to our sensitivity analysis of the PoA, since neither the demands nor the cost functions
	are parametric. In fact, we consider the most general case that both the cost functions
	and the demands may vary arbitrarily, and so cannot be parameterized \wulast{by a finite dimensional Euclidean space.}
Nevertheless, they are very inspiring, and pave a feasible way for future work on the differentiability
of the PoA.}

	\subsection{Outline of the paper}
	\label{subsec:Arrangement}
The paper is organized as follows. We develop our approach for general non-atomic congestion games but with
the terminology of traffic networks. These are introduced in Section~\ref{sec:Model_Preliminary}. 
Section~\ref{sec:Topology} defines the metric and the topological space for games with the same combinatorial structure.
Section~\ref{sec:Theory} then presents  our techniques and sensitivity analysis results. 
Section~\ref{sec:Applications} demonstrates
that \wu{our} results \wu{also} facilitate the analysis of the
convergence rate of the PoA
when the total demand tends to $0$ or $\infty.$ We conclude with a
short summary and discussion  in Section~\ref{sec:Summary}.

	\section{Model and preliminaries}
	\label{sec:Model_Preliminary}
	\subsection{The model}
	\label{subsec:Model}
	
	We define an arbitrary non-atomic congestion game with the \wusec{terminology} of traffic games (see, e.g.,
		\citet{Nisan2007,Roughgarden2000How}), since this is more intuitive.
	A {\em non-atomic congestion game}  $\Gamma$ is then
	associated with a traffic network $G=(V,A)$, and 
    consists of a tuple $(\K,\S,\tau,d)$ with components defined in G1--G5 below.
    \begin{itemize}
    		\item[\textbf{G1}] $\K$ is a finite non-empty set of \emph{groups} or \emph{populations} of 
    		\emph{users}. 
    		Every group $k\in\K$ corresponds to a (transport) \emph{origin-destination}
    		(O/D) pair in the network $G.$
    		We will write an \wu{O/D pair} $k\in\K$ as $(o_k,t_k)\in V^2$ when the origin $o_k$ and
    		the destination $t_k$ are specified.
    		\item[\textbf{G2}] $\S=\cup_{k\in\K}\S_k,$ \wuu{where} each 
    		$\S_k\subseteq 2^A\setminus\emptyset$ denotes 
    		a non-empty collection of \emph{$(o_k,t_k)$-paths} that are
    		\emph{(pure) strategies} available \emph{only} to users of \wu{O/D pair} $k$.
    		Here, a $(o_k,t_k)$-path is a non-empty subset of the arc set $A$ 
    		\wusec{that} is loop-free and leads from the origin $o_k$ to the destination $t_k.$
    		The sets $\S_k$ are then \emph{mutually \wusec{disjoint,}} i.e., 
    		$\S_k\cap \S_{k'}=\emptyset$ for \wu{any two distinct
    		O/D pairs $k,k'\in\K.$}
    		\item[\textbf{G3}] $\tau=(\tau_a)_{a\in A}$ is a \emph{cost function vector}, in which each 
    		$\tau_a:[0,\infty)\to [0,\infty)$ \wu{is} a \emph{continuous} and \emph{non-decreasing} 
    		\emph{latency} or \emph{cost} function of \wu{arc} $a\in A$
    		\wuu{that depends on the flow value of arc $a$ and includes also all other extra cost for using arc $a$ such as
    			tolls.}
    		\item[\textbf{G4}] $d=(d_k)_{k\in\K}$ is a \emph{demand vector}
    		 whose component $d_k\ge 0$ \wuu{denotes} the \emph{demand} to be transported by (users of) O/D pair $k\in\K$
    		 using paths in $\S_k.$
    		 So $\Gamma$ has the \emph{total (transport) demand} $T(d):=\sum_{k\in \K}d_k.$
    		 \item[\textbf{G5}] Users are
    		 \emph{non-cooperative}. Each user of
    		 an arbitrary O/D pair $k\in\K$
    		 is \emph{infinitesimal}, i.e., controls an \emph{infinitesimal} fraction
    		 of the demand $d_k$, and must satisfy that by choosing
    		 a path $s\in\S_k$. The demand $d_k$
    		  will then be \emph{arbitrarily split} among
    		 paths in $\S_k$ for each \wu{O/D pair} $k\in\K.$
    \end{itemize}

    A
    (pure) \emph{strategy profile} (or simply, \emph{profile})
    over all users is expressed as
    a \emph{multi-commodity (path) flow} $f=(f_s)_{s\in \S}=(f_s)_{s\in\S_k,k\in\K}$
    of the network $G$
    with  $\sum_{s' \in S_k}f_{s'}=d_k$
    for each \wu{O/D pair} $k\in\K.$
    Herein, the flow value $f_s\ge 0$
    is just the demand \wusec{transported along the} path $s\in\S.$
  The \emph{flow value} $f_a$ of an arc $a \in A$ is then obtained as $f_a := \sum_{s\in \S: a\in s} f_s$. Hence,
   an arc $a\in A$ has the cost $\tau_a(f_a),$ a path
   $s\in\S$ has the cost $\tau_s(f):=\sum_{a\in A: a\in s}\tau_a(f_a),$
   and all users have the \emph{total cost}
   $C(\Gamma,f):=\sum_{s\in \S}f_s\cdot\tau_s(f)
   =\sum_{a\in A}f_a\cdot\tau_a(f_a)$.

	We call the triple $(G,\K,\S)$ the \emph{combinatorial structure} of  $\Gamma$,
	and denote $\Gamma$ \wusec{by} simply  the pair $(\tau,d)$ when its combinatorial structure 
	$(G,\K,\S)$ is given. 
	
	Viewed as a general non-atomic congestion
	game, the arcs $a\in A$ and paths $s\in\S$ correspond to 
	\emph{resources} and (pure) \emph{strategies} of users, see, e.g., 
	\citet{Dafermos1969}, \citet{Rosenthal1973A} and \citet{Correa2005On}.
	Although we chose to use the
	\wusec{terminology} of traffic games, the analysis and results are independent of this view
	and carry over to general non-atomic congestion games.
	
	

	\subsection{Equilibria, optimality and the price of anarchy}
	\label{subsec:NE_SO_PoA}

A flow $f^*=(f^*_s)_{s\in\S}$ 
of $\Gamma$
is called \wusec{a} {\em social optimum} (SO)
if it has the minimum total cost, i.e., $C(\Gamma,f^*)\le C(\Gamma,f)$ for
an arbitrary flow $f$ of $\Gamma.$ 

A flow $\tilde{f}=(\tilde{f}_s)_{s\in\S}$ of $\Gamma$
is called a \emph{Wardrop equilibrium} (WE) if 
it fulfills \emph{Wardrop's first principle} (\citet{Wardrop1952ROAD}), i.e.,
	\begin{equation}\label{def:WE}
	\forall k\in \K \ \forall s,s'\in \S_k \Big(
	\tilde{f}_s>0\implies \tau_s(\tilde{f})\le 
	\tau_{s'}(\tilde{f})\Big).
	\end{equation}
Clearly, every WE flow $\tilde{f}$ of $\Gamma$ satisfies \wuu{condition~\eqref{eq:UserCost}:}
		\begin{equation}\label{eq:UserCost}
		\forall k\in \K\ \forall s\in \S_k: \quad 
		\tilde{f}_s>0\implies \tau_s(\tilde{f})=L_k(\tau,d):=\min_{s'\in \S_k}\tau_{s'}(\tilde{f}).
		\end{equation}
		We  call the constant $L_k(\tau,d)$ in condition~\eqref{eq:UserCost}
		the {\em user cost} of O/D pair $k\in \K.$ The total cost of  WE flows $\tilde{f}$ is then expressed
		equivalently by
		\begin{equation}\label{eq:WE-TotalCost-Equivalent}
			C(\Gamma,\tilde{f})=\sum_{s\in \S}\tilde{f}_s\cdot\tau_s(\tilde{f})=\sum_{k\in \K}L_k(\tau,d)\cdot d_k.
		\end{equation}
		
		Trivially, a flow $\tilde{f}$ of $\Gamma$ is a WE if and only if $\tilde{f}$ satisfies
		the \emph{variational inequality} 
		\begin{equation}\label{eq:WE-Variational-Inequality}
			\sum_{a\in A}\tau_a(\tilde{f}_a)\cdot (g_a-\tilde{f}_a)=\sum_{s\in \S}\tau_s(\tilde{f})\cdot (g_s-\tilde{f}_s)>0
		\end{equation}
		for an arbitrary flow $g=(g_s)_{s\in\S}$ of $\Gamma,$ see, e.g., \citet{Dafermos1980}.

		Since the cost functions $\tau_a(\cdot)$ are non-negative, continuous and non-decreasing, and \wuu{since} the users are infinitesimal,
		$\Gamma$ has \emph{essentially unique} WE flows,
		i.e., $\tau_a(\tilde{f}_a)=\tau_{a}(\tilde{f}'_a)$
		for all $a\in A$ for two arbitrary   WE flows
		$\tilde{f}$ and $\tilde{f}'$ of $\Gamma,$ see, e.g., \citet{Schmeidler1973Equilibrium}, \citet{Smith1979The}
	and \citet{Roughgarden2000How}. 
	In particular, WE flows of $\Gamma$ coincide with \emph{(pure) Nash equilibria} (NE) of $\Gamma$,
     see, e.g., \citet{Roughgarden2000How} for a definition of NE flows
     in non-atomic congestion games.


 When all 
	cost functions $\tau_a(\cdot)$ are
	 differentiable, an SO flow $f^*$
	is also a WE flow w.r.t.\ the {\em marginal \wu{cost} functions} 
	$
		c_a(x):=x\cdot \tau'_a(x)+\tau_a(x),
	$ and, vice versa, a WE flow is an SO flow w.r.t. the cost functions 
\begin{displaymath}
	\frac{1}{x} \int_0^{x} \tau_a(\xi)d\xi,
\end{displaymath}
	see, e.g., \citet{Beckmann1956} or \citet{Roughgarden2000How}.
	Hence, SO flows
coincide with WE flows when all cost functions $\tau_a(\cdot)$ are monomials of
the same degree $\beta\ge 0.$

    
    In general, SO flows and WE flows differ, see, e.g., \citet{Roughgarden2000How}.
    The ratio of the worst-case total cost of
    a WE flow over the total cost of an SO flow
    is known as the \emph{Price of Anarchy} (PoA, see \citet{Papadimitriou1999} and \citet{Papadimitriou2001Algorithms}), \wu{and}
    measures the inefficiency of WE flows.
	Formally, the PoA of $\Gamma$ is \wu{defined as}
	\begin{equation}\label{eq:PoA}
	\rho(\Gamma):=\frac{C(\Gamma,\tilde{f})}{C(\Gamma,f^*)}
	=\frac{\sum_{s\in \S} \tilde{f}_s\cdot \tau_s(\tilde{f})}{\sum_{s\in \S} f^*_s\cdot \tau_s(f^*)},
	\end{equation}
	where $\tilde{f}$ and $f^*$ are an \emph{arbitrary} WE flow and an \emph{arbitrary} SO flow
	of $\Gamma,$ respectively. Definition~\eqref{eq:PoA} is
	unambiguous since WE flows are essentially unique.
	
	\subsection{Potential functions and $\epsilon$-approximate equilibria}
	\label{subsec:Potential_epsilon_NE}
	A non-atomic congestion game
	$\Gamma$ is actually a {\em potential game},
	\wu{see, e.g.,} \citet{Sandholm2001}. The  \wuu{(Rosenthal)} {\em potential function}
	of $\Gamma$ is given by
	\begin{equation}\label{eq:PotentialFunction}
		\Phi(\Gamma,f)=\sum_{a\in A}\int_{0}^{f_a}\tau_a(x)\ dx,
	\end{equation}
	\wusec{and} reaches its global minimum 
	at its WE flows $\tilde{f}$, see, e.g., \citet{Roughgarden2000How}.
	
	A flow $f$ is an \emph{$\epsilon$-approximate WE flow} of $\Gamma$ for a small constant $\epsilon>0,$ if 
	\wuu{\begin{equation}\label{eq:Epsilon-Approximate-WE}
		\sum_{a\in A}\tau_a(f_a)\cdot (f_a-g_a)=\sum_{s\in\S}\tau_s(f)\cdot (f_s-g_s)=\sum_{k\in\K}\sum_{s\in\S_k}\tau_s(f)\cdot (f_s-g_s)\le \epsilon
	\end{equation} for  an arbitrary}
	flow $g=(g_s)_{s\in \S}$ of $\Gamma$.
	\WU{The variational inequality~\eqref{eq:Epsilon-Approximate-WE}
	implies that 
	\begin{equation}\label{eq:WE-approximate-second-def}
		\tau_s(f)\le \tau_{s'}(f)+\frac{\epsilon}{
			\min_{s''\in \S:\ f_{s''}>0}f_{s''}}
		=\tau_{s'}(f)+\Theta(\epsilon)
	\end{equation}
	for two arbitrary paths $s,s'\in\S_k$ with $f_s>0,$ \wuu{for}
	every O/D pair $k\in\K.$
	Inequality~\eqref{eq:WE-approximate-second-def} means that a unilateral change of paths 
	in an $\epsilon$-approximate WE flow $f$ 
	reduces the cost by \emph{at most} $\Theta(\epsilon),$
	and so $f$	indeed approximates a WE flow. 
	 In principle, we can alternatively define 
	an $\epsilon$-approximate WE flow directly by inequality~\eqref{eq:WE-approximate-second-def}
    } (as was done, e.g., by \citet{Roughgarden2000How}).
\WU{But this does not considerably influence
our analysis.} 
\WU{We thus stick to the variational inequality definition~\eqref{eq:Epsilon-Approximate-WE}, since
it, together with the variational inequality~\eqref{eq:WE-Variational-Inequality}, facilitates the cost comparison between an $\epsilon$-approximate
WE flow and a precise WE flow, see, e.g., Lemma~\ref{lem:EpsilonNE}b).}
	

	Lemma~\ref{lem:EpsilonNE} shows some useful properties of 
	$\epsilon$-approximate WE flows. \WU{Here, a real-valued function $h(\cdot)$ is {\em Lipschitz \wusec{continuous}} on an interval $I\subseteq [0,\infty)$
		with  \emph{Lipschitz constant} $M$ if 
		$|h(x)-h(y)|\le M\cdot |x-y|$ for all $x,y\in I.$}
	
	\begin{lemma}\label{lem:EpsilonNE}
		Consider an arbitrary non-atomic congestion game $\Gamma$
		with cost function vector $\tau$ and demand vector $d,$
		and a constant $\epsilon>0.$
		Let $f$ be an $\epsilon$-approximate WE flow, and let $\tilde{f}$ be a WE flow.
		Then $f$ and $\tilde{f}$ fulfill the
		following conditions.
		\begin{itemize}
			\item[a)] For each O/D pair $k\in \K,$ 
			$0\le  \sum_{s \in S_k}f_s\cdot \tau_s(f) -
				d_k\cdot \min_{s'\in \S_k}\tau_{s'}(f)<\epsilon.$
			Moreover, $\sum_{k\in \K}d_k\cdot \min_{s'\in \S_k}\tau_{s'}(f) \le  C(\Gamma,f)\le \sum_{k\in \K}d_k\cdot \min_{s'\in \S_k}\tau_{s'}(f)+\epsilon.$
			\item[b)] $0\le \sum_{a\in A} \tau_a(\tilde{f}_a)\cdot (f_a-\tilde{f}_a)\le  \Phi(\ga,f)-\Phi(\ga,\tilde{f})\le 
			\sum_{a\in A}\tau_a(f_a)\cdot (f_a-\tilde{f}_a)<\epsilon,$ and thus 
			$0\le \sum_{a\in A}|\tau_a(f_a)-\tau_a(\tilde{f}_a)|\cdot 
			|f_a-\tilde{f}_a|<\epsilon.$
			\item[c)] If every $\tau_a(\cdot)$
			is Lipschitz continuous on $[0,T(d)]$ with Lipschitz constant
				$M>0$, then 
			$|\tau_a(f_a)-\tau_a(\tilde{f}_a)|
			<\sqrt{M\cdot \epsilon}$
			for all \wu{arcs} $a\in A,$
			and $|L_k(\tau,d)- \min_{s'\in\S_k}\tau_{s'}(f)|\le |A|\cdot \sqrt{M\cdot\epsilon}$
			for all \wu{O/D pairs} $k\in\K.$ Furthermore, $|C(\Gamma,f)-C(\Gamma,\tilde{f})|\le |A|\cdot \sqrt{M\cdot \epsilon}\cdot T(d)
			+\epsilon.$
		\end{itemize}
	\end{lemma}

	Lemma~\ref{lem:EpsilonNE}a) follows
	trivially from inequality~\eqref{eq:Epsilon-Approximate-WE}.
	Lemma~\ref{lem:EpsilonNE}b) follows directly from
	\eqref{eq:WE-Variational-Inequality}, \eqref{eq:PotentialFunction}--\eqref{eq:Epsilon-Approximate-WE} and the fact that 
	\begin{equation}\label{eq:Lemma1-b}
	\tau_a(x)\cdot (y-x)\le \int_{x}^{y}\tau_a(z)\ dz
	\le \tau_a(y)\cdot (y-x)\quad\ \wu{\forall a\in A\ \forall x\in\R_{\ge 0}\ \forall y\in\R_{\ge 0}.}
	\end{equation}
	Herein, $\R_{\ge 0}:=[0,\infty).$ Inequality
	\eqref{eq:Lemma1-b} holds because every cost function 
	$\tau_a(\cdot)$ is non-decreasing, non-negative and continuous. 
	Lemma~\ref{lem:EpsilonNE}c) is a direct 
	\wu{consequence} of Lemma~\ref{lem:EpsilonNE}a)--b). It \wuu{yields} an approximation bound when all cost functions are Lipschitz continuous
	on the compact interval $[0,T(d)],$
	\wuu{which plays a pivotal role in the H{\"o}lder continuity
	analysis of the PoA in Section~\ref{subsec:Continuity_PoA}.
    Although this approximation bound is rather trivial, it is
    tight, see Example~\ref{example:Tightness_Of_epsilon_WE} below.}
	
	\wuu{\begin{example}\label{example:Tightness_Of_epsilon_WE} We illustrate the tightness of
		Lemma~\ref{lem:EpsilonNE}c) with Pigou's game
		(\citet{Pigou1920}). Pigou's game $\Gamma$
		has one unit of total demand and the simple traffic network shown in Figure~\ref{fig:Example-Tight-Epsilon-WE}.
		It thus has the unique WE flow 
		$\tilde{f}=(\tilde{f}_u,\tilde{f}_\ell)=(1,0),$
		where $u$ and $\ell$ denote the upper and lower arcs (paths), respectively.
		Let $\epsilon>0$ be an arbitrary
		small constant, and put 
		$f=(f_u,f_\ell)=(1-\sqrt{\epsilon},
		\sqrt{\epsilon}).$
		Then $f$ is an $\epsilon$-approximate WE flow, since  
		\begin{displaymath}
			\begin{split}
			\tau_u(f_u)\cdot (f_{u}-g_u)+&\tau_\ell(f_\ell)\cdot (f_{\ell}-g_\ell)
			=(1-\sqrt{\epsilon})\cdot (1-\sqrt{\epsilon}-g_u)+\sqrt{\epsilon}-g_\ell\\
			&=1-2\cdot \sqrt{\epsilon}+\epsilon-g_u+\sqrt{\epsilon}\cdot g_u+\sqrt{\epsilon}-g_\ell
			=\epsilon-\sqrt{\epsilon}\cdot (1-g_u)\le \epsilon.
			\end{split}
		\end{displaymath}
	for an arbitrary
	flow $g=(g_u,g_\ell).$
		Here, we use that $g_u+g_\ell=1$
		for each $g.$
		Furthermore, $|\tau_u(\tilde{f}_u)-\tau_u(f_u)|
		=|\tilde{f}_u-f_u|=\sqrt{\epsilon}$
		and $|C(\Gamma,f)-C(\Gamma,\tilde{f})|=
		\sqrt{\epsilon}-\epsilon\in \Theta(\sqrt{\epsilon}),$
		which shows that Lemma~\ref{lem:EpsilonNE}c)
		is tight.
		\begin{figure}[!htb]
			\centering
			\begin{tikzpicture}[
			>=latex
			]
				\node[scale=0.4,circle,color=black,fill=black,label=left:$o$](o){};
				\node[right = of o](m){};
				\node[scale=0.4,circle,color=black,fill=black,label=right:$t$,right = of m](t){};
				\node[above of= m](tau1){$x$};
				\node[below of= m](tau2){$1$};
				\draw[-,thick] (o) to [out =90, in= 180] (tau1);
				\draw[->,thick] (tau1) to [out=0,in=90] (t);
				\draw[-,thick] (o) to [out=-90,in=180] (tau2);
				\draw[->,thick] (tau2) to [out=0,in=-90] (t);
			\end{tikzpicture}
			\caption{The Pigou's game}
			\label{fig:Example-Tight-Epsilon-WE}
		\end{figure}
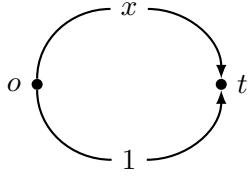
	\end{example}}

\wuu{Because of Lemma~\ref{lem:EpsilonNE}c),
we may thus want to find for a given flow $f$ an $\epsilon>0$ such that $f$ is an $\epsilon'$-approximate WE flow for 
	all $\epsilon'>\epsilon,$ but not for all $\epsilon'\in [0,\epsilon).$ We call such a constant $\epsilon\in (0,\infty)$ the
\emph{approximation threshold} of flow $f$ w.r.t. WE flows
$\tilde{f}$ of a non-atomic congestion game $\ga=(\tau,d)$, and denote it by
$\epsilon(\tau,d,f)$ to indicate its dependence on $\tau,$ $d$ and $f$.
Note that $\epsilon(\tau,d,f)=\sum_{k \in \K}\sum_{s \in S_k}[\tau_s(f)-L_k(\tau,d,f)]\cdot f_s$
for each flow $f,$
where $L_k(\tau,d,f):=\min_{s\in\S_k}\ \tau_s(f)$ is the minimum path cost
of O/D pair $k\in\K$ in flow $f.$ 
This follows since 
\begin{displaymath}
	\sum_{k \in \K}L_k(\tau,d,f)\cdot d_k
	=\min_{g\text{ is a flow of } \Gamma} 
	\sum_{k \in \K}\sum_{s \in S_k} \tau_s(f)\cdot g_s.
\end{displaymath}
Lemma~\ref{lem:EpsilonNE}c) with this approximation threshold $\epsilon(\tau,d,f)$ may then 
yield a tight upper bound
of $|C(\Gamma,f)-C(\Gamma,\tilde{f})|$ for arbitrary 
Lipschitz continuous cost functions on $[0,T(d)]$, see Example~\ref{example:Tightness_Of_epsilon_WE}. 
We will obtain a tight upper bound for this approximation threshold in Section~\ref{subsec:Continuity_PoA}.}

	\section{\wusec{The} topological space \wusec{of all} non-atomic congestion games with the same
	combinatorial structure}
	\label{sec:Topology}
	\wuu{In the sequel, we} fix an \emph{arbitrary} combinatorial structure 
	$(G,\K,\S)$ and consider only non-atomic congestion games
	with \wuu{this combinatorial structure.} 
	A non-atomic congestion game $\ga$ is then 
	simply \WU{specified} by the pair $(\tau,d).$

	\subsection{Assumptions} 
	 To avoid unnecessary \wusec{discussions,} we assume that the fixed combinatorial structure~$(G,\K,\S)$
	 satisfies Condition~1 below.
	\begin{description}
		\item[\textbf{Condition~1}] 
		Every arc $a\in A$ belongs to 
		some path $s\in\S=\cup_{k'\in\K}\S_{k'},$
		and $|\S_{k}|\ge 2$ for each \wu{O/D pair} $k\in\K$.
	\end{description} 

Condition~1 can be required w.l.o.g., as arcs $a\in A$ with 
$a\notin s$ for each $s\in\S$ play no role in a non-atomic congestion game $\ga=(\tau,d)$, and 
an O/D pair $k\in\K$ with a singleton path set $\S_k=\{s\}$ can be 
removed by using 
$\tau_a(x+d_k)$ instead of $\tau_a(x)$ for each arc $a$
belonging to the unique path $s$ in $\S_k.$

    The PoA $\rho(\ga)$ is obviously 
	$\frac{0}{0}$ when the total demand $T(d)=0.$
	In fact, even if $T(d)>0,$ the PoA may \wuu{still} be $\frac{0}{0}$
	when we allow $\tau_a(x)=0$ for some $a\in A$ and some 
	$x\in (0,\infty)$.
	To avoid these undefined cases of the PoA, we assume that an arbitrary
	non-atomic congestion game
	$\ga=(\tau,d)$ satisfies Condition~2 below.
	\begin{description}
		\item[\textbf{Condition~2}] $T(d)>0$ and $\tau_a(x)>0$ for 
		\wu{all $a\in A$ and all $x\in (0,T(d)].$}
	\end{description} 

   Lemma~\ref{lemma:SO_Cost_Positive} shows that $\rho(\Gamma)\in [1,\infty)$ is well defined
   for each non-atomic congestion game $\Gamma=(\tau,d)$ fulfilling Condition~2. 
   The proof of Lemma~\ref{lemma:SO_Cost_Positive} is trivial, and thus omitted.

\begin{lemma}\label{lemma:SO_Cost_Positive}
	Consider an arbitrary non-atomic congestion game $\Gamma=(\tau,d)$
	fulfilling Condition~2. Let $f, \tilde{f}$ and 
		$f^*$ be an arbitrary flow,
		a WE flow 
		and an SO flow of $\Gamma,$ respectively. Then $f_a\in [0,T(d)]$
	for all \wu{arcs} $a\in A,$  and
	\[
	0<\frac{T(d)}{|\S|}\cdot \min_{a\in A}\tau_{a}\Big(\frac{T(d)}{|\S|}\Big) \le  C(\Gamma,f^*) \le C(\Gamma,\tilde{f})\le |A|\cdot T(d)\cdot \max_{a\in A}
	\tau_{a}\big(T(d)\big).
	\]
	So \wuu{$\rho(\Gamma)\in [1, \frac{|A|\cdot |\S|
		\cdot \max_{a\in A} \tau_a(T(d))}{\min_{a\in A} \tau_a(T(d)/|\S|)}]\subseteq [1,\infty)$} and \wu{is thus} well defined.
\end{lemma}

As the definitions of cost functions $\tau_a(x)$ on
the unbounded interval $(T(d),\infty)$ play no role
in a non-atomic congestion game $\ga=(\tau,d),$ we define an \emph{equivalence} relation
between non-atomic congestion games as \wu{follows.}
\begin{description}
	\item[\textbf{Equivalence}]
	Two non-atomic congestion games 
	$\ga=(\tau,d)$ and $\ga'=(\sigma,d')$ are \emph{equivalent}, denoted by
	\wu{$\ga\simeq\ga',$}
	when \begin{equation}\label{eq:GameEquivalence}
	d_k=d'_k\ \forall k\in\K\quad \text{and}\quad 
	\tau_a(x)=\sigma_a(x)\ \wuu{\forall a\in A\ \forall x\in [0,T(d)]=[0,T(d')].}
	\end{equation}
\end{description}
While the cost function values $\tau_a(x)$ and $\sigma_a(x)$ might differ \wuu{largely} on the unbounded interval $ (T(d),\infty),$
the corresponding non-atomic congestion games $\ga=(\tau,d)$ and $\ga'=(\sigma,d)$ have the same game-theoretic properties when 
 \wu{$\ga\simeq \ga'$}.


	\subsection{The game space, \wu{the} metric and \wu{the} topology}
	\label{subsec:Formalization_PoA_Undefined}
	
	\wu{We now introduce a topology \wuu{for} the space of
	all non-atomic congestion games defined on the combinatorial
structure $(G,\K,S).$}
	\wu{All topological notions not explicitly defined here are standard, and
	we \wuu{recommend} \citet{Kelley1975Topology} as a standard reference
	\wuu{for them}.}
	
	We denote by $\G(G,\K,\S)$ the set of all non-atomic congestion games $\ga=(\tau,d)$ 
	\wu{that have} the fixed combinatorial structure $(G,\K,\S)$ and \wu{satisfy} 
	Condition~2. We call $\Gamespace$ the $(G,\K,\S)$-\emph{game space} (or, simply,
	\wu{the} \emph{game space}), and call each non-atomic congestion
	game $\ga=(\tau,d)\in\Gamespace$
	a \emph{game}, or a \WU{\emph{point}} of the game space. 
	
	Clearly, 
	\begin{equation}\label{eq:GeneralizedGameSpace}
	\G(G,\K,\S)\subsetneq \overline{\G}
	(G,\K,\S):=\mathcal{C}_+^{\uparrow}(\R_{\ge 0})^A\times \R_{\ge 0}^\K, 
	\end{equation}
	where $\mathcal{C}_+^{\uparrow}(\R_{\ge 0})$ denotes
	the set of all non-decreasing, non-negative and continuous functions
	defined on $\R_{\ge 0}=[0,\infty),$  $\mathcal{C}_+^{\uparrow}(\R_{\ge 0})^A:=\{(\tau_a)_{a\in A}:
	\tau_a\in \mathcal{C}_+^{\uparrow}(\R_{\ge 0})\ \forall a\in A\}$
	and $\R_{\ge 0}^\K:=\{d=(d_k)_{k\in\K}: d_k\in \R_{\ge 0}\ \forall k\in\K\}.$
	We call the super set $\overline{\G}(G,\K,\S)$ in~\eqref{eq:GeneralizedGameSpace} the \emph{generalized game space} and
	each element $\bar{\Gamma}=(\tau,d)\in \overline{\G}(G,\K,\S)$ a \emph{generalized game}.
	Trivially, $\GGamespace$ is the collection
	of all non-atomic congestion games \wu{with}
	the fixed combinatorial structure $(G,\K,\S).$
	\wuu{Note that} the PoA \wuu{may be} \emph{undefined} for some generalized games $\gga\in\GGamespace\setminus\Gamespace.$
	

	
	We now define \wu{a ``distance" on the generalized game space by} the binary operator $\text{Dist}: \GGamespace\times\GGamespace\to [0.\infty)$ with 
	\begin{equation}\label{eq:GeneralizedNorm}
		\text{Dist}(\gga,\gga'):=\max\left\{|\!|d-d'|\!|_{\infty},
		\max_{a\in A}\max_{x\in [0,\min\{T(d),T(d')\}]}|\tau_a(x)-\sigma_a(x)|
		,|\!|\tau(T(d))-\sigma(T(d'))|\!|_{\infty}\right\}
	\end{equation}
	for \wusec{two} arbitrary generalized games $\gga=(\tau,d)$ and
	$\gga'=(\sigma,d').$ Here, $\norm{y}:=\max_{i=1}^{n}|y_n|$ \wu{is} the \emph{$L^\infty$-norm}
	for an arbitrary vector $y=(y_1,\ldots,y_n)$ with an arbitrary length
	$n\in\N$, and
	$\tau(T(d))$ and $\sigma(T(d))$ \wu{are} the respective vectors $(\tau_a(T(d)))_{a\in A}$
	and $(\sigma_a(T(d)))_{a\in A}.$ 
	To \wu{simplify} notation, we denote by
	\begin{equation}\label{eq:restricted-function-norm}
		\norm{\tau_{|T(d)}-\sigma_{|T(d')}}:=\max\left\{
		\max_{a\in A}\max_{x\in [0,\min\{T(d),T(d')\}]}|\tau_a(x)-\sigma_a(x)|
		,|\!|\tau(T(d))-\sigma(T(d'))|\!|_{\infty}\right\}
	\end{equation}
	the ``distance'' between $\tau$ and $\sigma$ w.r.t.
	the restricted domains $[0,T(d)]$ and $[0,T(d')].$
	Then $\text{Dist}(\gga,\gga')$ in \eqref{eq:GeneralizedNorm} is expressed 
	equivalently as
	\begin{equation}\label{eq:Metric-Equivalence}
		\text{Dist}(\gga,\gga')=\max\left\{\norm{d-d'},\norm{\tau_{|T(d)}-\sigma_{|T(d')}}\right\}.
	\end{equation}
	\wu{Note that} $\text{Dist}(\cdot,\cdot)$ is \emph{consistent} with
	equivalence relation~\eqref{eq:GameEquivalence}, i.e., $\text{Dist}(\gga,\gga'')=\text{Dist}(\gga',\gga'')$
	for \wu{three arbitrary generalized games} $\gga,\gga',\gga''\in\GGamespace$
	with $\gga\simeq \gga'$.

	Lemma~\ref{lemma:Metric} shows that $\text{Dist}(\cdot,\cdot)$
	is actually a \emph{metric} on $\GGamespace$
	w.r.t. equivalence relation~\eqref{eq:GameEquivalence}.
	We thus denote $\text{Dist}(\gga,\gga')$ \emph{symbolically} by $\gnorm{\gga-\gga'}$
	for \wu{two} arbitrary \wu{generalized games} $\gga,\gga'\in \GGamespace,$ as
	this is a more intuitive way to denote the metric despite of
	\wusec{the} undefined \wusec{operator} $\gga-\gga'$.
	\begin{lemma}\label{lemma:Metric}
		Consider now \wu{three arbitrary generalized games} $\gga_1,\gga_2,\gga_3\in\GGamespace.$ \wu{Then:}
		\begin{itemize}
			\item[a)] $\text{Dist}(\gga_1,\gga_2)=\text{Dist}(\gga_2,\gga_1)\ge 0.$
			\item[b)] $\text{Dist}(\gga_1,\gga_2)=0$ if and only if 
			\WU{$\gga_1\simeq \gga_2$}.
			\item[c)] $\text{Dist}(\gga_1,\gga_2)\le \text{Dist}(\gga_1,\gga_3)+\text{Dist}(\gga_3,\gga_2).$
		\end{itemize}
	\end{lemma}
    \proof{Proof of Lemma~\ref{lemma:Metric}} Lemma~\ref{lemma:Metric}a)--\wu{\ref{lemma:Metric}b)} \wuu{are} trivial.
    We prove only Lemma~\ref{lemma:Metric}c).
    Assume, w.l.o.g., that $\gga_1=(\tau^{(1)}=(\tau^{(1)}_a)_{a\in A},d^{(1)}),$
    $\gga_2=(\tau^{(2)}=(\tau^{(2)}_a)_{a\in A},d^{(2)})$ and $\gga_3=(\tau^{(3)}=(\tau^{(3)}_a)_{a\in A},d^{(3)}).$
    Define 
    \begin{displaymath}
    \sigma^{(i)}:=(\sigma_a^{(i)})_{a\in A}\text{ with }
    	\sigma^{(i)}_a(x):=
    	\begin{cases}
    	\tau^{(i)}_a(x)&\text{if }x\le T(d^{(i)}),\\
    	\tau_a^{(i)}(T(d^{(i)})) &\text{if }x>T(d^{(i)}),
    	\end{cases}
    	\quad \forall x\in \R_{\ge 0}\ \forall a\in A\ \forall i=1,2,3.
    \end{displaymath}
    \wusec{Let} $\gga'_1=(\sigma^{(1)},d^{(1)}),$ 
    $\gga'_2=(\sigma^{(2)},d^{(2)})$ and 
    $\gga'_3=(\sigma^{(3)},d^{(3)}).$
    Then \wu{$\gga_i\simeq \gga'_i$} for $i=1,2,3.$
    
    \wusec{Let} $T_{\max}:=\max\{T(d^{(1)}),T(d^{(2)}),T(d^{(3)})\}.$
    Then 
    \begin{displaymath}
    	\norm{\sigma_{|T(d^{(i)})}^{(i)}-\sigma_{|T(d^{(j)})}^{(j)}}
    	=\max_{a\in A}\max_{x\in [0,T_{max}]}|\sigma_a^{(i)}(x)-\sigma_{a}^{(j)}(x)|
    	=\norm{\sigma_{|T_{max}}^{(i)}-\sigma_{|T_{\max}}^{(j)}}\quad
    	\forall i,j=1,2,3.
    \end{displaymath}
    Hence,
    \begin{displaymath}
    	\text{Dist}(\gga'_i,\gga'_j)=\max\left\{
    	\norm{d^{(i)}-d^{(j)}},\norm{\sigma_{|T_{max}}^{(i)}-\sigma_{|T_{\max}}^{(j)}}
    	\right\}\quad\forall i,j=1,2,3.
    \end{displaymath}
    \wusec{The triangle inequality} 
    $\text{Dist}(\gga'_1,\gga'_2)\le \text{Dist}(\gga'_1,\gga'_3)+\text{Dist}(\gga'_3,\gga'_2)$
    follows from \eqref{eq:Metric-Proof}.
    \begin{equation}\label{eq:Metric-Proof}
    	\begin{split}
    	&\norm{d^{(1)}-d^{(2)}}\le \norm{d^{(1)}-d^{(3)}}+\norm{d^{(3)}-d^{(2)}}\\
    	&\norm{\sigma_{|T_{max}}^{(1)}-\sigma_{|T_{\max}}^{(2)}}\le \norm{\sigma_{|T_{max}}^{(1)}-\sigma_{|T_{\max}}^{(3)}}+\norm{\sigma_{|T_{max}}^{(3)}-\sigma_{|T_{\max}}^{(2)}}
    	\end{split}
    \end{equation}

    Lemma~\ref{lemma:Metric}c) then follows \wu{since} the operator $\text{Dist}(\cdot,\cdot)$
    is consistent with equivalence relation~\eqref{eq:GameEquivalence}.
    
    \hfill$\square$
    
    \wuu{\begin{remark}\label{remark:NewMetric-IS-Impossible}
    	Note that we cannot substitute the cost function distance~\eqref{eq:restricted-function-norm} in the metric \eqref{eq:GeneralizedNorm} (equivalently, \eqref{eq:Metric-Equivalence})
    	by the $L^\infty$-norm
    	\begin{equation}\label{eq:NewCostNorm}
    		\gnorm{\tau_{|T_{max}}-\sigma_{|T_{max}}}_{\infty}:=
    		\max_{a\in A,\ x\in [0,T_{max}]} |\tau_a(x)-\sigma_a(x)|,\quad T_{max}:=\max\{T(d),T(d')\},
    	\end{equation}
    although this is more intuitive and has been applied to the auxiliary cost functions
    $\sigma^{(i)}$ in the proof of Lemma~\ref{lemma:Metric}c). The reason is that the resulting binary operator
    \begin{equation}\label{eq:NewMetric}
    D(\gga',\gga)=	\max\{\gnorm{d-d'}_\infty,\gnorm{\tau_{|T_{max}}-\sigma_{|T_{max}}}_{\infty}\},\quad 
    	\forall \gga,\gga'\in \GGamespace,
    \end{equation}
is inconsistent with equivalence relation \eqref{eq:GameEquivalence}, as 
$D(\gga',\gga'')=D(\gga,\gga'')$ need not hold for three arbitrary generalized games
$\gga,\ \gga',$ and $\gga''$ with 
$\gga\simeq \gga'.$
Moreover, $D(\cdot,\cdot)$
    does not fulfill the \emph{triangle inequality}
    in Lemma~\ref{lemma:Metric}c), and
    is thus neither a metric nor a pseudo-metric and so does not induce a metric space.
    To see this, we consider three arbitrary generalized games $\gga=(\tau,d),\ \gga'=(\sigma,d),\ \gga''=(\sigma,d')
    \in\GGamespace$ such that $T(d')>T(d)>0,$
    $\tau_a(x)=\sigma_a(x)$ for all $(a,x)\in [0,T(d)],$ and 
    $\tau_b(y)\ne \sigma_b(y)$ for all $(a,y)\in (T(d),T(d')].$
    Then $D(\gga,\gga')=0$ and $D(\gga',\gga'')=\gnorm{d-d''}_\infty.$
    When 
    $\tau_a(\cdot)$ and $\sigma_a(\cdot)$ differ more than 
    $\gnorm{d-d'}_\infty$ on the non-empty interval
    $(T(d),T(d')],$ then
    $D(\gga,\gga'')>\gnorm{d-d'}_\infty=D(\gga,\gga')+D(\gga',\gga'').$ 
    This follows since the binary operator~\eqref{eq:NewMetric} does not distinguish
    the cost functions of the two generalized games $\gga'=(\sigma,d)$ and $\gga''=(\sigma,d')$ when \eqref{eq:NewCostNorm} 
    is used to quantify the cost function distance. 
    Hence, the binary operator~\eqref{eq:NewMetric} does not
    fulfill the triangle inequality. 
   %
    Our definition~\eqref{eq:restricted-function-norm} of the distance of cost
    functions takes also the domains of cost functions into account. So, 
    $\gga'$ and $\gga''$ have different cost functions under our definition, and
    $\text{Dist}(\gga,\gga')+\text{Dist}(\gga',\gga'')=\text{Dist}(\gga',\gga'')=\max\{\gnorm{d-d'}_\infty,\gnorm{\sigma(T(d))-\sigma(T(d'))}_\infty\}=\text{Dist}(\gga,\gga'').$
    \end{remark}}
	
	Equipped with the metric~\eqref{eq:GeneralizedNorm}, $\GGamespace$ becomes
	a metric space with the topology 
	generated by \emph{open $\epsilon$-balls}
	of the form~\eqref{eq:Epsilon-Ball},
	\begin{equation}\label{eq:Epsilon-Ball}
		B_\epsilon(\gga):=\{\gga'\in\GGamespace:\ \gnorm{\gga-\gga'}=\text{Dist}(\gga,\gga')<\epsilon\},\quad \gga\in\GGamespace,\ \epsilon>0.
	\end{equation}

	\subsection{\wuu{The PoA is continuous}}
	\label{subsec:ContinuousMap_GameLimit}

	\wuu{The metric~\eqref{eq:GeneralizedNorm} induces the definition of convergence of games
		and of the continuity of real-valued maps
		on $\GGamespace$. A sequence
	$(\gga_n)_{n\in\N}\in \GGamespace^\N$
    \emph{converges} to a
    \emph{limit} $\gga\in\GGamespace,$
    denoted by $\lim_{n\to\infty}\gga_n=\gga,$
    if for each $\epsilon>0,$ there is an $N>0$ such that
   $\gga_n\in B_\epsilon(\gga)$ for all $n\ge N.$
   Trivially, $\lim_{n\to\infty}\gga_n=\gga$
   if and only if $\lim_{n\to \infty}\gnorm{\gga_n-\gga}=0.$
   Then a real-valued map $\bar{\varphi}: \GGamespace\to \R$ 
is \emph{continuous} if $\lim_{n\to \infty}\bar{\varphi}(\gga_n)
=\bar{\varphi}(\gga)$ for each sequence $(\gga_n)_{n\in\N}\in \GGamespace^\N$
and each $\gga\in\GGamespace$ with $\lim_{n\to \infty}\gga_n=\gga.$
In addition, a real-valued map $\varphi:\Gamespace\to\R$
is \emph{continuous} if $\lim_{n\to \infty}\varphi(\ga_n)
=\varphi(\ga)$ for each sequence $(\ga_n)_{n\in\N}\in \Gamespace^\N$
and each $\ga\in\GGamespace$ with $\lim_{n\to \infty}\ga_n=\ga.$}

	\wuu{Note that} every game $\ga=(\tau,d)\in \Gamespace$ has \wusec{the} unique total cost
	$C(\ga,f^*)$ for its 
	SO flows $f^*$. This defines an SO cost map $\C^*:\Gamespace\to \R_{\ge 0}$
	with $\C^*(\ga):=C(\ga,f^*)$ for each $\ga\in\Gamespace.$
	Similarly, we can define a WE cost map $\tilde{\C}:\Gamespace\to\R_{\ge 0}$ by putting $\tilde{\C}(\ga):=C(\ga,\tilde{f})$
	for each $\ga\in\Gamespace,$ where $\tilde{f}$
	is an arbitrary WE flow of $\ga.$
	Lemma~\ref{lemma:LimitFlow}b)--c)
	imply that both $\C^*(\cdot)$ and
	$\tilde{\C}(\cdot)$ are continuous maps
	on $\Gamespace$.
	\begin{lemma}\label{lemma:LimitFlow}
		Consider \wusec{a} convergent game sequence 
		$(\ga_n=(\tau^{(n)},d^{(n)}))_{n\in\N}\in\Gamespace^{\N}$ with $\lim_{n\to \infty}\ga_n=\ga$
		for a game $\ga=(\tau,d)\in\Gamespace.$
		\begin{itemize}
			\item[a)] Let $f$ be an arbitrary flow of $\ga.$
			Then there is a  sequence $(f_n)_{n\in\N}$ \wuu{of flows}
			such that each $f_n,$ \wu{$n\in\N,$} is a
			flow of $\ga_n$, and $f=\lim_{n\to\infty}f^{(n)}.$
			\item[b)] Let $(f^{*(n)}_{n})_{n\in\N}$
			be \wusec{a sequence of SO flows} $f^{*(n)}$
			of $\ga_n,$
			and let $(n_i)_{i\in\N}$ be an
			infinite subsequence with
			$\lim_{i\to\infty}f^{*(n_i)}=f^*$
			for a vector $f^*\in \R_{\ge 0}^{\S}.$
			Then $f^*$ is an SO flow of $\ga.$
			Moreover, $(f^*_n)_{n\in\N}$ converges
			to an SO flow of $\ga$
			when $\ga$ has a unique SO flow.
			In particular, $\lim_{n\to\infty}\C^*(\ga_n)=\C^*(\ga).$
			\item[c)] Let $(\tilde{f}_{n})_{n\in\N}$
			be a sequence \wusec{of WE flows} $\tilde{f}^{(n)}$
			of $\ga_n,$
			and let $(n_i)_{i\in\N}$ be an
			infinite subsequence with
			$\lim_{i\to\infty}\tilde{f}^{(n_i)}=\tilde{f}$
			for a vector $\tilde{f}\in \R_{\ge 0}^{\S}.$
			Then $\tilde{f}$ is a WE flow of $\ga.$
			Moreover, $(\tilde{f}_n)_{n\in\N}$
			converges to a WE flow of $\ga$ when
			$\ga$ has a unique WE flow.
			In particular, $\lim_{n\to\infty}\tilde{\C}(\ga_n)=\tilde{\C}(\ga).$
		\end{itemize}
		
	\end{lemma}
     \proof{Proof of Lemma~\ref{lemma:LimitFlow}}
     \proof{\wusec{Proof of} Lemma~\ref{lemma:LimitFlow}a):}
     Define $\delta:=\min_{s\in\S:f_s>0}f_s$ and 
     $s_k:=\arg\min_{s\in \S_k:f_s>0}f_s$ for each $k\in\K.$
     Then $f_s\ge \delta>0$ for all $s\in\S$
     with $f_s>0,$ and
     $f_{s'}\ge f_{s_k}\ge \delta$ for each $s'\in\S_k$
     with $f_{s'}>0$ for each \wu{O/D pair}
     $k\in\K.$
     Since $\lim_{n\to \infty}\ga_n=\ga,$ we have
     $
     	\delta>\gnorm{\ga_n-\ga}\ge \norm{d^{(n)}-d}=\max_{k\in \K}
     	|d_k^{(n)}-d_k|
     $ for each $n\ge N$ for some integer $N\in\N.$
     
     Define for each $n\ge N$ a vector $f^{(n)}=(f^{(n)}_{s})_{s\in\S}$
     with 
     \begin{displaymath}
     	f_s^{(n)}:=
     	\begin{cases}
     	f_s&\text{if }s\in \S_k\setminus\{s_k\},\\
     	f_s+d_k^{(n)}-d_k&\text{if }s=s_k,
     	\end{cases}
     	\quad \forall s\in\S_k\ \forall k\in\K.
     \end{displaymath}
     Then $f_s^{(n)}\ge 0$ for each $s\in\S,$
     since $\delta+d_k^{(n)}-d_k\ge 0$ for each \wu{O/D pair} $k\in\K$
     when $n\ge N.$
     Moreover, $\sum_{s \in S_k}f_s^{(n)}=\sum_{s \in S_k}
     f_s+d_k^{(n)}-d_k=d_k^{(n)}$
     for each \wu{O/D pair} $k\in\K.$
     So $f^{(n)}$ is a flow of $\ga_n$ for each $n\ge N.$
     
     $\lim_{n\to \infty}f^{(n)}=f$ follows immediately
     from the definition of $f^{(n)}$ and the fact that
     $\lim_{n\to\infty}\ga_n=\ga.$
     
     \proof{\wusec{Proof of} Lemma~\ref{lemma:LimitFlow}b):}
     Trivially, $f^*$ is a flow of $\ga.$
     Let $f$ be an arbitrary flow of $\ga.$
     Lemma~\ref{lemma:LimitFlow}a) implies that
     $f=\lim_{n\to \infty}f^{(n)}$ for a sequence
     $(f^{(n)})_{n\in\N}$ of flows of games $\ga_n.$
     Then we obtain by the continuity of cost functions
     that 
     \WU{$
     	C(\ga,f^*)=\lim_{i\to\infty}C(\ga_{n_i},f^{*(n_i)})
     	\le \lim_{i\to\infty}C(\ga_{n_i},f^{(n_i)})=C(\ga,f).
     $}
     This shows that $f^{*}$ is an SO flow of $\ga$ due to
     the arbitrary choice of $f.$
     The \WU{remainder} of Lemma~\ref{lemma:LimitFlow}b) then
     follows trivially.
     
     \proof{\wusec{Proof of} Lemma~\ref{lemma:LimitFlow}c):}
     Similarly, $\tilde{f}$ is a flow of $\ga.$
     Consider an arbitrary \wu{O/D pair} $k\in\K$ and
     arbitrary two paths $s,s'\in\S_k$ with
     $\tilde{f}_s>0.$ Then $\tilde{f}^{(n_i)}_s>0$
     when $i$ is large enough.
     Since $\tilde{f}^{(n_i)}$ is a WE flow of
     $\ga_{n_i}$ for each $i\in\N,$ we have
     $
     	\tau_{s}(\tilde{f})=\wu{\lim_{i\to\infty}\tau_s^{(n_i)}(\tilde{f}^{(n_i)})\le \lim_{i\to\infty}\tau_{s'}^{(n_i)}(\tilde{f}^{(n_i)})}
     	=\tau_{s'}(\tilde{f}).
     $
     This shows that $\tilde{f}$ is a WE flow of $\ga$ due to
     the arbitrary choice of $k,s$ and $s'.$
     The \WU{remainder} of Lemma~\ref{lemma:LimitFlow}c) then follows 
     trivially.
     
     \hfill$\square$
     
     Viewed as a \wuu{real-valued map} from $\Gamespace$ to $\R_{\ge 0},$
     the PoA $\rho(\cdot)$ is also \wuu{continuous,}
     since $\rho(\cdot)$
     is the \wusec{quotient} of two continuous maps
     \wusec{on $\Gamespace$}, i.e.,
     $\rho(\ga)=\frac{\tilde{\C}(\ga)}{\C^*(\ga)}$
     for each $\ga\in\Gamespace.$
     \wuu{Here, we recall that the PoA $\rho(\cdot)$ 
     is well defined on the whole game space $\Gamespace.$}
     
     We summarize all these continuity results in Theorem~\ref{thm:PoA_Continuity}.
     \begin{theorem}
     	\label{thm:PoA_Continuity}
     	The SO cost map $\C^*(\cdot),$ the
     	WE cost map $\tilde{\C}(\cdot)$ and
     	the PoA map $\rho(\cdot)$ are continuous
     	on the game space $\Gamespace$.
     \end{theorem}
 
    \wuu{Note that 
    	\citet{Hall1978} has proved that the user cost 
    $L_k(\tau,d)=\min_{s\in\S_k:\ \tilde{f}_s>0} \tau_s(\tilde{f})$ is a continuous function
of the demand vector $d$ when the cost function vector $\tau$ is fixed. This then implies directly that 
$\tilde{\C}(\cdot)$ is continuous when $\tau$ is fixed.  Theorem~\ref{thm:PoA_Continuity}  generalizes this continuity result to the game space $\Gamespace.$}
 
     Non-atomic congestion games $\gga$ \wuu{in} the gap $\GGamespace\setminus\Gamespace$
     may have \wusec{an} undefined PoA of $\frac{0}{0},$ \wusec{and} are thus
     \wusec{not considered} in our sequel analysis of the PoA map $\rho(\cdot)$.
     \wusec{One may of course} wonder if we could 
     include them
     in the analysis by constructing
     an \emph{extension} $\wu{\bar{\rho}}:\GGamespace\to\R_{\ge 0}$
     of the PoA map $\rho(\cdot)$ \wusec{to $\GGamespace.$}
     From a topological point of view, such an extension $\wu{\bar{\rho}}(\cdot)$
     should not only satisfy the condition that
     $\wu{\bar{\rho}}(\ga)=\rho(\ga)$ for each 
     $\ga\in\Gamespace,$ but also preserve the continuity
     of the PoA map $\rho(\cdot).$  When such a map $\wu{\bar{\rho}}(\cdot)$ exists, 
     Condition~2 would then no longer be needed,
     \wuu{which would considerably simplify the further analysis.}
     Unfortunately, Example~\ref{example:PoA_Discontinuity}
     illustrates that the PoA $\rho(\cdot)$ cannot be continuously extended
     to the generalized game space $\GGamespace.$
     Thus we \WU{need to exclude generalized games with an undefined
     PoA of $\frac{0}{0}$ and} have to accept the existence of this gap.
     \begin{example}
     	\label{example:PoA_Discontinuity}
     	Consider \wu{the} traffic network
     	$G$ shown in Figure~\ref{fig:PoA_Discontinuity}(a)--(b).
     	This network has two vertices
     	$o$ and $t$ with two
     	parallel paths (arcs). Denote by $\K$ and $\S$
     	the respective \wu{sets of O/D pairs and paths of} $G.$
     	Let $\gga=(\tau,d)\in\GGamespace$ be a generalized game
     	with total demand $T(d)=1$ and \wu{the} two cost functions $\tau_1(\cdot)$
     	and $\tau_2(\cdot)$
     	shown in Figure~\ref{fig:PoA_Discontinuity}(a). 
     	For each $n\in\N,$ let $\ga_n=(\tau^{(n)},d^{(n)})\in\Gamespace$ be a game again
     	with \wu{the same} total demand $T(d^{(n)})=1$ but \wusec{with} the two cost
     	functions $\tau_1^{(n)}(\cdot)$ and $\tau_2^{(n)}(\cdot)$ shown in Figure~\ref{fig:PoA_Discontinuity}(b).
     	\wusec{The} game sequence $(\ga_n)_{n\in\N}$ converges to $\gga$ 
     	for every \wusec{choice of} $\beta\in [0,\infty)$.
     	\begin{figure}[!htb]
     		\centering
     		\begin{subfigure}{0.49\textwidth}
     			\centering
     			\begin{tikzpicture}[
     			>=latex
     			]
     			\node[scale=0.4,circle,fill=black,label=left:$o$](o){};
     			\node[above right  =0.7of o](p1){$\tau_1(x)\equiv 0$};
     			\node[below right =0.7of o](p2){$\tau_2(x)\equiv  0$};
     			\node[scale=0.4,circle,fill=black,label=right:$t$,below right =0.7of p1](t){};
     			\draw[thick] (o) to [out=90,in=180] (p1);
     			\draw[->,thick] (p1) to [out=0,in=90] (t);
     			\draw[thick] (o) to [out=-90,in=180] (p2);
     			\draw[->,thick] (p2) to [out=0, in=-90] (t);
     			\end{tikzpicture}
     			\caption{Game $\gga\in\GGamespace$ with $T(d)=1$}
     		\end{subfigure}
     		\begin{subfigure}{0.49\textwidth}
     			\centering
     			\begin{tikzpicture}[
     			>=latex
     			]
     			\node[scale=0.4,circle,fill=black,label=left:$o$](o){};
     			\node[above right  =0.7of o](p1){$\tau_1^{(n)}(x)\equiv \frac{1}{n}$};
     			\node[scale=0.9,below right =0.7of o](p2){$\tau_2^{(n)}(x)=  \frac{x^{\beta}}{n}$};
     			\node[scale=0.4,circle,fill=black,label=right:$t$,below right =0.7of p1](t){};
     			\draw[thick] (o) to [out=90,in=180] (p1);
     			\draw[->,thick] (p1) to [out=0,in=90] (t);
     			\draw[thick] (o) to [out=-90,in=180] (p2);
     			\draw[->,thick] (p2) to [out=0, in=-90] (t);
     			\end{tikzpicture}
     			\caption{Game $\Gamma^{(n)}\in\Gamespace$
     				with $T(d^{(n)})=1$}
     		\end{subfigure}
     		\caption{Non-extensibility of the PoA}
     		\label{fig:PoA_Discontinuity}
     	\end{figure}
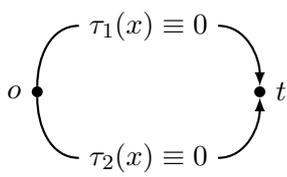
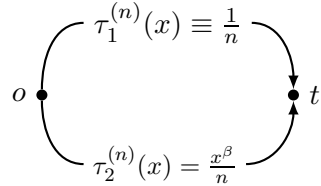
       \wusec{We now illustrate with this convergent sequence} that the PoA $\rho(\cdot)$
        can not be continuously extended to $\GGamespace.$
        We \wusec{do} this by contradiction, and thus
     	assume \wusec{that there is} a continuous extension $\wu{\bar{\rho}}:
     	\GGamespace\to\R_{\ge 0}$ of the PoA $\rho(\cdot).$ Then 
     	$\wu{\bar{\rho}}(\ga_n)=\rho(\ga_n)$ for each 
     	$n\in\N,$ and
     	$\wu{\bar{\rho}}(\gga)=\lim_{n\to\infty}\wu{\bar{\rho}}(\ga_n)=
     	\lim_{n\to\infty}\rho(\ga_n),$ since $\wu{\bar{\rho}}$ is continuous on
     	$\GGamespace$.
     	This means that the sequence $(\rho(\ga_n))_{n\in\N}=(\wu{\bar{\rho}}(\ga_n))_{n\in\N}$
     	has \wu{a} unique limit $\wu{\bar{\rho}}(\gga)$ \wu{that} is independent of $\beta$.
     	However, the limit of 
     	$(\rho(\ga_n))_{n\in\N}=(\wu{\bar{\rho}}(\ga_n))_{n\in\N}$
     	\wusec{depends} crucially on the value $\beta,$
     	\wusec{and yields} $\lim_{n\to \infty}\wu{\bar{\rho}}(\ga_n)=\lim_{n\to \infty}\rho(\ga_n)=1$
     	when $\beta=0,$ \wusec{and} $\lim_{n\to \infty}\wu{\bar{\rho}}(\ga_n)=\lim_{n\to \infty}\rho(\ga_n)=\frac{4}{3}$ when $\beta=1.$
     	Hence, there is no continuous extension $\bar{\rho}(\cdot)$
     	of the PoA $\rho(\cdot)$.
     \end{example}
 
     
     \subsection{\wuu{H{\"o}lder continuous maps}}
     
     \wuu{Theorem~\ref{thm:PoA_Continuity} implies that $|\rho(\ga)-\rho(\ga')|$
     	is small when the game $\ga'\in \Gamespace$ deviates 
     	only slightly from the game $\ga\in\Gamespace$.  Section~\ref{sec:Theory} below will
     	further quantify  the difference $|\rho(\ga)-\rho(\ga')|$ of the PoA
     	in terms of the metric $\gnorm{\ga-\ga'}$. To that end, we need
     	the notion of \emph{H{\"o}lder
     		continuity}.
     	
     	\begin{definition}
     		Consider a real-valued map
     		$\varphi:\Gamespace\to \R.$
     		\begin{itemize}
     			\item[$i)$] The map $\varphi$ is \emph{pointwise H{\"o}lder continuous
     				at a game} $\ga\in \Gamespace$
     			with a \emph{H{\"o}lder exponent} $\gamma_{\ga}>0$
     			(depending only on $\ga$), if
     			there are a \emph{H{\"o}lder constant} $\h_\ga>0$ (depending also
     			only on $\ga$) and a \emph{non-empty} 
     			open set $U_{\ga}\subseteq \Gamespace,$ s.t.,
     			$\ga\in U_{\ga}$ and 
     			$|\varphi(\ga)-\varphi(\ga')|\le \h_\ga\cdot \gnorm{\ga-\ga'}^{\gamma_{\ga}}$
     			for each $\ga'\in U_{\ga}.$ 
     			\item[$ii)$] The map $\varphi$ is \emph{pointwise H{\"o}lder continuous}
     			on the whole game space $\Gamespace$ with a (uniform) H{\"o}lder
     			exponent $\gamma>0$ when it is pointwise H{\"o}lder continuous
     			at \emph{every} $\ga\in\Gamespace$ with the H{\"o}lder exponent $\gamma$. 
     			\item[$iii)$] The map $\varphi$ is  
     			\emph{uniformly H{\"o}lder  continuous}
     			on the whole game space $\Gamespace$ with a (uniform) H{\"o}lder exponent
     			$\gamma>0$ and a (uniform) \emph{H{\"o}lder constant} $\h>0$
     			if $|\varphi(\ga)-\varphi(\ga')|\le \h\cdot \gnorm{\ga-\ga'}^{\gamma}$
     			for all $\ga',\ga\in\Gamespace$.
     			\item[$iv)$] The map $\varphi$ is 
     			\emph{Lipschitz continuous} on the whole game space $\Gamespace$ with 
     			a \emph{Lipschitz constant} $\h>0$ when $\varphi$ is uniformly H{\"o}lder  continuous on 
     			$\Gamespace$ 
     			with the H{\"o}lder exponent $\gamma=1$ and the
     			H{\"o}lder constant $\h$.
     		\end{itemize}
     	\end{definition}

     	Uniform H{\"o}lder continuity implies
     	pointwise H{\"o}lder continuity, but the converse need not
     	hold, since the game space $\Gamespace$ is not \emph{compact}. Moreover, the smaller
     	the H{\"o}lder exponent $\gamma_{\ga}$ of the PoA map $\rho(\cdot)$
     	at a game $\ga\in\Gamespace$, the more sensitive \wusec{is}
     	the PoA of the game $\ga$ w.r.t. small changes in its demands
     	and cost functions. 
     Hence, we can indeed quantify the sensitivity of the PoA
 by a H{\"o}lder continuity analysis of the map $\rho(\cdot)$.}

	\subsection{$\rho$-invariant operators}
	\label{subsec:Invariant_Kernel}
	
	\wuu{The H{\"o}lder continuity analysis in Section~\ref{sec:Theory} also involves the notion of
	$\rho$-invariant operators. Formally,}
	a continuous map $\Upsilon:\G(G,\K,\S)\to \G(G,\K,\S)$ is called
	a \emph{$\rho$-invariant operator}, if it is continuous and does not
	change the PoA. 
	i.e.,
	$\rho(\Gamma)=\rho \big(\Upsilon(\Gamma)\big)$
	for each game $\Gamma\in \G(G,\K,\S).$ 
%
	Examples are
	the \emph{cost} and \emph{demand} normalizations
	that
	have been used \wusec{by} \citet{Colini2017WINE,Colini2020OR}
	and \wusec{by} \citet{Wu2019}.
	
	A \emph{cost normalization} is an operator
	$\Psi:\Gamespace\to \Gamespace$
	with $\Psi(\ga)=\Psi(\tau,d)=(\frac{\tau}{\upsilon},d)\in\Gamespace$
	for each $\ga=(\tau,d)\in\Gamespace,$
	where $\upsilon>0$ is a constant \emph{factor} and 
	$\frac{\tau}{\upsilon}:=(\frac{\tau_a}{\upsilon})_{a\in A}.$
	We employ the notation $\Psi_\upsilon$ to denote a
	cost normalization with factor $\upsilon.$
	Cost normalizations $\Psi_\upsilon(\cdot)$ are \wuu{continuous}
	since $\lim_{n\to\infty}\Psi_\upsilon(\ga_n)=\Psi_\upsilon(\ga)$
	when $\lim_{n\to \infty}\gnorm{\ga_n-\ga}=0.$
	As the PoA map $\rho(\cdot)$ is the \wusec{quotient} of 
	the WE cost map $\tilde{\C}(\cdot)$ over the SO cost map $\C^*(\cdot)$, $\rho(\cdot)$ is then invariant w.r.t. arbitrary
	cost \wusec{normalizations} $\Psi_\upsilon(\cdot)$.
	
	\wusec{However,} a cost normalization does not leave the \wu{distance} \wusec{invariant.} In fact, we have
	\begin{equation}\label{eq:Cost-Norm-Inequ-Metric}
		\begin{split}
		\gnorm{\Psi_\upsilon(\ga)-\Psi_\upsilon(\ga')}&=
		\max\left\{\norm{d-d'},\frac{\norm{\tau_{|T(d)}-\sigma_{|T(d')}}}{\upsilon}\right\}\\
		&=
		\begin{cases}
		\norm{d-d'}&\text{if }\norm{d-d'}
		\ge \frac{\norm{\tau_{|T(d)}-\sigma_{|T(d')}}}{\upsilon},\\
		\frac{\norm{\tau_{|T(d)}-\sigma_{|T(d')}}}{\upsilon}&\text{if }\norm{d-d'}< \frac{\norm{\tau_{|T(d)}-\sigma_{|T(d')}}}{\upsilon},
		\end{cases}
		\end{split}
	\end{equation}
	for all $\ga=(\tau,d),\ga'=(\sigma,d')\in \Gamespace$
	and all $\upsilon>0,$
	which \wusec{may be different from} $\gnorm{\ga-\ga'}.$
	
	By~\eqref{eq:Cost-Norm-Inequ-Metric},
	$\gnorm{\Psi_\upsilon(\ga)-\Psi_\upsilon(\ga')}=\gnorm{\ga-\ga'}=\norm{d-d'}$ when
	\begin{displaymath}
		\norm{d-d'}\ge \max\left\{\norm{\tau_{|T(d)}-\sigma_{|T(d')}},\ \frac{\norm{\tau_{|T(d)}-\sigma_{|T(d')}}}{\upsilon}\right\}.
	\end{displaymath}
	In particular,  a cost normalization $\Psi_\upsilon$ will result in a
	\emph{scaling} \wu{of the distance} when $d=d',$ i.e., 
	\begin{equation}\label{eq:Metric-Scaling-Cost}
		\gnorm{\Psi_\upsilon(\ga)-\Psi_\upsilon(\ga')}
		=\frac{\gnorm{\ga-\ga'}}{\upsilon}\quad\forall
		\ga=(\tau,d),\  \ga'=(\sigma,d')\in\Gamespace
		\text{ with }d=d'.
	\end{equation}
	We will see in
	Section~\ref{sec:Theory} that  equation~\eqref{eq:Metric-Scaling-Cost}
	implies a rather unpleasant property of the map $\rho(\cdot).$

	A \emph{demand normalization} \wusec{is} an operator
	$\Lambda: \Gamespace\to\Gamespace$
	with $\Lambda(\ga)=\Lambda(\tau,d)=(\tau\circ \upsilon,\frac{d}{\upsilon})\in\Gamespace,$
	where $\upsilon>0$ is again a constant
	\emph{factor}, $\frac{d}{\upsilon}:=(\frac{d_k}{\upsilon})_{k\in\K}$
	and $\tau\circ \upsilon:=(\tau_a\circ \upsilon)_{a\in A}$
	with $\tau_a\circ \upsilon(x):=\tau_a(x\cdot \upsilon)$ for
	\wu{all $a\in A$ and all $x\in [0,\frac{T(d)}{\upsilon}].$}
	We employ the notation $\Lambda_{\upsilon}$ to denote
	\wusec{a} demand normalization with factor $\upsilon>0.$
	Similar \wusec{to cost normalizations,} demand normalizations are continuous. The PoA $\rho(\cdot)$
	is invariant \wusec{under demand normalizations},
	since $\tilde{f}$ and $f^*$ are a WE flow and an SO flow of $\ga$ if and only if
	$\frac{\tilde{f}}{\upsilon}$
	and $\frac{f^*}{\upsilon}$ are a WE flow and an SO flow of $\Lambda_{\upsilon}(\ga),$ \wusec{respectively,} and since 
	\WU{$C(\ga,f)=\upsilon\cdot C(\Lambda_{\upsilon}(\ga),\frac{f}{\upsilon})$}
	for an arbitrary flow $f$ of $\ga.$
	
	We will demonstrate in Section~\ref{sec:Applications}
	that \wuu{the} cost and demand
	normalizations \wuu{also} help to analyze the \WU{convergence rate}
	of the PoA \wuu{when the total demand tends to $0$ or $\infty.$}

	\section{H{\"o}lder continuity
	 of the PoA}
	\label{sec:Theory}

	Initial H{\"o}lder continuity results have been obtained
	by \citet{EnglertFraOlb2010,TakallooKwon2020}
	and \citet{Cominetti2019}.
	They considered the H{\"o}lder continuity of the PoA
	on \emph{subspaces} of $\Gamespace,$ in which 
	all games have the \emph{same cost functions}.

	Consider now an arbitrary cost function vector $\tau=(\tau_a)_{a\in A}$
	\wusec{defined} on $[0,\infty)$ and 
	satisfying Condition~2.
	We call the subspace \WU{$\Gamespace_{|\tau}:=\{\ga'=(\sigma,d)\in\Gamespace: \sigma_{a}(x)=\tau_{a}(x)\ \forall a\in A\ \forall x\in [0,T(d)]\}$}
	of $\Gamespace$
	a \emph{cost slice} w.r.t. the cost function vector $\tau.$
	Trivially, two arbitrary  games $\ga_1=(\sigma^{(1)},d^{(1)})$
	and $\ga_2=(\sigma^{(2)},d^{(2)})$ from the \WU{same} cost slice 
	$\Gamespace_{|\tau}$ satisfy that
	\begin{equation}\label{eq:Cost-Slice-Metric}
	\gnorm{\ga_1-\ga_2}=
	\max\left\{\norm{d^{(1)}-d^{(2)}},\
	\norm{\sigma^{(1)}(T(d^{(1)}))-\sigma^{(2)}(T(d^{(2)}))}
	\right\}.
	\end{equation}
	
	\citet{EnglertFraOlb2010} considered the H{\"o}lder
	continuity of the PoA on a cost slice $\Gamespace_{|\tau}$
	with \wuu{polynomial} of degree at most $\beta$  cost functions $\tau_a(\cdot)$ 
	on networks with only one O/D pair.
	They showed that  
	\begin{displaymath}
		\rho(\ga')-\rho(\ga)\le \big((1+\epsilon)^{\beta}-1\big)\cdot
		\rho(\ga)
	\end{displaymath}
	for two arbitrary games $\ga=(\tau,d)$ and $\ga'=(\tau,d')$
	\wusec{of} the cost slice $\Gamespace_{|\tau}$
	with $d'=(1+\epsilon)\cdot d$
	\wusec{for an arbitrary constant} $\epsilon>0$.
	
	\citet{TakallooKwon2020} generalized the results of \citet{EnglertFraOlb2010} and considered the H{\"o}lder continuity
	of the PoA on a cost slice $\Gamespace_{|\tau}$
	with \wu{again} polynomial cost functions $\tau_a(\cdot)$
	of degree at most $\beta,$ but for \wu{networks with multiple
	O/D pairs.}
	They showed for this \wu{more} general case that
	\begin{equation}\label{eq:Known-Results}
	-O(\epsilon)=\big(\frac{1}{(1+\epsilon)^{\beta}}-1\big)\cdot 
	\rho(\ga)\le \rho(\ga')-\rho(\ga)\le \big((1+\epsilon)^{\beta}-1\big)\cdot
	\rho(\ga)= O(\epsilon)
	\end{equation}
    for two arbitrary  games $\ga=(\tau,d)$ and $\ga'=(\tau,d')$
    \wusec{of} the cost slice $\Gamespace_{|\tau}$
    with $d'=(1+\epsilon)\cdot d$
    \wusec{for an arbitrary constant} $\epsilon>0$.
	
	\wusec{Their} result implies that 
	\WU{$|\rho(\ga)-\rho(\ga')|\in O(\rho(\ga)\cdot \epsilon)$}
	when the cost functions $\tau_a(\cdot)$
	are polynomials, \wuu{and when} $\ga=(\tau,d)$ and $\ga'=(\tau,(1+\epsilon)\cdot d)$
	for a constant $\epsilon>0,$ 
	see \eqref{eq:Known-Results}.
	This together with \eqref{eq:Cost-Slice-Metric}
	then yields that 
	\wu{$|\rho(\ga)-\rho(\ga')|< \h_{\ga}\cdot\gnorm{\ga-\ga'}$
	for a H{\"o}lder constant 
	$\h_{\ga}>0$ depending on $\ga$
    when $\ga=(\tau,d)$ and $\ga'=(\tau,(1+\epsilon)\cdot d)$
    for a constant $\epsilon>0,$}
	so with a H{\"o}lder exponent of $1$
	\WU{though at the cost of the restrictive
	condition ``$d'=(1+\epsilon)\cdot d$".}
	This result is quite inspiring
	and implies for each polynomial cost function
		vector $\tau$ and each demand vector 
		$d$ with $T(d)>0$ that the PoA map $\rho(\cdot)$
		is pointwise H{\"o}lder continuous
		with H{\"o}lder exponent
		$1$ on the resulting one-dimensional affine subspace $
		\{\ga=(\tau,d')\in\Gamespace_{|\tau}:
	d'=\alpha\cdot d,\ \alpha>0\}$ of
   the cost slice $\Gamespace_{|\tau}$.
	
	\citet{Cominetti2019} \wusec{also} considered the H{\"o}lder
	continuity on a cost slice $\Gamespace_{|\tau},$
	but \wu{on networks with one O/D pair and}
	with cost functions $\tau_a(\cdot)$
	that are affine \wusec{linear} or have strictly positive
	derivatives.
	Unlike \citet{EnglertFraOlb2010}, \citet{Cominetti2019}
	focused on the \WU{differentiability} of the \wusec{resulting}  PoA map
	$\rho(\cdot)$ on \wusec{the} cost slice $\Gamespace_{|\tau}.$
	They showed for this case
	that the PoA map $\rho(\cdot)$
	is differentiable at each demand level except for a finite set of $\mathcal{E}$-breakpoints.
	This implies that  
	the PoA map 
	is pointwise H{\"o}lder continuous with H{\"o}lder
	exponent $1$ on the cost slice $\Gamespace_{|\tau}$
  except for a finite set of points.
	This is the first relatively
	complete result on the H{\"o}lder continuity of the PoA
	as it  needs no longer the condition 
	``$d=(1+\epsilon)\cdot d'$". But the restriction
	to one O/D pair and cost slice only
	\WU{is still strong}.
	
	
	We now generalize these results by analyzing
	the H{\"o}lder
	continuity of the PoA map on the \wusec{whole}
	game space $\Gamespace$ \wu{and for an arbitrary
	finite set $\K$ of O/D pairs.} 
	\wusec{Our H{\"o}lder continuity} analysis \wusec{is thus not restricted 
	to a cost slice, but quantifies} the changes of the PoA when \emph{both the cost functions
	and the demands change.}
	As we consider the most general case,  \wusec{one cannot}
	\wusec{expect similar results on the} differentiability \wusec{of
	the PoA} as in 
	\citet{Cominetti2019}.

	\subsection{The PoA is not uniformly H{\"o}lder 
	continuous}
	\label{subsec:NotGlobalContinuity}

We show first that the PoA map
	$\rho(\cdot)$ is not uniformly H{\"o}lder  continuous on 
	$\Gamespace$. This \wuu{also} means  that the PoA map is not
	Lipschitz continuous on the whole game space
	$\Gamespace.$ 
	We assume by contradiction that $\rho(\cdot)$ is uniformly H{\"o}lder
	 continuous on the whole game space $\Gamespace$ with \wuu{a} uniform H{\"o}lder exponent $\gamma>0$
	and \wuu{a} uniform H{\"o}lder constant $\h>0,$
	\wuu{i.e., we assume for all $\ga,\ga'\in\Gamespace$ that}
	\begin{equation}\label{eq:GlobalHoelderContinuity}
		|\rho(\ga)-\rho(\ga')|\le \h\cdot \gnorm{\ga-\ga'}^{\gamma}.
	\end{equation}
	We now choose two arbitrary games $\ga=(\tau,d)$ and $\ga'=(\sigma,d')$
	with $\rho(\ga)\ne \rho(\ga')$ and $d=d'.$
	Note that there are indeed such two games in 
	\wuu{the \emph{demand slice} $\Gamespace_{|d}:=\{(\sigma,d'')\in \Gamespace:
		d''=d\},$}
	as \wu{every O/D pair has at least two paths, i.e.,} $|\S_k|\ge 2$ for each $k\in\K,$
	see Condition~1.
	Let $\upsilon>1$ be an \wusec{arbitrary} factor.
	By \eqref{eq:GlobalHoelderContinuity}, 
	\eqref{eq:Metric-Scaling-Cost}, and a \wusec{repeated application}
	of the cost normalization $\Psi_\upsilon,$ we obtain for
	each $n\in\N$ that
	\begin{equation}\label{eq:MetricShrinking}
		\begin{split}
		|\rho(\ga)-\rho(\ga')|&=|\rho(\Psi_\upsilon(\ga))-\rho(\Psi_\upsilon(\ga'))|
		=\cdots =|\rho(\underbrace{\Psi_\upsilon\circ\cdots \circ \Psi_\upsilon}_{n}(\ga))-\rho(\underbrace{\Psi_\upsilon\circ\cdots \circ \Psi_\upsilon}_n(\ga'))|\\
		&\le \h\cdot \gnorm{\Psi_\upsilon\circ\cdots \circ \Psi_\upsilon(\ga)-\Psi_\upsilon\circ\cdots \circ \Psi_\upsilon(\ga')}^{\gamma}=\h\cdot \frac{\gnorm{\ga-\ga'}^{\gamma}}{\wuu{\upsilon^{n\cdot \gamma}}},
		\end{split}
	\end{equation} 
	which
	implies $|\rho(\ga)-\rho(\ga')|=0$
	by letting $n\to \infty$ on both sides, and so contradicts with
	the fact that $\rho(\ga)\ne \rho(\ga').$ 
	
	Hence, a uniform
	H{\"o}lder constant $\h>0$ and a uniform
	H{\"o}lder exponent $\gamma>0$
	cannot exist simultaneously. The H{\"o}lder
	continuity results of \citet{EnglertFraOlb2010} and \citet{TakallooKwon2020}
	\wusec{seemingly} indicate \wu{that there is} \wusec{a uniform H{\"o}lder exponent but only with} a pointwise H{\"o}lder constant, see \eqref{eq:Known-Results}.
	\wusec{We thus also focus} on
	a uniform H{\"o}lder exponent $\gamma$
	\wusec{with a pointwise} H{\"o}lder constant $\h_\ga$
	in our analysis.

	\wusec{Given a} game $\ga\in\Gamespace$
	and a H{\"o}lder exponent 
	$\gamma>0,$ we call an \emph{open} subset 
	$U_{\ga}\subseteq \Gamespace$ a \emph{H{\"o}lder (continuity) neighborhood of order $\gamma$} \wusec{(in short, a 
		\emph{$\gamma$-neighborhood})}
	of $\ga$ if $\ga\in U_{\ga},$ and if
	there is a (pointwise) H{\"o}lder constant 
	$\h_\ga>0$ s.t. $|\rho(\ga')-\rho(\ga)|
	\le \h_\ga\cdot \gnorm{\ga'-\ga}^{\gamma}$ for
	each $\ga'\in U_{\ga}.$
	Here, we employ \wusec{the} convention that the empty set
	$\emptyset$ is a $\gamma$-neighborhood of every game $\ga\in\Gamespace$.
	Then $\rho(\cdot)$ is H{\"o}lder continuous
	at a game $\ga\in\Gamespace$ if and only if 
	$\ga$ has a \emph{non-empty} $\gamma$-neighborhood
	$U_\ga\subseteq \Gamespace$ for some H{\"o}lder
	exponent $\gamma>0.$
	
	We now show that every \wusec{$\gamma$-neighborhood}
	$U_\ga$
	is a \emph{proper} subset of $\Gamespace,$
	i.e., 
	$U_\ga \neq \Gamespace.$ This means that the H{\"o}lder
	continuity of the PoA map $\rho(\cdot)$ \wusec{can hold only  \emph{locally}} at a game $\ga\in\Gamespace$, even when
	the H{\"o}lder constant $\h_\ga$ is pointwise, i.e.,
	\wu{may depend} \wusec{on} the game $\ga.$
	
	We \wuu{again} show this by contradiction, and thus assume that there
	is a game $\ga=(\tau,d)\in\Gamespace$ whose
	$\gamma$-neighborhood is
	the entire space
	$\Gamespace$ for some H{\"o}lder exponent $\gamma>0.$
	Then there is a H{\"o}lder constant
	$\h_{\ga}>0$ such that
	\begin{equation}\label{eq:Global-Hoelder-Neighbor}
		|\rho(\ga)-\rho(\ga')|\le \h_{\ga}\cdot \gnorm{\ga-\ga'}
		^{\gamma}\quad \forall \ga'\in\Gamespace.
	\end{equation}
	
	To obtain a contradiction, we consider now an \emph{arbitrary}
	\wuu{game} $\ga'=(\sigma,d)$
	\wuu{from the same demand slice $\Gamespace_{|d}$ of}  $\ga.$
	Let $\upsilon>0$ be an \emph{arbitrary} factor. 
	\wusec{Then} the cost normalization 
	$\Psi_\upsilon(\cdot)$ and \eqref{eq:Global-Hoelder-Neighbor}
	\wusec{yield}
	\begin{equation}\label{eq:Global-Hoelder-Neighbor-Normalization}
	|\rho(\ga)-\rho(\ga')|=|\rho(\ga)-\rho(\Psi_\upsilon(\ga'))|\le \h_{\ga}\cdot \gnorm{\ga-\Psi_\upsilon(\ga')}
	^{\gamma}.
	\end{equation}
	Note that 
	\begin{displaymath}
		\gnorm{\ga-\Psi_\upsilon(\ga')}=
		\norm{\tau_{|T(d)}-\frac{\sigma_{|T(d)}}{\upsilon}}
		=\max_{a\in A,x\in [0,T(d)]}\left|\tau_{a}(x)-\frac{\sigma_{a}(x)}{\upsilon}\right|\to \max_{a\in A,x\in [0,T(d)]}\tau_a(x)\quad \text{ as }\upsilon
		\to \infty.
	\end{displaymath}
	\wusec{Inequality} \eqref{eq:Global-Hoelder-Neighbor-Normalization} then
	implies that
	\begin{displaymath}
		|\rho(\ga)-\rho(\ga')|\le \h_{\ga}\cdot \max_{a\in A,x\in [0,T(d)]}\tau_a(x)^{\gamma}=:\h_{\ga}\cdot\norm{\tau_{|T(d)}}^{\gamma}.
	\end{displaymath}
	\wusec{Since $\ga'$ is an arbitrary game of
		the demand slice $\Gamespace_{|d}=\{\ga'=(\sigma,d')\in\Gamespace:d'=d\},$} 
	the PoA map $\rho(\cdot)$ has a
	uniform \emph{finite} upper bound
	$\rho(\ga)+\h_{\ga}\cdot\norm{\tau_{|T(d)}}^{\gamma}$  on the demand slice
	$\Gamespace_{|d}.$

	However, the demand slice $\Gamespace_{|d}$
	contains games with polynomial cost functions
	of  degree at most $\beta$ for arbitrary $\beta>0$. 
	The PoA map $\rho(\cdot)$ is then \emph{unbounded} on 
	the demand slice $\Gamespace_{|d},$
	since these games 
	have a \emph{tight} upper bound \wuu{$\Theta(\beta/\ln \beta)$}
	tending to $\infty$ as $\beta\to\infty,$
	see \citet{Roughgarden2015Intrinsic}.
	Here, we notice that $|\S_k|\ge 2$ for each \wu{O/D pair} $k\in\K,$
	see Condition~1, and thus there is
	a game $\ga'_{\beta}=(\sigma,d)
	\in\Gamespace_{|d}$ 
	for each $\beta>0,$ who performs similarly \wusec{to}
	Pigou's game (\citet{Pigou1920})
	with cost functions $x^{\beta}$ and $1,$ 
	and whose PoA  \wuu{reaches the upper bound $\Theta(\beta/\ln\beta).$}
	

	We summarize this in Theorem~\ref{thm:NotUniformlyHoelder}.
	
	\begin{theorem}\label{thm:NotUniformlyHoelder}
		\begin{itemize}
			\item[a)] There are no constants
			$\gamma>0$ and  $\h>0$ s.t. the PoA map $\rho(\cdot)$ is uniformly 
			H{\"o}lder continuous with H{\"o}lder
			exponent 
			$\gamma$ and H{\"o}lder 
			constant $\h.$
			\item[b)] For every open subset $U\subseteq \Gamespace$
			and \wu{every} arbitrary H{\"o}lder exponent $\gamma>0,$
			if $U$ is a \wusec{$\gamma$-neighborhood} of a game $\ga\in\Gamespace,$
			then $U\ne \Gamespace.$
		\end{itemize}
	\end{theorem}

\wuu{
    \begin{remark}\label{remark:Cost-Slice}
    	We actually have proved 
    	that 
    	$\rho(\cdot)$ is neither  uniformly 
    	H{\"o}lder continuous on a demand slice, nor 
    	has a $\gamma$-neighborhood including
    	a demand slice $\Gamespace_{|d}$ as a subspace. However,
    	the PoA map $\rho(\cdot)$ may be Lipschitz continuous on a cost slice $\Gamespace_{|\tau}.$ 
    	To see this, we consider a vector $\tau=(\tau_a)_{a\in A}$ of cost functions with 
    	$\tau_a(x)\equiv c_a$ for some constant 
    	$c_a>0$ and all $(a,x)\in A\times [0,\infty).$
    	Then $\rho(\ga)\equiv 1$ for all $\ga=(\tau,d)\in\Gamespace_{|\tau},$
    	and so $\rho(\cdot)$ is Lipschitz continuous on $\Gamespace_{|\tau}$ when we restrict 
    	the map $\rho(\cdot)$ onto $\Gamespace_{|\tau}.$ In fact,
    	there is even a $1$-neighborhood including
    	the cost slice $\Gamespace_{|\tau}$ as a subspace.
    	To show this, pick an arbitrary
    	$\ga=(\tau,d)\in \Gamespace_{|\tau}$
    	and an arbitrary constant $\kappa>0.$ Then
    	\begin{displaymath}
    		\Gamespace_{|\tau}\setminus\{\ga\}\subseteq V_\ga(\kappa):=\{
    		\ga'\in\Gamespace:\ 
    		|\rho(\ga')-\rho(\ga)|-\kappa\cdot \gnorm{\ga-\ga'}
    		< 0\},
    	\end{displaymath}
    	and $V_\ga(\kappa)$ is open since 
    	$\rho(\cdot)$ is continuous.
    	Theorem~\ref{thm:Finer_PoA_Approx_Particular}a)
    	in Section~\ref{subsubsec:Finer_PoA_Approximation_Particular}
    	shows that there are a H{\"o}lder 
    	constant $\h_{\ga}>0$ and a nonempty $1$-neighborhood
    	$U_\ga$ with 
    	$|\rho(\ga')-\rho(\ga)|
    	\le \h_{\ga}\cdot \gnorm{\ga-\ga'}$
    	for each $\ga'\in U_\ga.$ 
    	Thus 
    	\begin{displaymath}
    	    |\rho(\ga')-\rho(\ga)|
    	    \le \max\{\kappa,\h_{\ga}\}
    	    \cdot \gnorm{\ga'-\ga}\quad 
    	    \forall \ga'\in U_\ga\cup V_\ga(\kappa).
    	\end{displaymath}
    	Clearly, $U_\ga\cup V_\ga(\kappa)$
    	is a $1$-neighborhood of $\ga$ that includes
    	the whole cost slice $\Gamespace_{|\tau}$ as
    	a subspace.
    	Hence, the H{\"o}lder continuity
    	of the PoA 
    	may differ largely
    	on the two types of slices.
    	We thus need separate discussions for them when
    	we analyze the pointwise H{\"o}lder continuity
    	of the PoA.
    \end{remark}}

	
	\subsection{\wuu{Pointwise H{\"o}lder continuity of the PoA}}
	\label{subsec:Continuity_PoA}
	Because of Theorem~\ref{thm:NotUniformlyHoelder},
	we now consider H{\"o}lder continuity of the PoA map
	$\rho(\cdot)$ \emph{pointwise} and
	\emph{locally} at each game $\ga\in\Gamespace$.
	
	Theorem~\ref{thm:HalfHoelderContinuity} \wusec{shows} that
	the PoA is H{\"o}lder continuous with H{\"o}lder
	exponent $\gamma=\frac{1}{2}$ at every game $\ga=(\tau,d)\in\Gamespace$ whose cost functions $\tau_a(\cdot)$ are Lipschitz continuous on the compact
	interval $[0,T(d)].$ Note that games satisfying
	these assumptions are \emph{dense} in $\Gamespace,$
	i.e., every game $\ga'\in\Gamespace$ is the limit of a convergent sequence of 
	such games. Thus every non-empty open subset $U$ of $\Gamespace$ contains 
	a non-empty $\frac{1}{2}$-neighborhood,
	although Theorem~\ref{thm:NotUniformlyHoelder}b) 
	implies that this neighborhood
	might not be large.
	\begin{theorem}\label{thm:HalfHoelderContinuity}
		Consider an arbitrary game $\ga=(\tau,d)\in\Gamespace$
		whose cost functions $\tau_a(\cdot)$ are Lipschitz
		continuous on the compact interval $[0,T(d)].$
		Then
		the PoA map $\rho(\cdot)$ is H{\"o}lder continuous at  
		game $\ga$ 
		with H{\"o}lder exponent $\gamma=\frac{1}{2}$ within
		a $\gamma$-neighborhood $B_{\epsilon_\ga}(\ga)$ for a small constant 
		$\epsilon_{\ga}>0$ depending only on the game $\ga.$
	\end{theorem}

Theorem~\ref{thm:HalfHoelderContinuity} presents
the first pointwise H{\"o}lder continuity result
of the PoA map $\rho(\cdot)$ on the whole game space
$\Gamespace$.
\wuu{It} applies to the most general case that
the cost functions and the demands change simultaneously.

    \wuu{To prove Theorem~\ref{thm:HalfHoelderContinuity}, we will now analyze the H{\"o}lder continuity of the PoA 
    on cost and demand slices separately, since it may differ
    on cost and demand slices, see Remark~\ref{remark:Cost-Slice}.}
   Theorem~\ref{thm:HalfHoelderContinuity} then follows by combining the results appropriately.
   
   \subsubsection{\wuu{H{\"o}lder continuity of the PoA on demand slices}}
    
    Lemma~\ref{lemma:PoA_Lipschiz} presents the \emph{first} results about
    H{\"o}lder continuity of the PoA on a demand slice 
    $\Gamespace_{|w}$ for an arbitrary demand vector
    $w=(w_k)_{k\in\K}$ with $T(w)=\sum_{k \in \K}w_k>0.$
    Lemma~\ref{lemma:PoA_Lipschiz}a) shows that
    the SO map $\C^*(\cdot)$ is Lipschitz continuous 
    with Lipschitz constant $|A|\cdot T(w)$ on the demand 
    slice $\Gamespace_{|w}.$
    Lemma~\ref{lemma:PoA_Lipschiz}b) shows a similar continuity result
    for the potential \wuu{function values} of WE flows of games in 
    $\Gamespace_{|w}.$
    Lemma~\ref{lemma:PoA_Lipschiz}c) shows that the WE flows
    of two arbitrary games $\ga_1$ and $\ga_2$ in $\Gamespace_{|w}$
    \wuu{are} mutually
    \wuu{$|A|\cdot T(w)\cdot \gnorm{\ga_1-\ga_2}$-}approximate WE flows of each other.
    Finally, \wuu{with Lemma~\ref{lem:EpsilonNE}c) and Lemma~\ref{lemma:PoA_Lipschiz}c),} Lemma~\ref{lemma:PoA_Lipschiz}d) shows that,
    \wusec{when restricted to the demand slice 
    	$\Gamespace_{|w},$} both the 
    WE cost map $\tilde{\C}(\cdot)$ and the PoA map
    $\rho(\cdot)$ are \wu{pointwise} H{\"o}lder continuous with 
    H{\"o}lder exponent $\frac{1}{2}$ at each game 
    $\ga_1\in\Gamespace_{|w}$ whose cost functions are 
    Lipschitz continuous on the compact interval $[0,T(w)].$
    \begin{lemma}
    	\label{lemma:PoA_Lipschiz}
    	Consider an arbitrary demand vector $w=(w_k)_{k\in\K}$
    	with $T(w)>0,$ and two arbitrary games
    	$\ga_1=(\pi^{(1)},w)$ and 
    	$\ga_2=(\pi^{(2)},w)$ \wusec{of} the demand slice $\Gamespace_{|w}.$
    	Let $\tilde{f}$ and
    	$\tilde{g}$ be WE flows
    	of $\ga_1$ and $\ga_2,$
    	respectively.
    	Then, \wu{the following statements hold.}
    	\begin{itemize}
    		\item[a)] $|\C^*(\ga_1)-\C^*(\ga_2)|\le |A|\cdot T(w)\cdot 
    		\gnorm{\ga_1-\ga_2}.$
    		\item[b)] $|\Phi(\ga_1,f)-
    		\Phi(\ga_2,f)|\le |A|\cdot T(w)\cdot \gnorm{\ga_1-\ga_2}$
    		for every flow $f.$
    		Moreover, 
    		$|\Phi(\ga_1,\tilde{f})-
    		\Phi(\ga_2,\tilde{g}
    		)|\le |A|\cdot T(w)\cdot \gnorm{\ga_1-\ga_2},$
    		$0\le \Phi(\ga_1,\tilde{g})-\Phi(\ga_1,\tilde{f})
    		\le
    		2\cdot |A|\cdot T(w)\cdot \gnorm{\ga_1-\ga_2},$
    		and $0\le \Phi(\ga_2,\tilde{f})-
    		\Phi(\ga_2,\tilde{g})\le 
    		2\cdot |A|\cdot T(w)\cdot \gnorm{\ga_1-\ga_2}.$
    		\item[c)] $\tilde{f}$ is an $|A|\cdot T(w)\cdot \gnorm{\ga_1-\ga_2}$-approximate WE flow of the game $\ga_2,$
    		and $\tilde{g}$ is an $|A|\cdot T(w)\cdot \gnorm{\ga_1-\ga_2}$-approximate WE flow of the game $\ga_1.$ 
    		\item[d)] If $\pi_a^{(1)}(\cdot)$
    		is Lipschitz continuous on $[0,T(w)]$ with Lipschitz constant
    		$M>0$ for each $a\in A,$ then
    		$
    		|\tilde{\C}(\ga_1)-\tilde{\C}(\ga_2)|
    		\le 
    		(\sqrt{M\cdot |A|\cdot T(w)}+2)\cdot |A|\cdot T(w)\cdot\max\{\sqrt{\gnorm{\ga_1-\ga_2}},\
    		\gnorm{\ga_1-\ga_2}\},
    		$
    		and so $|\rho(\ga_1)-\rho(\ga_2)|\le \mb_{\ga_1}\cdot \max\{\sqrt{\gnorm{\ga_1-\ga_2}},\ \gnorm{\ga_1-\ga_2}\}$
    		with
    		\begin{displaymath}
    		\mb_{\ga_1}:= 2\cdot \frac{\rho(\ga_1)+\sqrt{M\cdot |A|\cdot T(w)}+2}{\C^*(\ga_1)}\cdot |A|\cdot T(w)
    		\end{displaymath}
    		when $\gnorm{\ga_1-\ga_2}\le \frac{\C^*(\ga_1)}{2\cdot |A|\cdot T(w)}.$
    	\end{itemize}
    \end{lemma}
    \proof{Proof of Lemma~\ref{lemma:PoA_Lipschiz}}
    Let $f^*$ and $g^*$ be an SO flow of
    $\ga_1$ and an SO flow of $\ga_2,$
    respectively.
    Note that $\ga_1$ and $\ga_2$ have 
    the same set of flows, since both
    belong to the demand slice $\Gamespace_{|w}.$
    So $\tilde{g}$ and $g^*$
    are also flows of $\ga_1,$ and $\tilde{f}$
    and $f^*$ are also flows of $\ga_2.$
    
    \proof{Proof of Lemma~\ref{lemma:PoA_Lipschiz}a):}
    Note that 
    \begin{displaymath}
    \begin{split}
    \C^*(\ga_1)&=C(\ga_1,f^*)\le C(\ga_1,g^*)
    =\sum_{a\in A}\pi^{(1)}_a(g_a^*)\cdot g_a^*
    \le \sum_{a\in A}\pi^{(2)}_a(g_a^*)\cdot g_a^*+
    \sum_{a\in A}g_a^*\cdot \gnorm{\ga_1-\ga_2}\\
    &=\C^*(\ga_2)+\sum_{a\in A}g_a^*\cdot \gnorm{\ga_1-\ga_2}
    \le \C^*(\ga_2)+|A|\cdot T(w)\cdot \gnorm{\ga_1-\ga_2}.
    \end{split}
    \end{displaymath}
    Here, we \wusec{have used} that
    $g_a^*\le T(w)$ for each $a\in A,$ and that 
    \begin{displaymath}
    |\pi_a^{(1)}(x)-\pi^{(2)}_a(x)|\le \gnorm{\ga_1-\ga_2}
    =\max_{b\in A,y\in [0,T(w)]}|\pi_b^{(1)}(y)-\pi^{(2)}_b(y)|\quad 
    \forall a\in A\ \forall x\in [0,T(w)].
    \end{displaymath}
    Similarly, we have $\C^*(\ga_2)\le \C^*(\ga_1)+|A|\cdot T(w)\cdot \gnorm{\ga_1-\ga_2}.$
    Then Lemma~\ref{lemma:PoA_Lipschiz}a) follows.
    
    \proof{Proof of Lemma~\ref{lemma:PoA_Lipschiz}b):}
    Consider an arbitrary flow $f$ for the demand vector $w.$
    \wusec{Definition} \eqref{eq:PotentialFunction} of the potential function $\Phi(\cdot,\cdot)$ 
    implies that
    \begin{displaymath}
    \Phi(\ga_1,f)=\sum_{a\in A}\int_{0}^{f_a}\pi_a^{(1)}(x)dx
    \le \sum_{a\in A}\int_{0}^{f_a}\pi_a^{(2)}(x)dx+\sum_{a\in A}f_a\cdot \gnorm{\ga_1-\ga_2}
    \le \Phi(\ga_2,f)+|A|\cdot T(w)\cdot \gnorm{\ga_1-\ga_2},
    \end{displaymath}
    and, similarly \wusec{that} $\Phi(\ga_2,f)\le \Phi(\ga_1,f)+|A|\cdot T(w)\cdot \gnorm{\ga_1-\ga_2}.$
    Thus we have $|\Phi(\ga_1,f)-\Phi(\ga_2,f)|\le |A|\cdot T(w)\cdot \gnorm{\ga_1-\ga_2}$
    for an arbitrary flow $f.$
    
    \wusec{Hence}
    \begin{displaymath}
    \Phi(\ga_1,\tilde{f})\le \Phi(\ga_1,\tilde{g})\le \Phi(\ga_2,\tilde{g})+|A|\cdot T(w)\cdot \gnorm{\ga_1-\ga_2},
    \end{displaymath}
    and, similarly, $\Phi(\ga_2,\tilde{g})\le \Phi(\ga_1,\tilde{f})+|A|\cdot T(w)\cdot \gnorm{\ga_1-\ga_2}.$
    Here, we \wusec{have used} the fact that WE flows of a game minimize
    the potential \wusec{function} of that game.
    So $|\Phi(\ga_1,\tilde{f})-\Phi(\ga_2,\tilde{g})|
    \le |A|\cdot T(w)\cdot \gnorm{\ga_1-\ga_2}.$
    
    \wusec{Hence} 
    \begin{displaymath}
    \begin{split}
    &0\le \Phi(\ga_1,\tilde{g})-\Phi(\ga_1,\tilde{f})
    =\Phi(\ga_1,\tilde{g})-\Phi(\ga_2,\tilde{g})+\Phi(\ga_2,\tilde{g})-\Phi(\ga_1,\tilde{f})\le 2\cdot 
    |A|\cdot T(w)\cdot \gnorm{\ga_1-\ga_2},\\
    &0\le \Phi(\ga_2,\tilde{f})-\Phi(\ga_2,\tilde{g})
    \le \Phi(\ga_2,\tilde{f})-\Phi(\ga_1,\tilde{f})+\Phi(\ga_1,\tilde{f})-\Phi(\ga_2,\tilde{g})\le 
    2\cdot 
    |A|\cdot T(w)\cdot \gnorm{\ga_1-\ga_2}.
    \end{split}
    \end{displaymath}
    
    This completes the proof of Lemma~\ref{lemma:PoA_Lipschiz}b).
    
    \proof{Proof of Lemma~\ref{lemma:PoA_Lipschiz}c):}
    Consider again an arbitrary flow $f$ for the demand vector $w.$
    Since $\tilde{f}$ is a WE flow of $\ga_1=(\pi^{(1)},w),$ we have 
    \begin{equation}\label{eq:Epsilon-WE-For-G1}
    \begin{split}
    \sum_{a\in A}\pi_a^{(2)}(\tilde{f}_a)\cdot (\tilde{f}_a-f_a)&=
    \sum_{a\in A}\pi_a^{(1)}(\tilde{f}_a)\cdot (\tilde{f}_a-f_a)+
    \sum_{a\in A}(\pi_a^{(2)}(\tilde{f}_a)-\pi_a^{(1)}(\tilde{f}_a))\cdot (\tilde{f}_a-f_a)\\
    &\le \sum_{a\in A}(\pi_a^{(2)}(\tilde{f}_a)-\pi_a^{(1)}(\tilde{f}_a))\cdot (\tilde{f}_a-f_a)
    \le |A|\cdot T(w)\cdot \gnorm{\ga_1-\ga_2}.
    \end{split}
    \end{equation}
    Similarly,
    \begin{equation}\label{eq:Epsilon-WE-For-G2}
    \sum_{a\in A}\pi_a^{(1)}(\tilde{g}_a)\cdot (\tilde{g}_a-f_a)
    \le \sum_{a\in A}(\pi_a^{(1)}(\tilde{g}_a)-\pi_a^{(2)}(\tilde{g}_a))\cdot (\tilde{g}_a-f_a)
    \le |A|\cdot T(w)\cdot \gnorm{\ga_1-\ga_2}.
    \end{equation}
    Then Lemma~\ref{lemma:PoA_Lipschiz}c) follows from
    \eqref{eq:Epsilon-WE-For-G1}--\eqref{eq:Epsilon-WE-For-G2} and
    the definition of $\epsilon$-approximate WE flows.
    
    \proof{Proof of Lemma~\ref{lemma:PoA_Lipschiz}d):}
    Lemma~\ref{lem:EpsilonNE}c) and Lemma~\ref{lemma:PoA_Lipschiz}c) together
    imply that 
    \begin{equation}\label{eq:Demand-Slice-WE-cost-diff}
    \begin{split}
    |\tilde{\C}(\ga_1)-C(\ga_1,\tilde{g})|&=|C(\ga_1,\tilde{f})-C(\ga_1,\tilde{g})|\\
    &\le 
    |A|\cdot T(w)\cdot \sqrt{M\cdot |A|\cdot T(w)\cdot \gnorm{\ga_1-\ga_2}}+|A|\cdot T(w)\cdot \gnorm{\ga_1-\ga_2}.
    \end{split}
    \end{equation}
    Trivially,
    \begin{equation}\label{eq:Demand-Slice-SameWE-differ-cost}
    \begin{split}
    |\tilde{\C}(\ga_2)-C(\ga_1,\tilde{g})|
    &=|C(\ga_2,\tilde{g})-C(\ga_1,\tilde{g})|
    \le \sum_{a\in A}|\pi_a^{(2)}(\tilde{g}_a)-\pi_a^{(1)}(\tilde{g}_a)|\cdot \tilde{g}_a\\
    &\le |A|\cdot T(w)\cdot \gnorm{\ga_1-\ga_2}.
    \end{split}
    \end{equation}
    Hence
    \begin{equation}\label{eq:Demand-Slice-WE-cost-diff-final}
    \begin{split}
    |\tilde{\C}(\ga_1)-\tilde{\C}(\ga_2)|&\le |A|\cdot T(w)\cdot \sqrt{M\cdot |A|\cdot T(w)\cdot \gnorm{\ga_1-\ga_2}}+2\cdot |A|\cdot T(w)\cdot \gnorm{\ga_1-\ga_2}\\
    &\le
    \begin{cases}
    (\sqrt{M\cdot |A|\cdot T(w)}+2)\cdot |A|\cdot T(w)\cdot\sqrt{\gnorm{\ga_1-\ga_2}}&\text{if }
    \gnorm{\ga_1-\ga_2}\le 1,\\
    (\sqrt{M\cdot |A|\cdot T(w)}+2)\cdot |A|\cdot T(w)\cdot\gnorm{\ga_1-\ga_2} &\text{if }\gnorm{\ga_1-\ga_2}> 1.
    \end{cases}
    \end{split}
    \end{equation}
    This together with Lemma~\ref{lemma:PoA_Lipschiz}a) implies that
    \begin{displaymath}
    \begin{split}
    &\rho(\ga_1)-\rho(\ga_2)
    \le \frac{\tilde{\C}(\ga_1)}{\C^*(\ga_1)}-\frac{\tilde{\C}(\ga_1)-(\sqrt{M\cdot |A|\cdot T(w)}+2)\cdot |A|\cdot T(w)\cdot\max\{\sqrt{\ga_1-\ga_2},\ \gnorm{\ga_1-\ga_2}\}}{\C^*(\ga_1)+|A|\cdot T(w)\cdot \gnorm{\ga_1-\ga_2}}\\
    &=\frac{\rho(\ga_1)\cdot |A|\cdot T(w)\cdot \gnorm{\ga_1-\ga_2}+(\sqrt{M\cdot |A|\cdot T(w)}+2)\cdot |A|\cdot T(w)\cdot\max\{\sqrt{\ga_1-\ga_2},\ \gnorm{\ga_1-\ga_2}\} }{\C^*(\ga_1)+|A|\cdot T(w)\cdot \gnorm{\ga_1-\ga_2}}\\
    &\le \frac{\rho(\ga_1)+\sqrt{M\cdot |A|\cdot T(w)}+2}{\C^*(\ga_1)+|A|\cdot T(w)\cdot \gnorm{\ga_1-\ga_2}}\cdot |A|\cdot T(w)\cdot \max\{\sqrt{\ga_1-\ga_2},\ \gnorm{\ga_1-\ga_2}\},
    \end{split}
    \end{displaymath}
    and that
    \begin{align}
    \rho(\ga_2)-\rho(\ga_1)&\le \frac{\tilde{\C}(\ga_1)+(\sqrt{M\cdot |A|\cdot T(w)}+2)\cdot |A|\cdot T(w)\cdot\max\{\sqrt{\gnorm{\ga_1-\ga_2}},\ \gnorm{\ga_1-\ga_2}\}}{\C^*(\ga_1)-|A|\cdot T(w)\cdot \gnorm{\ga_1-\ga_2}}-\frac{\tilde{\C}(\ga_1)}{\C^*(\ga_1)}\notag\\
    &\le \frac{\rho(\ga_1)+\sqrt{M\cdot |A|\cdot T(w)}+2}{\C^*(\ga_1)-|A|\cdot T(w)\cdot \gnorm{\ga_1-\ga_2}}\cdot |A|\cdot T(w)\cdot \max\{\sqrt{\ga_1-\ga_2},\ \gnorm{\ga_1-\ga_2}\}\label{eq:Demand-Slice-LipschitzCost-PoA}
    \end{align}
    when $\gnorm{\ga_1-\ga_2}\le \frac{\C^*(\ga_1)}{|A|\cdot T(w)}.$
    Therefore,
    $
    |\rho(\ga_1)-\rho(\ga_2)|
    \le \mb_{\ga_1}\cdot \max\{\sqrt{\gnorm{\ga_1-\ga_2}},\ \gnorm{\ga_1-\ga_2}\}
    $
    with 
    \begin{displaymath}
    \mb_{\ga_1}:=2\cdot \frac{\rho(\ga_1)+\sqrt{M\cdot |A|\cdot T(w)}+2}{\C^*(\ga_1)}\cdot |A|\cdot T(w)
    \end{displaymath}
    when $\gnorm{\ga_1-\ga_2}\le \frac{\C^*(\ga_1)}{2\cdot |A|\cdot T(w)}.$
    This completes the proof of Lemma~\ref{lemma:PoA_Lipschiz}d).

    \hfill$\square$
    
    \wuu{\begin{remark}\label{remark:Exponent-LischpitzCost-Cannot-Improve}
    		In the proof of Lemma~\ref{lemma:PoA_Lipschiz}d),
    		we have used Lemma~\ref{lem:EpsilonNE}c) to bound
    		$|C(\ga_1,\tilde{f})-C(\ga_1,\tilde{g})|,$
    		see~\eqref{eq:Demand-Slice-WE-cost-diff}, 
    		since Lemma~\ref{lemma:PoA_Lipschiz}c) has shown that
    		$\tilde{g}$ is an $|A|\cdot T(w)\cdot \gnorm{\ga_1-\ga_2}$-approximate
    		WE flow of $\ga_1.$ Note that $|A|\cdot T(w)\cdot \gnorm{\ga_1-\ga_2}\in \Theta(\gnorm{\ga_1-\ga_2})$
    		is already a tight upper bound
    		on the approximation threshold 
    		$\epsilon(\pi^{(1)},w,\tilde{g})$
    		of the flow $\tilde{g}$ for Lipschitz continuous
    		cost functions on $[0,T(w)]$.
    		We illustrate this with the two games $\ga_1$ and $\ga_2$ shown 
    		in~Figure~\ref{fig:Lipschitz-Demand-Slice-Tightness}(a)--(b).
    		Clearly, $\gnorm{\ga_1-\ga_2}=\epsilon$
    		since $\ga_1$ and $\ga_2$ have the same demand 
    		vector $w=(1).$ When viewed as a flow of $\ga_1,$ 
    		the WE flow $\tilde{g}=(0.5,0.5)$ of $\ga_2$
    		has the approximation threshold
    		$
    			\epsilon(\pi^{(1)},w,\tilde{g})=\tilde{g}_\ell\cdot [\pi_\ell^{(1)}(\tilde{g}_\ell)-\pi_u^{(1)}(\tilde{g}_u)]=0.5\cdot \epsilon\in \Theta(\gnorm{\ga_1-\ga_2}),
    		$ which shows that
    		the upper bound $|A|\cdot T(w)\cdot \gnorm{\ga_1-\ga_2}$
    		is tight (w.r.t.\ the magnitude and the exponent of 
    		$\gnorm{\ga_1-\ga_2}$). This 
   		means that the exponent $\frac{1}{2}$ in the right-hand of the inequality~\eqref{eq:Demand-Slice-WE-cost-diff} cannot be improved when the cost
   		functions are Lipschitz continuous on the interval 
   		$[0,T(w)],$ and when we use
  		Lemma~\ref{lem:EpsilonNE}c)  to bound
  		$|C(\ga_1,\tilde{f})-C(\ga_1,\tilde{g})|$, 
  		since Example~\ref{example:Tightness_Of_epsilon_WE} has also shown 
  		the tightness of Lemma~\ref{lem:EpsilonNE}c).
  		Hence, to improve the H{\"o}lder exponent, we need
  		a finer analysis, which we will develop
  		for
  		cost functions with special properties in Section~\ref{subsubsec:Finer_PoA_Approximation_Particular}.
    		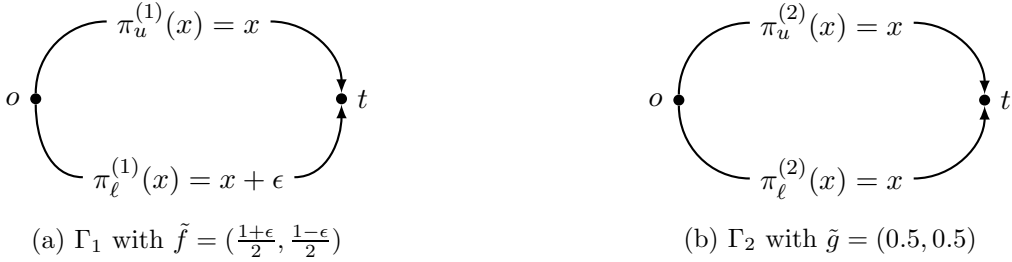
\begin{figure}[!htb]
    			\centering
    			\begin{subfigure}{0.49\textwidth}
    				\centering
    				\begin{tikzpicture}[
    				>=latex
    				]
    				\node[scale=0.4,circle,color=black,fill=black,label=left:$o$](o){};
    				\node[right =1.8of o](m){};
    				\node[scale=0.4,circle,color=black,fill=black,label=right:$t$,right =1.8of m](t){};
    				\node[above =0.5of m](c1){$\pi^{(1)}_u(x)= x$};
    				\node[below =0.5of m](c2){$\pi^{(1)}_\ell(x) = x+\epsilon$};
    				\draw[-,thick] (o) to [out=90,in=180] (c1);
    				\draw[-,thick] (o) to [out=-90,in=180] (c2);
    				\draw[->,thick] (c1) to [out=0,in=90] (t);
    				\draw[->,thick] (c2) to [out=0,in=-90] (t);
    				\end{tikzpicture}
    				\subcaption{$\ga_1$ with $\tilde{f}=(\frac{1+\epsilon}{2},\frac{1-\epsilon}{2})$}
    			\end{subfigure}
    		    \begin{subfigure}{0.49\textwidth}
    		    	\centering
    		    	\begin{tikzpicture}[
    		    	>=latex
    		    	]
    		    	\node[scale=0.4,circle,color=black,fill=black,label=left:$o$](o){};
    		    	\node[right =1.8of o](m){};
    		    	\node[scale=0.4,circle,color=black,fill=black,label=right:$t$,right =1.8of m](t){};
    		    	\node[above =0.5of m](c1){$\pi^{(2)}_u(x)= x$};
    		    	\node[below =0.5of m](c2){$\pi^{(2)}_\ell(x) = x$};
    		    	\draw[-,thick] (o) to [out=90,in=180] (c1);
    		    	\draw[-,thick] (o) to [out=-90,in=180] (c2);
    		    	\draw[->,thick] (c1) to [out=0,in=90] (t);
    		    	\draw[->,thick] (c2) to [out=0,in=-90] (t);
    		    	\end{tikzpicture}
    		    	\subcaption{$\ga_2$ with $\tilde{g}=(0.5,0.5)$}
    		    \end{subfigure}
    	        \caption{Games with total demand of $1$}
    	        \label{fig:Lipschitz-Demand-Slice-Tightness}
    		\end{figure}
    \end{remark}}
    

    \subsubsection{\wuu{H{\"o}lder continuity of the PoA on cost slices}}
    Lemma~\ref{lemma:HalfHoelderCostSlice} shows
    results similar \wusec{to} Lemma~\ref{lemma:PoA_Lipschiz} for
    two arbitrary games $\ga_1=(\pi,w)$ and
    $\ga_2=(\pi,w')$ from the same cost slice 
    $\Gamespace_{|\pi}$ for a cost function vector
    $\pi$ \wusec{defined} on $[0,\infty)$ and satisfying Condition~2. 
   \wuu{Using Lemma~\ref{lemma:PoA_Lipschiz}d),} Lemma~\ref{lemma:HalfHoelderCostSlice}a) shows
    that $|\tilde{\C}(\ga_1)-\tilde{\C}(\ga_2)|
    \le \tilde{\mb}_{\ga_1}\cdot \sqrt{\gnorm{\ga_1-\ga_2}}$
    for a constant $\tilde{\mb}_{\ga_1}>0$
    depending only on $\ga_1,$
    when the cost functions $\pi_a(\cdot)$ are Lipschitz continuous
    on the compact interval $[0,T(w)],$ $T(w')<T(w)$ and $\gnorm{\ga_1-\ga_2}$
    is small. \wuu{Using Lemma~\ref{lemma:PoA_Lipschiz}a),}
    Lemma~\ref{lemma:HalfHoelderCostSlice}b)
    shows with similar \wusec{assumptions} that 
    $|\C^*(\ga_1)-\C^*(\ga_2)|\le 
    \mb^*_{\ga_1}\cdot \gnorm{\ga_1-\ga_2}$
    for a constant $\mb^*_{\ga_1}>0$ depending
    only on $\ga_1.$
    Lemma~\ref{lemma:HalfHoelderCostSlice}c) then 
    presents \wusec{an upper} bound
    \wusec{for} $|\rho(\ga_1)-\rho(\ga_2)|$.
    Note that the Lipschitz constant 
    $M_{\ga_1}$ is required to be \wusec{at least} $1$ in Lemma~\ref{lemma:HalfHoelderCostSlice}.
    \wusec{However, this} is not an additional restriction, as 
    every Lipschitz continuous function always has a Lipschitz constant \wusec{of at least}
    $1.$
    \begin{lemma}\label{lemma:HalfHoelderCostSlice}
    	Consider an arbitrary game $\ga_1=(\pi,w)\in\Gamespace$
    	such that the cost functions $\pi_a(\cdot)$ are Lipschitz continuous
    	on the compact interval $[0,T(w)]$ with a Lipschitz constant
    	$M_{\ga_1}\ge 1.$ 
    	Then the following statements hold.
    	\begin{itemize}
    		\item[a)] For each game $\ga_2=(\pi,w')\in\Gamespace$
    		with $T(w')\le T(w),$
    		\begin{displaymath}
    		\begin{split}
    		|\tilde{\C}(\ga_2)-\tilde{\C}(\ga_1)|
    		&\le 2\cdot \Big((\sqrt{M_{\ga_1}\cdot |A|\cdot T(w)}+2)\cdot |A|\cdot T(w)\\
    		&\hspace{0.5cm}+|A|\cdot \pi_{max}(T(w))\cdot |\K|\Big)
    		\cdot \sqrt{M_{\ga_1}}\cdot \max\{\sqrt{\gnorm{\ga_1-\ga_2}},\ \sqrt{M_{\ga_1}}\cdot\gnorm{\ga_1-\ga_2}\}\\
    		&=:\tilde{\mb}_{\ga_1}\cdot \max\{\sqrt{\gnorm{\ga_1-\ga_2}},\ \sqrt{M_{\ga_1}}\cdot\gnorm{\ga_1-\ga_2}\}
    		\end{split}
    		\end{displaymath}
    		when $\gnorm{\ga_1\!-\!\ga_2}\!<\!\frac{T(w)}{|\K|},$
    		where
    		$
    		\tilde{\mb}_{\ga_1}:=2\cdot \big(\!\big(\sqrt{M_{\ga_1}\!\cdot\! |A|\!\cdot\! T(w)}\!+\!2\big)\cdot |A|\cdot T(w)\!+\!|A|\cdot |\K|\cdot \pi_{max}(T(w))\! \big)
    		\cdot \sqrt{M_{\ga_1}}
    		$
    		and $\pi_{max}(T(w))\!:=\!\max_{a\in A}\pi_a(T(w)).$
    		\item[b)] For each game $\ga_2=(\pi,w')\in\Gamespace$
    		with $T(w')\le T(w),$ 
    		\begin{displaymath}
    		|\C^*(\ga_1)-\C^*(\ga_2)|\le 2\cdot \big(|A|\cdot |\K|\cdot \pi_{max}(T(w))+|A|\cdot T(w)\cdot M_{\ga_1}\big)\cdot \gnorm{\ga_1-\ga_2}=:\mb^*_{\ga_1}\cdot \gnorm{\ga_1-\ga_2}
    		\end{displaymath}
    		when $\gnorm{\ga_1-\ga_2}<\frac{T(w)}{|\K|},$
    		where $\mb^*_{\ga_1}:=2\cdot \big(|A|\cdot |\K|\cdot \pi_{max}(T(w))+|A|\cdot T(w)\cdot M_{\ga_1}\big).$
    		\item[c)] For each game $\ga_2=(\pi,w')\in\Gamespace$
    		with $T(w')\le T(w),$ it holds that
    		$
    		|\rho(\ga_1)-\rho(\ga_2)|
    		< \mb'_{\ga_1}\cdot \max\{\sqrt{\gnorm{\ga_1-\ga_2}},\sqrt{M_{\ga_1}}\cdot \gnorm{\ga_1-\ga_2}\}
    		$
    		when 
    		$
    		\mb'_{\ga_1}:=\frac{2\cdot 
    			\rho(\ga_1)\cdot \mb^*_{\ga_1}+2\cdot 
    			\tilde{\mb}_{\ga_2}}{\C^*(\ga_1)}
    		$
    		and 
    		$
    		\gnorm{\ga_1-\ga_2}\le \min\{\frac{T(w)}{|\K|},\
    		\frac{\C^*(\ga_1)}{2\cdot \mb^*_{\ga_1}}\}.
    		$
    	\end{itemize}
    \end{lemma}
    \proof{Proof of Lemma~\ref{lemma:HalfHoelderCostSlice}}
    Define an auxiliary demand vector
    $\hat{w}=(\hat{w}_k)_{k\in\K}$
    with $\hat{w}_k=\min\{w_k,w'_k\}\ge 0$
    for each $k\in \K.$ Denote by $\hat{\ga}$
    the pair $\hat{\ga}=(\pi,\hat{w}).$ 
    Since $T(w)>0,$ 
    $
    T(\hat{w})\ge 
    T(w)-|\K|\cdot \gnorm{\ga_1-\ga_2}>0
    $ and
    $\hat{\ga}$ is a game, when the condition $\gnorm{\ga_1-\ga_2}<\frac{T(w)}{|\K|}$
    is fulfilled.
    Here, we use that
    \begin{displaymath}
    \gnorm{\ga_1-\ga_2}=\max\{\norm{w-w'},
    \norm{\pi(T(w))-\pi(T(w'))}\}\ge \norm{w-w'}=
    \max_{k\in \K}|w_k-w'_k|,
    \end{displaymath} 
    and so 
    $0\le T(w)-T(\hat{w})\le |\K|\cdot  \gnorm{\ga_1-\ga_2},$ 
    $0\le T(w')-T(\hat{w})\le |\K|\cdot  \gnorm{\ga_1-\ga_2},$ $0\le w_k-\hat{w}_k\le \gnorm{\ga_1-\ga_2}$ and $0\le w'_k-\hat{w}_k\le \gnorm{\ga_1-\ga_2}$ for each $k\in\K.$
    
    We assume now that \wusec{$\hat{\ga}$ is a game, i.e.,} $\gnorm{\ga_1-\ga_2}< \frac{T(w)}{|\K|}.$
    
    Since the cost functions $\pi_a(\cdot)$ are Lipschitz
    continuous on the compact interval $[0,T(w)]$ with Lipschitz 
    constant $M_{\ga_1},$ they are also Lipschitz continuous
    on the compact interval $[0,T(\hat{w})]\subseteq [0,T(w)]$
    with the same Lipschitz constant $M_{\ga_1}.$
    Then Lemma~\ref{lemma:PoA_Lipschiz}a) and d) yield,
    respectively, that
    \begin{equation}\label{eq:SO-demand-slice}
    |\C^*(\hat{\ga})-\C^*(\ga_3)|
    \le |A|\cdot T(\hat{w})\cdot \gnorm{\ga_3-\hat{\ga}}
    \quad \forall \ga_3=(\hat{\pi},\hat{w})\in
    \Gamespace_{|\hat{w}},
    \end{equation}
    and that
    \begin{equation}\label{eq:WE-demand-slice}
    |\tilde{\C}(\hat{\ga})-\tilde{\C}(\ga_3)|
    \le (\sqrt{M_{\ga_1}\cdot |A|\cdot T(\hat{w})}+2)\cdot |A|\cdot T(\hat{w})\cdot 
    \max\{\sqrt{\gnorm{\ga_3-\hat{\ga}}},\ \gnorm{\ga_3-\hat{\ga}}\}
    \end{equation}
    for every game $\ga_3=(\hat{\pi},\hat{w})\in
    \Gamespace_{|\hat{w}}.$
    
    We now prove Lemma~\ref{lemma:HalfHoelderCostSlice}
    with inequalities \eqref{eq:SO-demand-slice} \wusec{and} \eqref{eq:WE-demand-slice}.
    
    \proof{Proof of Lemma~\ref{lemma:HalfHoelderCostSlice}a):}
    
    Consider an arbitrary WE flow $\tilde{f}$ of the game 
    $\ga_1=(\pi,w).$
    Since $\hat{w}_k\le w_k$ for each $k\in\K,$ there
    is a vector $\mu=(\mu_s)_{s\in\S_k}$ such that 
    $0\le \mu_s\le \tilde{f}_s$ for all 
    $s\in\S,$ and $\sum_{s' \in S_k}(\tilde{f}_{s'}-\mu_{s'})=
    \hat{w}_k$ for each $k\in\K.$
    Then $\tilde{f}-\mu=(\tilde{f}_s-\mu_s)_{s\in\S_k}$
    is a WE flow of the game 
    $\ga'_1:=(\pi_{|\mu},\hat{w})\in\Gamespace_{|\hat{w}},$
    where $\pi_{|\mu}:=(\pi_{a|\mu})_{a\in A}$
    with $\pi_{a|\mu}(x):=\pi_a(x+\mu_a)$
    and $\mu_a:=\sum_{s\in\S:a\in s}\mu_s$
    for each $(a,x)\in A\times [0,T(\hat{w})].$
    This follows since the cost
    $\pi_{s|\mu}(\tilde{f}-\mu):=
    \sum_{a\in A: a\in s}\pi_{a|\mu}(\tilde{f}_a-\mu_a)=\sum_{a\in A: a\in s}\pi_{a}(\tilde{f}_a)=\pi_{s}(\tilde{f})$
    \wusec{remains} unchanged
    for each path $s\in\S.$
    
    Inequality \eqref{eq:WE-demand-slice} then yields that
    \begin{equation}\label{eq:WE-demand-slice-int1}
    |\tilde{\C}(\hat{\ga})-\tilde{\C}(\ga'_1)|
    \le  (\sqrt{M_{\ga_1}\cdot |A|\cdot T(\hat{w})}+2)\cdot |A|\cdot T(\hat{w})\cdot 
    \max\{\sqrt{\gnorm{\ga'_1-\hat{\ga}}},\ \gnorm{\ga'_1-\hat{\ga}}\}.
    \end{equation}
    Note that
    \begin{equation}\label{eq:WE-demand-slice-int2}
    \begin{split}
    |\tilde{\C}(\ga'_1)-\tilde{\C}(\ga_1)|
    &=|C(\ga'_1,\tilde{f}-\mu)-C(\ga_1,\tilde{f})|
    =\sum_{k\in\K}L_k(\ga_1)\cdot (w_k-\hat{w}_k)\\
    &\le L_{\max}(\ga_1)\cdot |\K|\cdot \gnorm{\ga_1-\ga_2}\le L_{\max}(\ga_1)\cdot |\K|\cdot \max\{\sqrt{\gnorm{\ga_1-\ga_2}},\ \gnorm{\ga_1-\ga_2}\}\\
    &\le |A|\cdot \pi_{max}(T(w))\cdot |\K|\cdot \max\{\sqrt{\gnorm{\ga_1-\ga_2}},\ \gnorm{\ga_1-\ga_2}\},
    \end{split}
    \end{equation}
    where we recall that $L_k(\ga_1)$
    is the user cost of O/D pair $k\in\K$ of 
    $\ga_1$ in the WE flow $\tilde{f},$
    and $L_{\max}(\ga_1):=\max_{k\in\K}\ L_k(\ga_1)\le |A|\cdot \max_{a\in A}\pi_a(T(w))=|A|\cdot \pi_{max}(T(w)).$
    Note that 
    \begin{displaymath}
    \gnorm{\ga'_1-\hat{\ga}}=
    \max_{a\in A,x\in [0,T(\hat{w})]}
    |\pi_a(x)-\pi_a(x+\mu_a)|
    \le M_{\ga_1}\cdot \max_{a\in A}\ \mu_a
    \le M_{\ga_1}\cdot \norm{w-w'}
    \le M_{\ga_1}\cdot \gnorm{\ga_1-\ga_2}.
    \end{displaymath}
    This together with \eqref{eq:WE-demand-slice-int1} \wusec{and} \eqref{eq:WE-demand-slice-int2} yields that
    \begin{equation}\label{eq:WE-demand-slice-Core1}
    \begin{split}
    |\tilde{\C}(\hat{\ga})-\tilde{\C}(\ga_1)|
    &\le \Big((\sqrt{M_{\ga_1}\cdot |A|\cdot T(w)}+2)\cdot |A|\cdot T(w)\\
    &\hspace{0.5cm}+|A|\cdot \pi_{max}(T(w))\cdot |\K|\Big)
    \cdot \sqrt{M_{\ga_1}}\cdot \max\{\sqrt{\gnorm{\ga_1-\ga_2}},\ \sqrt{M_{\ga_1}}\cdot \gnorm{\ga_1-\ga_2}\},
    \end{split}
    \end{equation}
    since $T(\hat{w})\le T(w).$ Here, we observe that $M_{\ga_1}\ge 1.$
    
    Similarly, we have 
    \begin{equation}\label{eq:WE-demand-slice-Core2}
    \begin{split}
    |\tilde{\C}(\hat{\ga})-\tilde{\C}(\ga_2)|
    &\le \Big((\sqrt{M_{\ga_1}\cdot |A|\cdot T(w)}+2)\cdot |A|\cdot T(w)\\
    &\hspace{0.5cm}+|A|\cdot \pi_{max}(T(w))\cdot |\K|\Big)
    \cdot \sqrt{M_{\ga_1}}\cdot \max\{\sqrt{\gnorm{\ga_2-\ga_1}},\ \sqrt{M_{\ga_1}}\cdot\gnorm{\ga_2-\ga_1}\},
    \end{split}
    \end{equation}
    since $T(w')\le T(w).$
    Inequalities \eqref{eq:WE-demand-slice-Core1}--\eqref{eq:WE-demand-slice-Core2}
    then yield
    \begin{displaymath}
    \begin{split}
    |\tilde{\C}(\ga_2)-\tilde{\C}(\ga_1)|
    &\le 2\cdot \Big((\sqrt{M_{\ga_1}\cdot |A|\cdot T(w)}+2)\cdot |A|\cdot T(w)\\
    &\hspace{0.5cm}+|A|\cdot \pi_{max}(T(w))\cdot |\K|\Big)
    \cdot \sqrt{M_{\ga_1}}\cdot \max\{\sqrt{\gnorm{\ga_1-\ga_2}},\ \sqrt{M_{\ga_1}}\cdot\gnorm{\ga_1-\ga_2}\}.
    \end{split}
    \end{displaymath} 
    
    \proof{Proof of Lemma~\ref{lemma:HalfHoelderCostSlice}b):}
    
    Consider an arbitrary SO flow $f^*$ of $\ga_1.$
    \wusec{Similar to the proof of Lemma~\ref{lemma:HalfHoelderCostSlice}a),} there is a vector $\lambda=(\lambda_s)_{s\in\S}$
    such that $0\le \lambda_s\le f_s^*$ for
    each $s\in\S,$ and 
    $\sum_{s'\in \S_k}(f_{s'}^*-\lambda_{s'})=\hat{w}_k$
    for each $k\in\K.$
    Redefine $\ga'_1:=(\pi_{|\lambda},\hat{w})$
    with $\pi_{|\lambda}:=(\pi_{a|\lambda})_{a\in A},$
    $\pi_{a|\lambda}(x):=\pi_a(x+\lambda_a)$ and 
    $\lambda_a:=\sum_{s\in\S: a\in s}\lambda_s$ 
    for each $(a,x)\in A\times [0,T(\hat{w})].$
    
    \wusec{Then} $\ga'_1$ is a game when $\gnorm{\ga_1-\ga_2}
    <\frac{T(w)}{|\K|},$
    \wusec{and} $f^*-\lambda=(f^*_s-\lambda_s)_{s\in\S}$
    is a flow of $\ga'_1,$ but need not be an SO flow of $\ga'_1$.
    
    Let $f$ be a flow of $\ga_1$ such that $f-\lambda$
    is an SO flow of $\ga'_1.$ Such a flow $f$ exists
    since $f^{*'}+\lambda$ is a flow of $\ga_1$
    when $f^{*'}$ is an SO flow of $\ga'_1.$
    \wusec{Then}
    \begin{displaymath}
    \begin{split}
    \C^*(\ga_1)\le C(\ga_1,f)
    &=\sum_{a\in A}f_a\cdot \pi_a(f_a)
    = \sum_{a\in A}\lambda_a\cdot \pi_a(f_a)+\sum_{a\in A}
    (f_a-\lambda_a)\cdot \pi_{a}(f_a)\\
    &\le |A|\cdot |\K|\cdot \pi_{max}(T(w))\cdot \gnorm{\ga_1-\ga_2}+\sum_{a\in A}
    (f_a-\lambda_a)\cdot \pi_{a}(f_a)\\
    &=|A|\cdot |\K|\cdot \pi_{max}(T(w))\cdot \gnorm{\ga_1-\ga_2}+\sum_{a\in A}
    (f_a-\lambda_a)\cdot \pi_{a|\lambda}(f_a-\lambda_a)\\
    &=|A|\cdot |\K|\cdot \pi_{max}(T(w))\cdot \gnorm{\ga_1-\ga_2}+
    \C^*(\ga'_1)\\
    &\le |A|\cdot |\K|\cdot \pi_{max}(T(w))\cdot \gnorm{\ga_1-\ga_2}+C(\ga'_1,f^*-\lambda_a)\\
    &\le |A|\cdot |\K|\cdot \pi_{max}(T(w))\cdot \gnorm{\ga_1-\ga_2}+C(\ga_1,f^*)\\
    &=|A|\cdot |\K|\cdot \pi_{max}(T(w))\cdot \gnorm{\ga_1-\ga_2}+\C^*(\ga_1).
    \end{split}
    \end{displaymath}
    Here, we \wusec{have used} $\lambda_a=\sum_{k\in \K}\sum_{s\in\S_k:a\in s}\lambda_s
    \le \sum_{k\in \K}\sum_{s\in\S_k}\lambda_s=\sum_{k\in \K} (w_k-\hat{w}_k)
    \le |\K|\cdot \norm{w-\hat{w}}\le |\K|\cdot \norm{w-w'}\le 
    |\K|\cdot\gnorm{\ga_1-\ga_2}.$
    Then we \wusec{obtain}
    \begin{equation}\label{eq:SO-cost-slice-Import1}
    0\le \C^*(\ga_1)-\C^*(\ga'_1) \le |A|\cdot |\K|\cdot \pi_{max}(T(w))\cdot \gnorm{\ga_1-\ga_2}.
    \end{equation}
    
    Inequality \eqref{eq:SO-demand-slice} yields
    \begin{equation}\label{eq:SO-cost-slice-Import2}
    |\C^*(\ga'_1)-\C^*(\hat{\ga})|<|A|\cdot T(w)\cdot M_{\ga_1}\cdot \gnorm{\ga_1-\ga_2},
    \end{equation}
    \wusec{where we observe that} $T(\hat{w})\le T(w)$ and 
    $\gnorm{\ga'_1-\hat{\ga}}\le M_{\ga_1}\cdot \gnorm{\ga_1-\ga_2}.$
    Inequalities \eqref{eq:SO-cost-slice-Import1} \wusec{and} \eqref{eq:SO-cost-slice-Import2}
    together imply
    \begin{equation}\label{eq:SO-cost-slice-Core1}
    |\C^*(\hat{\ga}-\C^*(\ga_1)|\le \big(|A|\cdot |\K|\cdot \pi_{max}(T(w))+|A|\cdot T(w)\cdot M_{\ga_1}\big)\cdot \gnorm{\ga_1-\ga_2}.
    \end{equation}
    \wusec{Similar to the proof of Lemma~\ref{lemma:HalfHoelderCostSlice}a),} we have
    \begin{equation}\label{eq:SO-cost-slice-Core2}
    |\C^*(\hat{\ga})-\C^*(\ga_2)|\le \big(|A|\cdot |\K|\cdot \pi_{max}(T(w))+|A|\cdot T(w)\cdot M_{\ga_1}\big)\cdot \gnorm{\ga_1-\ga_2},
    \end{equation}
    since $T(w')\le T(w).$ Lemma~\ref{lemma:HalfHoelderCostSlice}b) then
    follows immediately from \eqref{eq:SO-cost-slice-Core1} \wusec{and} \eqref{eq:SO-cost-slice-Core2}.
    
    \proof{Proof of Lemma~\ref{lemma:HalfHoelderCostSlice}c):} Lemma~\ref{lemma:HalfHoelderCostSlice}a)--b)
    imply that
    \begin{displaymath}
    \begin{split}
    |\rho(\ga_1)-\rho(\ga_2)|
    &\le \frac{\tilde{\C}(\ga_1)
    	+\tilde{\mb}_{\ga_1}\cdot \max\{\sqrt{\gnorm{\ga_1-\ga_2}},\sqrt{M_{\ga_1}}\cdot\gnorm{\ga_1-\ga_2}\}}{\C^*(\ga_1)-\mb^*_{\ga_1}\cdot 
    	\gnorm{\ga_1-\ga_2}}
    -\frac{\tilde{\C}(\ga_1)}{\C^*(\ga_1)}\\
    &=\frac{\rho(\ga_1)\cdot \mb^*_{\ga_1}\cdot 
    	\gnorm{\ga_1-\ga_2}+\tilde{\mb}_{\ga_1}\cdot \max\{\sqrt{\gnorm{\ga_1-\ga_2}},\sqrt{M_{\ga_1}}\cdot\gnorm{\ga_1-\ga_2}\}}{\C^*(\ga_1)-\mb^*_{\ga_1}\cdot 
    	\gnorm{\ga_1-\ga_2}}\\
    &\le \frac{2\cdot \rho(\ga_1)\cdot \mb^*_{\ga_1}+2\cdot \tilde{\mb}_{\ga_1}}{\C^*(\ga_1)}
    \cdot \max\{\sqrt{\gnorm{\ga_1-\ga_2}},\sqrt{M_{\ga_1}}\cdot\gnorm{\ga_1-\ga_2}\}
    \end{split}
    \end{displaymath}
    when $\mb^*_{\ga_1}\cdot 
    \gnorm{\ga_1-\ga_2}\le \frac{\C^*(\ga_1)}{2}$
    and $\gnorm{\ga_1-\ga_2}<\frac{T(w)}{|\K|}.$
    
    This completes the proof of Lemma~\ref{lemma:HalfHoelderCostSlice}.
    
    \hfill$\square$

    
    While Lemma~\ref{lemma:HalfHoelderCostSlice}
    \wusec{shows} a \wusec{weaker} H{\"o}lder exponent
    $\frac{1}{2}$ than 
    \wusec{the exponent $1$} of \citet{EnglertFraOlb2010} and \citet{TakallooKwon2020},  it applies to more general cases.
    On \wuu{the} one hand, 
    two games $\ga_1=(\pi,w)$
    and $\ga_2=(\pi,w')$
    from the same cost slice $\Gamespace_{|\pi}$
    \wusec{need only} satisfy \wusec{the weaker} condition
    $T(w)\ge T(w')$ \wusec{instead of} the stronger condition
    ``$w=(1+\epsilon)\cdot w'$''.
    On the other hand, Lemma~\ref{lemma:HalfHoelderCostSlice} applies 
    to a cost slice $\Gamespace_{|\pi}$ with arbitrary continuously differentiable cost functions,
    as it
    assumes only Lipschitz continuity of the cost functions
    on the compact interval $[0,T(w)],$ but not \wusec{on} the whole
    unbounded interval $[0,\infty).$
    
    \wuu{The proof of Lemma~\ref{lemma:HalfHoelderCostSlice}
    	essentially builds on Lemma~\ref{lemma:PoA_Lipschiz}. Since we cannot improve the H{\"o}lder exponent $\frac{1}{2}$ of Lemma~\ref{lemma:PoA_Lipschiz}
    	for demand slices  (see Remark~\ref{remark:Exponent-LischpitzCost-Cannot-Improve}),  
    	we are presently also unable to do that
    	for cost slices in Lemma~\ref{lemma:HalfHoelderCostSlice}, although 
    the PoA might have stronger H{\"o}lder continuity properties on cost slices than demand slices (see Remark~\ref{remark:Cost-Slice}).
While \citet{EnglertFraOlb2010}, \citet{TakallooKwon2020}
and \citet{Cominetti2019} have already proposed independent techniques to
analyze the H{\"o}lder continuity
of the PoA on cost slices for some inspiring cases,
we are still eager for a general technique to independently analyze the
H{\"o}lder continuity of the PoA on cost slices with 
arbitrary Lipschitz continuous cost functions on $[0,T(d)].$
Nonetheless, we will see in Section~\ref{subsubsec:Finer_PoA_Approximation_Particular}
that the current technical framework used
in the proof of Lemma~\ref{lemma:HalfHoelderCostSlice} yields
a stronger H{\"o}lder exponent of $1$ for cost slices when the cost functions
have good properties similar to those in \citet{EnglertFraOlb2010}, \citet{EnglertFraOlb2010}
and \citet{Cominetti2019}.}

    \subsubsection{\wuu{Proof of Theorem~\ref{thm:HalfHoelderContinuity}}}
    \wuu{We now combine} Lemma~\ref{lemma:PoA_Lipschiz} and Lemma~\ref{lemma:HalfHoelderCostSlice}
    \wuu{to prove} Theorem~\ref{thm:HalfHoelderContinuity}.
    %
	Consider an arbitrary game $\ga=(\tau,d)\in\Gamespace$
	whose cost functions $\tau_a(x)$ are Lipschitz 
	continuous on $[0,T(d)]$ with a Lipschitz constant $M_{\ga}>0.$
	
	For an arbitrary game $\ga'=(\sigma,d')\in\Gamespace,$
	we define an auxiliary game $\ga''=(\hat{\tau},d')$
	with
	\begin{equation}\label{eq:IntermediateCost}
		\hat{\tau}_a(x)=
		\begin{cases}
		\tau_a(x)&\text{if }x\le T_{min}:=\min\{T(d),T(d')\}\\
		\tau_a(T_{\min})&\text{if }x>T_{min}.
		\end{cases}
		\quad \forall a\in A.
	\end{equation}
	Trivially, each auxiliary cost function $\hat{\tau}_a(x)$ is
	Lipschitz continuous on $[0,T(d')]$ with the \wusec{same} Lipschitz constant
	$M_{\ga},$ since $\tau_a(\cdot)$
	is non-decreasing. Moreover, we obtain by \eqref{eq:GeneralizedNorm} and \eqref{eq:Metric-Equivalence} that
	\begin{equation}\label{eq:IntermediateMetric}
		\begin{split}
		&\gnorm{\ga-\ga''}=\max\left\{
		\norm{d-d'}, \norm{\tau(T_{min})-\tau(T(d))}
		\right\}\\
		&\gnorm{\ga''-\ga'}
		=\max\left\{\max_{a\in A,x\in [0,T_{min}]}|\tau_a(x)-\sigma_a(x)|,
		\norm{\tau(T_{min})-\sigma(T(d'))}\right\}.
		\end{split}
	\end{equation}
	Again by \eqref{eq:GeneralizedNorm} and \eqref{eq:Metric-Equivalence},
	we have $\norm{\tau(T_{min})-\tau(T(d))}\le M_{\ga}\cdot |T(d')-T(d)|
	\le M_{\ga}\cdot |\K| \cdot \norm{d-d'}\le M_{\ga}\cdot |\K| \cdot\gnorm{\ga-\ga'},$ and 
	\begin{displaymath}
		\norm{\tau(T_{min})-\sigma(T(d'))}\le 
		\norm{\tau(T_{min})-\tau(T(d))}+\norm{\tau(T(d))-\sigma(T(d'))}
		\le (M_{\ga}\cdot |\K|+1)\cdot \gnorm{\ga-\ga'}.
	\end{displaymath}
	This together with \eqref{eq:IntermediateMetric} yields
	\begin{equation}\label{eq:IntermedianMetric-Relation}
		\gnorm{\ga-\ga''}\le \max\{M_{\ga}\cdot |\K|,1\}\cdot \gnorm{\ga-\ga'}
		\text{ and }
		\gnorm{\ga'-\ga''}\le (M_{\ga}\cdot |\K|+1)\cdot \gnorm{\ga-\ga'}.
	\end{equation}
	So both $\gnorm{\ga-\ga''}$ and $\gnorm{\ga'-\ga''}$ converge to
	$0$ as $\gnorm{\ga-\ga'}$ \wusec{tends to} $0.$
	
	Since $\ga'=(\sigma,d')$ and $\ga''=(\hat{\tau},d')$ belong the same demand slice 
	$\Gamespace_{|d'},$ and since the auxiliary cost functions
	$\hat{\tau}$ are Lipschitz continuous 
	on $[0,T(d')]$ with Lipschitz constant 
	$M_{\ga},$ Lemma~\ref{lemma:PoA_Lipschiz}d) implies that
	\begin{equation}\label{eq:First-Part-OneHalf-Hoelder}
		|\rho(\ga'')-\rho(\ga')|\le \mb_{\ga''}\cdot 
		\max\{\sqrt{\gnorm{\ga''-\ga'}},\gnorm{\ga''-\ga'}\}
	\end{equation}
	with a constant
	$
	\mb_{\ga''}>0
	$ depending only on $\ga'',$ 
	when $\gnorm{\ga'-\ga''}$ is small enough.
	Moreover, Lemma~\ref{lemma:PoA_Lipschiz}d) and Theorem~\ref{thm:PoA_Continuity} together imply that 
	the constant $\mb_{\ga''}$ converges to a constant 
	$\mb'_{\ga}>0$ depending only on $\ga$
	as $\ga'\to \ga$ (which implies 
	$\ga''\to \ga$ by \eqref{eq:IntermedianMetric-Relation}), since the constant
	$\mb_{\ga''}$ depends only on $M_{\ga},$
	$T(d'),$ $\C^*(\ga''),$ and $\rho(\ga''),$
	and since the Lipschitz constant $M_{\ga}$ does not change in the limit.
	This together with \eqref{eq:IntermedianMetric-Relation} implies that there is a small constant 
	$\epsilon_{1,\ga}>0$ such that
	\begin{equation}\label{eq:HalfHoelder-I}
		|\rho(\ga')-\rho(\ga'')|<2\cdot \mb'_{\ga}\cdot 
		\sqrt{(M_{\ga}\cdot |\K|+1)\cdot \gnorm{\ga-\ga'}}
	\end{equation}
	when $\gnorm{\ga-\ga'}<\epsilon_{1,\ga}.$

   When $T(d')\ge T(d),$ we obtain 
   by the definition \eqref{eq:IntermediateCost} of 
   the auxiliary cost functions $\hat{\tau}_a$
   that $\ga=(\tau,d)=(\hat{\tau},d)$ w.r.t.
   equivalence relation \eqref{eq:GameEquivalence}.
   Then $\ga$ and $\ga''$ are from the same cost slice 
   $\Gamespace_{|\hat{\tau}}$ with $T(d)\le T(d').$
   Lemma~\ref{lemma:HalfHoelderCostSlice}c) then yields
   \begin{equation}\label{eq:Part-II-I-Half-Hoelder}
   	|\rho(\ga)-\rho(\ga'')|\le \mb'_{\ga''}\cdot 
   	\max\{\sqrt{\gnorm{\ga-\ga''}},\gnorm{\ga-\ga''}\}
   \end{equation}
   for a constant $\mb'_{\ga''}>0$ depending on $\ga'',$
   when $\gnorm{\ga-\ga''}$ is small enough.
   Similar \wusec{to the previous arguments}, we obtain by Theorem~\ref{thm:PoA_Continuity}
   and Lemma~\ref{lemma:HalfHoelderCostSlice}c) that 
   this constant  $\mb'_{\ga''}$ converges also to a constant
   $\mb''_{\ga}$ as $\ga'\to\ga,$ and
   so
   \begin{equation}\label{eq:HoelderHalf-II}
   	|\rho(\ga)-\rho(\ga'')|\le 2\cdot \mb''_{\ga}\cdot 
   	\sqrt{\max\{M_{\ga}\cdot |\K|,1\}\cdot \gnorm{\ga-\ga'}}
   \end{equation}
   when $\gnorm{\ga-\ga'}<\epsilon_{2,\ga}$
   for a small constant $\epsilon_{2,\ga}>0.$
   
   When $T(d')<T(d),$ then $\ga''=(\hat{\tau},d')=(\tau,d')$ by equivalence relation~\eqref{eq:GameEquivalence}. Then 
   $\ga$ and $\ga''$ are from the same cost slice 
   $\Gamespace_{|\tau}$ with $T(d')<T(d).$
   Lemma~\ref{lemma:HalfHoelderCostSlice}c) then yields
   \begin{equation}\label{eq:Part-II-II-Half-Hoelder}
   |\rho(\ga)-\rho(\ga'')|\le \mb'''_{\ga}\cdot 
   \max\{\sqrt{\gnorm{\ga-\ga''}},\gnorm{\ga-\ga''}\}
   \le \mb'''_{\ga}\cdot\sqrt{\max\{M_{\ga}\cdot |\K|,1\}\cdot \gnorm{\ga-\ga'}}
   \end{equation}
   for a constant $\mb'''_{\ga}>0$ depending only on 
   $\ga,$ when $\gnorm{\ga-\ga'}<\epsilon_{3,\ga}$ for a small 
   constant $\epsilon_{3,\ga}>0$.

   Inequalities \eqref{eq:HalfHoelder-I}, \eqref{eq:HoelderHalf-II}  
   and \eqref{eq:Part-II-II-Half-Hoelder} together then imply
   Theorem~\ref{thm:HalfHoelderContinuity},
   since $|\rho(\ga)-\rho(\ga')|
   \le |\rho(\ga)-\rho(\ga'')|+|\rho(\ga'')-\rho(\ga')|$ and
   \wusec{the} arbitrary
   choice of $\ga'.$
  
   This completes the proof of Theorem~\ref{thm:HalfHoelderContinuity}.
   
   \hfill$\square$

    \subsection{\wuu{H{\"o}lder continuity of the PoA for special cost functions}}
    \label{subsubsec:Finer_PoA_Approximation_Particular}
    
    Although Theorem~\ref{thm:NotUniformlyHoelder} shows that
    the PoA map $\rho(\cdot)$ is not Lipschitz continuous on the whole
    game space $\Gamespace,$
    the differentiability results of \citet{Cominetti2019} \wusec{seemingly suggest} that 
    the map $\rho(\cdot)$ might be \emph{pointwise Lipschitz 
    continuous} (i.e., pointwise H{\"o}lder continuous with H{\"o}lder exponent $1$) 
    at each game $\ga=(\tau,d)$ whose cost functions 
    $\tau_a(\cdot)$ have strictly positive derivatives
    on the compact interval $[0,T(d)].$
    Theorem~\ref{thm:Finer_PoA_Approx_Particular} \wusec{confirms this.}
    \begin{theorem}
    	\label{thm:Finer_PoA_Approx_Particular}
    	Consider an arbitrary game $\ga=(\tau,d)\in\Gamespace.$
    	\begin{itemize}
    		\item[a)] If 
    		$\tau_a(x)$ \wusec{is constant} 
    		for all $a\in A$ and $x\in [0,T(d)],$ then there
    		are a small $\epsilon_{\ga}>0$ and 
    		a H{\"o}lder constant $\h_\ga>0$
    		depending only on $\ga$ such that
    		$|\rho(\ga)-\rho(\ga')|<\h_{\ga}\cdot \gnorm{\ga-\ga'}$
    		for every game $\ga'\in\Gamespace$
    		with $\gnorm{\ga-\ga'}<\epsilon_{\ga}.$
    		\item[b)] If $\tau_a(\cdot)$
    		is continuously differentiable on $[0,T(d)]$
    		and
    		$\tau'_a(x)\ge m_{\ga}$
    		for all $a\in A$ and $x\in [0,T(d)]$
    		for a constant $m_{\ga}>0$ depending only on $\ga,$
    		then there
    		are a small $\epsilon_{\ga}>0$ and 
    		a H{\"o}lder constant $\h_\ga>0$
    		depending only on $\ga$ such that
    		$|\rho(\ga)-\rho(\ga')|<\h_{\ga}\cdot \gnorm{\ga-\ga'}$
    		for every game $\ga'\in\Gamespace$
    		with $\gnorm{\ga-\ga'}<\epsilon_{\ga}.$
    		\end{itemize}
    \end{theorem}
    \proof{Proof of Theorem~\ref{thm:Finer_PoA_Approx_Particular}}
    \proof{Proof of Theorem~\ref{thm:Finer_PoA_Approx_Particular}a):}
    Assume that $\tau_a(x)\equiv \tau_a(0)$
    for \wu{all $a\in A$ and $x\in [0,T(d)].$}
    Trivially, \wusec{the} cost functions $\tau_a(\cdot)$ are Lipschitz \wusec{continuous}
    on $[0,T(d)]$ with Lipschitz constant 
    $M_{\ga}=1.$
    
    Let $\ga'=(\sigma,d')\in\Gamespace$
    be an arbitrary game. 
    Define an auxiliary cost function vector 
    $\hat{\tau}=(\hat{\tau}_{a})_{a\in A}$ with
    $
    	\hat{\tau}_a(x):=\tau_a(0)
    $ for all $a\in A$ and $x\in [0,\infty).$
    Let $\ga''=(\hat{\tau},d')$ be the corresponding
    auxiliary game.
    Then $\rho(\ga)=\rho(\ga'')=1,$ inequality \eqref{eq:IntermedianMetric-Relation} holds and $\ga''\to \ga$ as $\gnorm{\ga-\ga'}\to 0.$
    
    Lemma~\ref{lemma:PoA_Lipschiz}a) and \eqref{eq:IntermedianMetric-Relation} yield that
    \begin{equation}\label{eq:ConstCost-Demand-Slice-SO}
    	|\C^*(\ga'')-\C^*(\ga')|<|A|\cdot T(d')\cdot 
    	(|\K|+1)\cdot \gnorm{\ga-\ga'}\le 2\cdot |A|\cdot T(d)\cdot 
    	(|\K|+1)\cdot \gnorm{\ga-\ga'}
    \end{equation}
    when $T(d')\le 2\cdot T(d).$
    
    Let $\tilde{f}$ and $\tilde{g}$ be 
    a WE flow of $\ga''=(\hat{\tau},d')$ and a WE flow of $\ga'=(\sigma,d'),$
    respectively.
    We then have 
    \begin{equation}\label{eq:ConstCost-Demand-Slice-WE}
    	\begin{split}
    	\tilde{\C}(\ga')&-\tilde{\C}(\ga'')
    	=C(\ga',\tilde{g})-C(\ga'',\tilde{f})
    	=\sum_{a\in A}\sigma_a(\tilde{g}_a)\cdot 
    	\tilde{g}_a-\sum_{a\in A}\tau_a(0)\cdot \tilde{f}_a\\
    	&\le \sum_{a\in A}\big(\sigma_a(\tilde{g}_a) 
    	-\tau_a(0)\big)\cdot \tilde{f}_a
    	\le |A|\cdot T(d')\cdot \gnorm{\ga'-\ga}< 2\cdot |A|\cdot T(d)\cdot 
    	(|\K|+1)\cdot \gnorm{\ga-\ga'}
    	\end{split}
    \end{equation}
    when $T(d')\le 2\cdot T(d).$
    Inequalities \eqref{eq:ConstCost-Demand-Slice-SO} \wusec{and} \eqref{eq:ConstCost-Demand-Slice-WE} together imply that
    \begin{equation}\label{eq:ConstCost-PoA}
    	\begin{split}
    	0\le \rho(\ga')-\rho(\ga)&=\rho(\ga')-\rho(\ga'')
    	= \frac{\tilde{\C}(\ga')}{\C^*(\ga')}
    	-\frac{\tilde{\C}(\ga'')}{\C^*(\ga'')}\\
    	&\le \frac{\tilde{\C}(\ga'')+2\cdot |A|\cdot T(d)\cdot 
    		(|\K|+1)\cdot \gnorm{\ga-\ga'}}{\C^*(\ga'')-2\cdot |A|\cdot T(d)\cdot 
    		(|\K|+1)\cdot \gnorm{\ga-\ga'}}-\frac{\tilde{\C}(\ga'')}{\C^*(\ga'')}\\
    	&< \frac{8\cdot |A|\cdot T(d)\cdot 
    	(|\K|+1)}{\C^*(\ga)}\cdot \gnorm{\ga-\ga'}
    	\end{split}
    \end{equation}
    when $\gnorm{\ga-\ga'}$ is small.
    Here, we notice that $T(d')<2\cdot T(d)$
    holds when $\gnorm{\ga-\ga'}$ is small.
    This completes the proof of Theorem~\ref{thm:Finer_PoA_Approx_Particular}a).
    
    \proof{Proof of Theorem~\ref{thm:Finer_PoA_Approx_Particular}b):}
    Assume now that $\tau_a(\cdot)$
    is continuously differentiable on $[0,T(d)]$
    and
    $\tau'_a(x)\ge m_{\ga}$
    for all $a\in A$ and $x\in [0,T(d)]$
    for a constant $m_{\ga}>0$ depending only on $\ga.$
    Trivially, $\tau_a(\cdot)$ is Lipschitz continuous
    on $[0,T(d)]$ with Lipschitz constant
    $M_{\ga}:=\max_{b\in A}\max_{x\in [0,1]}\ \tau'_b(x)\ge m_{\ga}>0$
    for each arc $a\in A.$
    
    Let $\ga'=(\sigma,d')\in\Gamespace$
    be an arbitrary game. Define an auxiliary cost function vector 
    $\hat{\tau}$ with
    \begin{equation}\label{eq:PosiAuxCostFunction-d>d'}
        \hat{\tau}_a(x):=\tau_a(x)\quad \forall 
        a\in A\ \forall x\in [0,T(d')]
    \end{equation}
    when $T(d)\ge T(d'),$
    and 
    \begin{equation}\label{eq:PosiAuxCostFunction-d<=d'}
    	\hat{\tau}_a(x):=
    	\begin{cases}
    	\tau_a(x)&\text{if }x\le T(d)\\
    	\tau_a(T(d))+\tau'_a(T(d))\cdot (x-T(d))&\text{if }x\in (T(d),T(d')]
    	\end{cases}
    	\quad \forall a\in A\ \forall x\in [0,T(d')]
    \end{equation}
    when $T(d')>T(d).$
    Denote by $\ga''$ the game $(\hat{\tau},d').$
    Then the cost functions $\hat{\tau}_a$
    are continuously differentiable on 
    $[0,T(d')],$ and  $0< m_{\ga}\le \min_{a\in A,x\in [0,T(d')]}
    \hat{\tau}'_a(x)\le 
    \max_{a\in A,x\in [0,T(d')]}\hat{\tau}'_a(x)\le 
    M_{\ga}.$
    Similar \wusec{to} \eqref{eq:IntermedianMetric-Relation}, we have 
    \begin{equation}\label{eq:PosiAuxiCost-Metric-Relation}
    	\begin{split}
    	\gnorm{\ga''-\ga'}&=\norm{\sigma_{|T(d')}-\hat{\tau}_{|T(d')}}
    	\le \gnorm{\ga-\ga'}+M_{\ga}\cdot |T(d')-T(d)|\le 
    	(1+M_{\ga}\cdot |\K|)\cdot \gnorm{\ga-\ga'}\\
    	\gnorm{\ga-\ga''}&=\max\{\norm{d-d'},\norm{\tau_{|T(d)}-\hat{\tau}_{|T(d')}}\}
    	\le \gnorm{\ga-\ga'}+M_{\ga}\cdot |T(d')-T(d)|\\
    	&\hspace{2cm}\le 
    	(1+M_{\ga}\cdot |\K|)\cdot \gnorm{\ga-\ga'}.
    	\end{split}
    \end{equation}
    
    Lemma~\ref{lemma:PoA_Lipschiz}a) and inequality
    \eqref{eq:PosiAuxiCost-Metric-Relation} then yield
    \begin{equation}\label{eq:PosiAnalytic-Demand-Slice-SO}
    |\C^*(\ga'')-\C^*(\ga')|<|A|\cdot T(d')\cdot 
    (M_{\ga}\cdot |\K|+1)\cdot \gnorm{\ga-\ga'}\le 2\cdot |A|\cdot T(d)\cdot 
    (M_{\ga}\cdot |\K|+1)\cdot \gnorm{\ga-\ga'},
    \end{equation}
     when 
    $\gnorm{\ga-\ga'}$ is small, since 
    the cost functions $\hat{\tau}_a$ are
    \WU{Lipschitz} continuous on $[0,T(d')]$
    with \WU{Lipschitz} constant $M_{\ga},$ and 
    since both $\ga'=(\sigma,d')$ and 
    $\ga''=(\hat{\tau},d')$ are from the
    \wu{same} demand slice $\Gamespace_{|d'}.$
    
    Let $\tilde{f}$ and $\tilde{g}$ be a WE flow of
    $\ga''$ and a WE flow of $\ga',$ respectively.
    Lemma~\ref{lemma:PoA_Lipschiz}c) and Lemma~\ref{lem:EpsilonNE}b) together yield that 
    \begin{equation}\label{eq:Mutual-Epsilon-WE}
    	\begin{split}
    	&0\le \sum_{a\in A}\hat{\tau}_a(\tilde{f}_a)\cdot (\tilde{g}_a-\tilde{f}_a)
    	\le \sum_{a\in A}\hat{\tau}_a(\tilde{g}_a)\cdot (\tilde{g}_a-\tilde{f}_a)
    	\le |A|\cdot T(d')\cdot \gnorm{\ga'-\ga''},\\
    	&0\le \sum_{a\in A}\sigma_{a}(\tilde{g}_a)\cdot (\tilde{f}_a-\tilde{g}_a)
    	\le \sum_{a\in A}\sigma_{a}(\tilde{f}_a)\cdot (\tilde{f}_a-\tilde{g}_a)\le |A|\cdot T(d')\cdot \gnorm{\ga'-\ga''},
    	\end{split}
    \end{equation}
    which in turn implies that
    \begin{equation}
    	0\le \sum_{a\in A} (\hat{\tau}_a(\tilde{f}_a)-\sigma_{a}(\tilde{g}_a))\cdot 
    	(\tilde{g}_a-\tilde{f}_a)\le \sum_{a\in A} (\hat{\tau}_a(\tilde{g}_a)-\sigma_{a}(\tilde{f}_a))\cdot 
    	(\tilde{g}_a-\tilde{f}_a)\le 2\cdot |A|\cdot T(d')\cdot \gnorm{\ga'-\ga''}.\label{eq:f+g-WE}
    \end{equation}
    Inequalities~\eqref{eq:f+g-WE} and \eqref{eq:PosiAuxiCost-Metric-Relation} imply that
    \begin{equation}\label{eq:Magic-Regular-Rule}
    	\begin{split}
    	0\le m_{\ga}\cdot \sum_{a\in A}
    	|\tilde{g}_a-\tilde{f}_a|^2&\le \sum_{a\in A}\big(\hat{\tau}_a(\tilde{g}_a)-\hat{\tau}_a(\tilde{f}_a)\big)
    	\cdot \big(\tilde{g}_a-\tilde{f}_a\big)\\
    	&\le \sum_{a\in A}(\hat{\tau}_a(\tilde{g}_a)-\sigma_{a}(\tilde{g}_a))
    	\cdot (\tilde{g}_a-\tilde{f}_a)
    	\le \gnorm{\ga'-\ga''}\cdot \sum_{a\in A}|\tilde{g}_a-\tilde{f}_a| \\
    	&\le (1+|\K|\cdot M_{\ga})\cdot \gnorm{\ga'-\ga}\cdot \sum_{a\in A}|\tilde{g}_a-\tilde{f}_a|\\
    	&\le (1+|\K|\cdot M_{\ga})\cdot \gnorm{\ga'-\ga}\cdot \sqrt{\sum_{a\in A}|\tilde{g}_a-\tilde{f}_a|^2}.
    	\end{split}
    \end{equation}
    Inequality \eqref{eq:Magic-Regular-Rule} yields 
    immediately that 
    \begin{equation}\label{eq:PosiCost-Differ}
    	|\hat{\tau}_a(\tilde{f}_a)-\hat{\tau}_{a}(\tilde{g}_a)|
    	\le \frac{M_{\ga}\cdot (1+|\K|\cdot M_{\ga})}{m_{\ga}}\cdot \gnorm{\ga'-\ga}\quad \forall a\in A.
    \end{equation}
    
    We then obtain by \eqref{eq:PosiAuxiCost-Metric-Relation}, \eqref{eq:Mutual-Epsilon-WE} and \eqref{eq:PosiCost-Differ} that
    \begin{equation}\label{eq:PosiCost-Differ-WE}
    \begin{split}
    	\big|\tilde{\C}(\ga')-&\tilde{\C}(\ga'')\big|
    = \big|\sum_{a\in A}\big(\sigma_{a}(\tilde{g}_a)
    \cdot \tilde{g}_a-\hat{\tau}_a(\tilde{f}_a)
    \cdot \tilde{f}_a\big)\big|\\
    &= \big|\sum_{a\in A}\big(\sigma_{a}(\tilde{g}_a)-\hat{\tau}_a(\tilde{g}_a)\big)
    \cdot \tilde{g}_a+\sum_{a\in A}
    \big(\hat{\tau}_a(\tilde{g}_a)\cdot \tilde{g}_a-\hat{\tau}_a(\tilde{f}_a)
    \cdot \tilde{f}_a\big)\big|\\
    &=\sum_{a\in A}\big|\sigma_{a}(\tilde{g}_a)-\hat{\tau}_a(\tilde{g}_a)\big|
    \cdot \tilde{g}_a+\sum_{a\in A}
    \big|\hat{\tau}_a(\tilde{g}_a)-\hat{\tau}_a(\tilde{f}_a)\big|\cdot \tilde{g}_a+\sum_{a\in A}\hat{\tau}_a(\tilde{f}_a)
    \cdot \big(\tilde{g}_a-\tilde{f}_a\big)\\
    &\le 2\cdot |A|\cdot T(d')\cdot \gnorm{\ga'-\ga''}
    +|A|\cdot T(d')\cdot \frac{M_{\ga}\cdot (1+|\K|\cdot M_{\ga})}{m_{\ga}}\cdot \gnorm{\ga'-\ga}\\
    &< 2\cdot \left(2+\frac{M_{\ga}}{m_{\ga}}\right)\cdot |A|\cdot 
    T(d)\cdot (1+|\K|\cdot M_{\ga})\cdot \gnorm{\ga'-\ga}
    \end{split}
    \end{equation}
    when $\gnorm{\ga-\ga'}$ is small.
    
    Inequalities \eqref{eq:PosiAnalytic-Demand-Slice-SO}
    and \eqref{eq:PosiCost-Differ-WE} together then \wusec{give}
    \begin{equation}\label{eq:Posi-PoA-differ-Demand-Slice}
    	|\rho(\ga')-\rho(\ga'')|<\frac{
    		4+4\cdot \big(2+\frac{M_{\ga}}{m_{\ga}}\big)\cdot 
    		\rho(\ga)
    	}{\C^*(\ga)}\cdot |A|\cdot 
    T(d)\cdot (1+|\K|\cdot M_{\ga})\cdot \gnorm{\ga'-\ga}
    \end{equation}
    when $\gnorm{\ga-\ga'}$ is small.
    
    By an argument similar with that for Lemma~\ref{lemma:HalfHoelderCostSlice}, we obtain that
    \begin{equation}\label{eq:WE-Posit-diff-cost-slice}
    	\begin{split}
    	|\tilde{\C}(\ga)-\tilde{\C}(\ga'')|&\le 2\cdot \left(2+\frac{M_{\ga}}{m_{\ga}}\right)\cdot |A|\cdot 
    	T(d)\cdot (1+|\K|\cdot M_{\ga})\cdot \gnorm{\ga'-\ga}\\
    	&\hspace{2cm}+2\cdot |A|\cdot \tau_{max}(T(d))\cdot |\K|\cdot (1+|\K|\cdot M_{\ga})\cdot \gnorm{\ga'-\ga},
    	\end{split}
    \end{equation}
    when $\gnorm{\ga-\ga'}$ is small,
    where $\tau_{\max}(T(d)):=\max_{a}\tau_a(T(d)).$
    Here, we use \eqref{eq:PosiCost-Differ-WE}
    instead of \eqref{eq:WE-demand-slice-int1}.
    Moreover, Lemma~\ref{lemma:HalfHoelderCostSlice}b) yields
    that
    \begin{equation}\label{eq:Posi-SO-diff}
    	|\C^*(\ga)-\C^*(\ga'')|
    	<4\cdot \big(|A|\cdot |\K|\cdot \tau_{max}(T(d))+|A|\cdot T(d)\cdot M_{\ga_1}\big)\cdot (1+|\K|\cdot M_{\ga})\cdot \gnorm{\ga'-\ga}
    \end{equation}
    when $\gnorm{\ga-\ga'}$ is small.
    Inequalities \eqref{eq:WE-Posit-diff-cost-slice} \wusec{and} \eqref{eq:Posi-SO-diff}
    imply that 
    \begin{equation}\label{eq:Posi-PoA-Cost-Slice}
    	|\rho(\ga)-\rho(\ga'')|
    	<\mb_{\ga}(M_{\ga},m_{\ga},T(d),\tau_{\max}(T(d)))\cdot 
    	\gnorm{\ga-\ga'}
    \end{equation}
    when $\gnorm{\ga-\ga'}$ is small.
    Here $\mb_{\ga}(M_{\ga},m_{\ga},T(d),\tau_{\max}(T(d)))>0$
    is a constant depending only on $M_{\ga},$ $m_{\ga},$ $T(d),$ and $\tau_{\max}(T(d))$
    of $\ga.$ 
    
    Inequalities \eqref{eq:Posi-PoA-differ-Demand-Slice}
    and \eqref{eq:Posi-PoA-Cost-Slice} together 
    imply Theorem~\ref{thm:Finer_PoA_Approx_Particular}b).

    This completes the proof of Theorem~\ref{thm:Finer_PoA_Approx_Particular}.
    
    \hfill$\square$
    
    \wuu{We have been able to obtain a stronger H{\"o}lder exponent
    in Theorem~\ref{thm:Finer_PoA_Approx_Particular}a)--b) since 
we no longer need Lemma~\ref{lem:EpsilonNE}c) in the
H{\"o}lder continuity analysis of the PoA on demand slices when the
cost functions are constants or have strictly
positive first-order derivatives.
For constant cost functions, we have used inequality
\eqref{eq:ConstCost-Demand-Slice-WE} instead of Lemma~\ref{lem:EpsilonNE}c).
For cost functions with strictly positive first-order derivatives,
we have used inequalities \eqref{eq:f+g-WE}--\eqref{eq:PosiCost-Differ-WE}  
instead.
Unfortunately, these inequalities do not hold for 
arbitrary Lipschitz continuous cost functions on $[0,T(d)].$
So far, we are still lack a unified technique to replace
Lemma~\ref{lem:EpsilonNE}c) in the H{\"o}lder analysis of the PoA 
on demand slices with  arbitrary Lipschtiz continuous cost functions on $[0,T(d)].$}  

    \subsection{Open questions}\label{subsec:OpenQuestions}
    
    \wuu{We have shown 
    for Lipschitz continuous cost functions on $[0,T(w)]$ 
    that the PoA is pointwise H{\"o}lder continuous
    with H{\"o}lder exponent
    $\frac{1}{2}$ on a demand slice $\Gamespace_{|w},$ see
    Lemma~\ref{lemma:PoA_Lipschiz}d).
    At present, we are unable to improve
    this exponent, since we need
Lemma~\ref{lem:EpsilonNE}c) to bound $|C(\ga_1,\tilde{f})-C(\ga_2,\tilde{g})|$
for two games $\ga_1=(\pi^{(1)},w)$ and $\ga_2=(\pi^{(2)},w)$
from the same demand slice $\Gamespace_{|w},$
see \eqref{eq:Demand-Slice-WE-cost-diff}, and since
both Lemma~\ref{lem:EpsilonNE}c) and the upper bound on the approximation threshold in Lemma~\ref{lemma:PoA_Lipschiz}c) are tight, see Example~\ref{example:Tightness_Of_epsilon_WE} and Remark~\ref{remark:Exponent-LischpitzCost-Cannot-Improve}.  
Hence, it is unclear to us if the exponent $\frac{1}{2}$ is tight
for arbitrary Lipschitz continuous cost functions  on 
$[0,T(w)].$ We leave this as an open question.
\begin{open}\label{OpenQuestion:TightnessOf1/2}
	Is the H{\"o}lder exponent $\frac{1}{2}$ in Lemma~\ref{lemma:PoA_Lipschiz}d) tight
	for  Lipschitz continuous cost functions
	on $[0,T(w)]$?
\end{open}

Although we can neither affirm nor negate
this question, there is some evidence
for a negative answer.
Both Example~\ref{example:Tightness_Of_epsilon_WE}
and Remark~\ref{remark:Exponent-LischpitzCost-Cannot-Improve}
have only used constant cost functions or polynomial cost functions with strictly positive first-order derivatives, for which 
we have shown an improved H{\"o}lder exponent in Theorem~\ref{thm:Finer_PoA_Approx_Particular}.
Hence, cost functions making Lemma~\ref{lem:EpsilonNE}c)
tight need not make the approximation threshold upper bound
in Lemma~\ref{lemma:PoA_Lipschiz}c) tight, and vice versa.
Nonetheless, it will be challenging to show an improved H{\"o}lder exponent for arbitrary Lipschitz continuous cost functions 
on $[0,T(w)].$ This will require an alternative approach  to bound
$|C(\ga_1,\tilde{f})-C(\ga_2,\tilde{g})|$, even different from our special approaches for
cost functions that are constants or have strictly
positive first-order derivatives in Section~\ref{subsubsec:Finer_PoA_Approximation_Particular}.}

    %
    
    
    Theorem~\ref{thm:Finer_PoA_Approx_Particular}
    implies directly that the \wusec{\emph{upper derivative}}
    \begin{equation}\label{eq:Upper-Drivatives}
    	\overline{D}_\rho(\ga):=\overline{\lim}_{\ga'\to\ga}
    	\frac{|\rho(\ga')-\rho(\ga)|}{\gnorm{\ga'-\ga}}
    \end{equation}
    of the PoA map $\rho(\cdot)$ at a game $\ga=(\tau,d)$
    exists and
    is bounded from above by a finite H{\"o}lder constant
    $\h_{\ga}>0$ when the cost functions
    $\tau_a(\cdot)$ have strictly positive derivatives
    on the compact interval $[0,T(d)].$
    While \citet{Cominetti2019}
    obtained a \wusec{stronger} differentiability result of the resulting 
    PoA function on
    a \emph{cost slice} $\Gamespace_{|\tau}$
    for \wu{networks with one O/D pair,} we are unable
    to generalize \wusec{their} result to the whole game space
    $\Gamespace$ \wu{on networks with multiple O/D pairs,} as \wu{our} PoA map
    $\rho(\cdot)$ is not an ordinary
    real-valued function, but a functional
    on the game space $\Gamespace$. In particular, it is \wuu{also} unclear
    to us \wuu{under which conditions} the upper derivative $\overline{D}_\rho(\ga)$
    coincides with the \emph{lower derivative}
    \begin{equation}\label{eq:Lower-Derivatives}
    	\underline{D}_\rho(\ga):=
    	\underline{\lim}_{\ga'\to\ga}
    	\frac{|\rho(\ga')-\rho(\ga)|}{\gnorm{\ga'-\ga}},
    \end{equation}
    although it is clear that $0\le \underline{D}_\rho(\ga)\le \overline{D}_\rho(\ga).$
    \wuu{We leave this also} as an open question.
    \begin{open}\label{open:Differentiability}
    	\WU{Which condition on $\ga=(\tau,d)$ implies that $\underline{D}_\rho(\ga)= \overline{D}_\rho(\ga)$}?
    \end{open}


\wulast{Addressing Open Question~\ref{open:Differentiability} on the game space $\Gamespace$ might be too ambitious due to 
the extremely complicated structure of its topology. 
Inspired by \citet{Patriksson2004}, \citet{Patriksson2007},
\citet{Lu2017}, and \citet{Klimm2021},
a promising first step is to
	consider the differentiability of the PoA on a particular subspace of 
	$\Gamespace$ that can be parameterized and is homomorphic 
	to some finite dimensional Euclidean space. 
Then classic Calculus techniques for finite dimensional Euclidean spaces apply. This may
facilitate the differentiability analysis of the PoA. 
Nevertheless, we will not continue this direction in the current paper, and would like to 
leave it for future work.}

    \section{\wuu{An application to the convergence rate of the PoA}}
    \label{sec:Applications}
    As an application of \wu{our}
    H{\"o}lder continuity results, we now demonstrate that they
    help \wusec{to analyze} the convergence rate of the PoA
    in non-atomic congestion games for both growing and decreasing demands.

    \wusec{The} \emph{convergence analysis} of the PoA \wusec{investigates}
    the limit of the PoA sequence $\big(\rho(\ga_n)\big)_{n\in\N}$
    of a sequence $(\ga_n)_{n\in\N}$ \wusec{of games}
    \wu{when} all components $\ga_n=(\tau,d^{(n)})$
    \wu{belong to} the same cost slice $\Gamespace_{|\tau}$ \wu{and} \wusec{the} total demand
    $T(d^{(n)})$ \wusec{tends to zero or infinity, i.e.,}
    $\lim_{n\to\infty}T(d^{(n)})= 0$ or $\infty.$
    When  $\lim_{n\to\infty}\rho(\ga_n)= 1$
    for an \emph{arbitrary} sequence $(\ga_n)_{n\in\N}
    \in \Gamespace_{|\tau}^{\N}$
    with $\lim_{n\to \infty}T(d^{(n)})= \infty,$ then 
    \wusec{we say that} the cost slice $\Gamespace_{|\tau}$
    \emph{behaves well for growing demands.} Notice that 
    this \WU{notion has been introduced by 
    \citet{Colini2016On} and}
    corresponds to \WU{the notion of \emph{asymptotically
    	well designed} \wusec{introduced} by \citet{Wu2019}.}
    
    Similarly, when  $\lim_{n\to\infty}\rho(\ga_n)=1$
    for an \emph{arbitrary} sequence $(\ga_n)_{n\in\N}
    \in \Gamespace_{|\tau}^{\N}$
    with $\lim_{n\to \infty}T(d^{(n)})= 0,$ then 
    \wusec{we say that} the cost slice $\Gamespace_{|\tau}$
    \emph{behaves well for decreasing demands.} 
    In particular, the cost slice $\Gamespace_{|\tau}$
    is said to \emph{behave well in limits}
    when it behaves well for both decreasing and growing
    demands.  When a cost slice $\Gamespace_{|\tau}$
    behaves well in limits, then the PoA map $\rho(\cdot)$
    has a \emph{tight} and \emph{finite} upper bound on $\Gamespace_{|\tau},$ although the cost slice 
    $\Gamespace_{|\tau}$ is not compact w.r.t. the topology
    \wu{from} Section~\ref{sec:Topology}.  
    This follows since the PoA \wuu{is continuous and thus} has a tight and finite
    upper bound on each \wuu{compact} subspace 
    $\{\ga'=(\tau,d)\in \Gamespace_{|\tau}: N_1\le \norm{d}\le N_2\}$ of the cost slice $\Gamespace_{|\tau}$ 
    for two arbitrary  positive reals $N_1$ and 
    $N_2$ with $N_1<N_2.$


   \subsection{The state of the art}
    \label{subsec:ConvergenceAnalysis-Reference-Summary}
    Recent results by \wuu{\citet{Colini2017WINE,Colini2020OR} and}
    \citet{Wu2019} have already shown that 
    every cost slice $\Gamespace_{|\tau}$
    with \emph{regularly varying} (\citet{Bingham1987Regular}) cost functions
    behaves well for \emph{growing} demands.
    This applies directly to 
    cost slices $\Gamespace_{|\tau}$
    with cost functions $\tau_a(\cdot)$
    that are arbitrary polynomials, arbitrary logarithms, or 
    products of polynomials and logarithms,
    and thus confirms \wusec{the} \wu{earlier} observed
    convergence of the empirical PoA for
    growing demands by 
    \citet{Youn2008Price,Harks2015Computing,O2016Mechanisms}
    and \citet{Monnot2017}.
    
    When the cost functions $\tau_a(\cdot)$
    are \wuu{of the form
    $\tau_a(x)=\sum_{n\in\N}\xi_{a,n}\cdot x^n$}
    for each \WU{$a\in A$ and each $x\in [0,\infty)$,} \wu{then} \citet{Colini2017WINE,Colini2020OR}
    \wu{have shown} \wusec{that}
    $\lim_{n\to \infty}\rho(\ga_n)=1$
    for each sequence $(\ga_n)_{n\in\N}\in\Gamespace_{|\tau}^{\N}$
    with a demand sequence $(d^{(n)})_{n\in\N}$ satisfying
    $\lim_{n\to \infty}T(d^{(n)})=0$ and the condition that
    \begin{equation}\label{eq:Colini-Cond}
    	\varliminf_{n\to\infty}\frac{d_k^{(n)}}{T(d^{(n)})}>0\quad
    	\forall k\in\K.
    \end{equation}
    This is \wuu{presently} the \emph{only} known convergence result  for decreasing demands. 

    For polynomial cost functions
    	$\tau_a(\cdot)$ and a sequence $(\ga_n)_{n\in\N}$ 
    of games $\ga_n=(\tau=(\tau_a)_{a\in A},d^{(n)})$ 
    satisfying the condition that
    \begin{equation}\label{eq:Colini-Cond-2}
    \frac{d_k^{(n)}}{T(d^{(n)})}=: d_k>0\quad\forall
    k\in\K\ \forall n\in\N,
    \end{equation}   \citet{Colini2017WINE,Colini2020OR} \wuu{have shown} further that $\rho(\ga_n)=1+O(\frac{1}{T(d^{(n)})})$
    \wusec{when} $\lim_{n\to \infty}T(d^{(n)})=\infty,$
    and that $\rho(\ga_n)=1+O(T(d^{(n)}))$
    \wusec{when} $\lim_{n\to \infty}T(d^{(n)})=0.$
    \wusec{Here,}
    the rate $d_k$ in \eqref{eq:Colini-Cond-2} is a \emph{constant} independent of $n$ 
    for each \wu{O/D pair} $k\in\K.$ 
    
    \wuu{With a different technique,} \citet{Wu2019} \wuu{have shown} that condition
    \eqref{eq:Colini-Cond-2} \wusec{can} be removed
    when the cost functions possess certain good properties.
    In particular, they proved that
     $\rho(\ga)=1+o(\frac{1}{T(d)^{\beta}})$
    for each game $\ga=(\tau,d)\in\Gamespace_{|\tau}$
    when the total demand $T(d)$
    of the game $\ga$ is large and the cost functions $\tau_a(\cdot)$
    are of the \emph{BPR type} (\citet{BPR}) and have the same degree $\beta>0.$
    \wu{Moreover,} \citet{Wu2019} \wuu{have} illustrated on
    \wusec{an example} game $\ga=(\tau,d)$ with BPR cost functions that
    the rate at which $\rho(\ga)$
    converges to $1$ depends crucially on the \emph{\wuu{growth}
    pattern} of the demands.
    \wuu{For} each constant exponent $\theta\in <\beta+1,2\cdot \beta>,$
    there is a game sequence $(\ga_n)_{n\in\N}\in\Gamespace_{|\tau}^{\N}$
    such that $\lim_{n\to\infty}T(d^{(n)})=\infty$
    and $\rho(\ga_n)=1+\Theta(\frac{1}{T(d^{(n)})^{\theta}})$
    for large $n.$
    Here, $<v_1,v_2>$
    denotes the closed interval 
    $[\min\{v_1,v_2\},\max\{v_1,v_2\}]$
    for arbitrary two real numbers
    $v_1$ and $v_2.$ This negates a conjecture proposed by \citet{O2016Mechanisms}.
    
    
    \subsection{Convergence rates of the PoA \wusec{for decreasing demands}}
    \label{subsec:Light_traffic}
    
   Consider now \wusec{a} vector
   $\tau=(\tau_a)_{a\in A}$ \wusec{of cost functions
   $\tau_a(\cdot)$ that are defined} on the unbounded
   interval $[0,\infty),$ \wusec{strictly positive} and Lipschitz \wusec{continuous} on
   a compact interval $[0,b]$ with 
   Lipschitz constant $M_{\tau}>0$ for a constant $b\in (0,\infty).$
   \wusec{We now show with Theorem~\ref{thm:Finer_PoA_Approx_Particular}a)} that the cost slice 
   $\Gamespace_{|\tau}$ \wusec{of $\tau$} behaves well
   for decreasing demands.
   
   \wusec{Let} $\sigma_a(x)\equiv \tau_a(0)>0$ for each
   $a\in A$ and $x\in [0,\infty).$
   For each game \wu{$\ga=(\tau,d)\in \Gamespace_{|\tau},$}
   we define an auxiliary game \wu{$\ga':=(\sigma,\frac{d}{T(d)})\in\Gamespace.$}
   Clearly, $\rho(\ga')\equiv 1$
   for an arbitrary demand vector $d,$ since the
   cost functions $\sigma_a(\cdot)$ are \wusec{constant.}
   By \eqref{eq:GeneralizedNorm}, we obtain
   that
   \begin{displaymath}
   	\gnorm{\ga'-\Lambda_{T(d)}(\ga)}=\max_{a\in A,x\in [0,1]}
   	|\tau_a(T(d)\cdot x)-\sigma_a(x)|=
   	\max_{a\in A}\
   	|\tau_a(T(d))-\tau_a(0)|\le M_\tau\cdot T(d)
   \end{displaymath}
   when $T(d)\le b.$
   Here, we \wusec{use} that 
   $\Lambda_{T(d)}(\ga)\in \Gamespace$
   is  the game \wusec{resulting from}
   the demand normalization
   to $\ga$ with factor $T(d),$
   and that both $\Lambda_{T(d)}(\ga)$
   and $\ga'$ are from the same demand slice
   $\Gamespace_{|\frac{d}{T(d)}}$
   and so have the same total demand 
   $T(\frac{d}{T(d)})= 1.$
   
   \wusec{Let} $\tau_{\min}(0):=\min_{a\in A}\tau_a(0)>0.$
   Then $\C^*(\ga')\ge \tau_{\min}(0)$
   for every demand vector $d$ of $\ga,$
   since $\ga'$ has the total demand 
   $T(\frac{d}{T(d)})= 1.$
   \wusec{Inequality}~\eqref{eq:ConstCost-PoA}
   then applies and yields
   \begin{displaymath}
   	\begin{split}
   	|\rho(\ga')-\rho(\ga)|=|\rho(\Lambda_{T(d)}(\ga))-\rho(\ga')|=
   	\rho(\Lambda_{T(d)}(\ga))-1
   	\le \frac{8\cdot |A|\cdot 
   		(|\K|+1)}{\tau_{\min}(0)}\cdot M_{\tau}\cdot T(d)
   	\end{split}
   \end{displaymath}
   when $T(d)$ is small, which tends to 
   \wu{$0$} as $T(d)\to 0.$
   Here, we \wusec{use} that both the lower bound 
   $\tau_{\min}(0)$
   of the SO cost $\C^*(\ga')$
   and the total demand 
   $T(\frac{d}{T(d)})$ ($\equiv 1$)
   of the game $\ga'$ do not
   depend on the demand vector $d$ of $\ga$
   though the game $\ga'$ does, and that 
   the \emph{pointwise}
   H{\"o}lder constant in \eqref{eq:ConstCost-PoA} 
   depends only on the SO cost and
   the total demand, and is thus bounded
   from above by \wusec{the} constant $\frac{8\cdot |A|\cdot 
   	(|\K|+1)}{\tau_{\min}(0)}$
   independent of the demand vector
   $d$ of $\ga$.
   
   Hence, $\Gamespace_{|\tau}$ behaves well for decreasing
   demands.
   We summarize in Corollary~\ref{corollary:ConvergencePoA_Light}. 
    \begin{corollary}\label{corollary:ConvergencePoA_Light}
    	Consider an arbitrary vector
    	$\tau=(\tau_a)_{a\in A}$ \wusec{of cost functions
    	$\tau_a(\cdot)$ that are defined} on the unbounded
    	interval $[0,\infty),$ strictly positive and Lipschitz \wusec{continuous} on
    	a compact interval $[0,b]$ with 
    	Lipschitz constant $M_{\tau}>0$ for a constant $b\in (0,\infty).$
    	Then $\rho(\ga)\le 1+ \frac{8\cdot M_{\tau}\cdot |A|\cdot 
    		(|\K|+1)}{\tau_{\min}(0)}\cdot M_{\tau}\cdot T(d)$ for every game $\ga=(\tau,d)\in\Gamespace_{|\tau}$
    	when the total demand $T(d)$ is small
    	and $\tau_{\min}(0):=\min_{a\in A}\tau_a(0)>0.$
    	In particular, the cost slice 
    	$\Gamespace_{|\tau}$ behaves well for decreasing demands.
    \end{corollary}

    Corollary~\ref{corollary:ConvergencePoA_Light}
    \wusec{and} the convergence results in \citet{Wu2019} \wusec{imply}
    immediately \wuu{for} every slice $\Gamespace_{|\tau}$
    whose cost functions $\tau_a(\cdot)$ are strictly
    positive,
    differentiable and regularly varying \wuu{that the PoA}
    \wusec{behaves} well in limits. Theorem~\ref{thm:PoA_Continuity}
    then implies that the PoA has a finite and tight upper bound
    on each of these cost slices $\Gamespace_{|\tau}.$
    In particular, Corollary~\ref{corollary:ConvergencePoA_Light}
    and \citet{Wu2019} together confirm the observed \wusec{behavior}
    of the empirical PoA for Beijing traffic data,
    which starts at $1$, then increases
    at a relatively mild rate with \wusec{growing} total demand, and eventually decays
    rapidly to $1$ after the total demand
    \wusec{reaches a certain threshold,} see
    \citet{Wu2019}.

    \subsection{Convergence rates of the PoA \wusec{for growing demands}} 
    \label{subsec:Heavy-traffic}
    \citet{Colini2017WINE,Colini2020OR}
    have demonstrated 
    that the PoA may diverge \wusec{for growing} total demand 
    when 
    the cost functions are not regularly varying.
    \citet{Wu2019} showed the convergence of
    the PoA to $1$ \wusec{for growing total demand} for arbitrary regularly varying cost functions, \wusec{but concrete}
    convergence rates of the PoA for arbitrary regularly varying cost functions (other than polynomials)
    \wusec{are} still \wu{missing.} 
    \wusec{Our H{\"o}lder continuity 
    results \wu{prove} to be very helpful here and yield} the first convergence
    rate \wu{for growing total demand} for certain regularly varying cost functions
    that are not polynomials.
    
    

    Consider \wusec{a} vector $\tau=(\tau_a)_{a\in A}$ \wusec{of}
    regularly varying \wusec{cost functions $\tau_a(\cdot)$ that
    have} the same regular
    variation index $\beta>0$ and satisfy the condition
    \begin{equation}\label{eq:Salience}
    \begin{split}
    \lim_{T\to\infty}\frac{\tau_a(T)}{\tau_b(T)}=:
    \lambda_{a,b}\in (0,\infty)\quad \forall
    a,b\in A.
    \end{split}
    \end{equation} 
    Karamata's Characterization Theorem (\citet{Bingham1987Regular})  \wusec{yields} that these cost
    functions $\tau_a(\cdot)$ have the form
    $\tau_a(x)=Q_a(x)\cdot x^{\beta}$
    for a (\emph{slowly varying}) function $Q_a(x)$ satisfying
    the condition that
    $\lim_{T\to \infty}\frac{Q_a(T\cdot x)}{Q_a(T)}=1$
    for all $x>0.$ 
    Condition~\eqref{eq:Salience} then
    implies \wusec{that} $\lim_{T\to\infty}\frac{Q(T)}{Q_b(T)}=\lambda_{a,b}$
    for \wu{all $a,b\in A.$}
    
    \wusec{Consider now} an \emph{arbitrary} game \wu{$\ga=(\tau,d)$}
    from the 
    cost slice $\Gamespace_{|\tau}$
    \wusec{of $\tau,$}
    and an \emph{arbitrary} arc $b\in A.$ 
    Then $\rho(\ga)=\rho(\hat{\ga})$
    when $\hat{\ga}:=(\hat{\tau},\frac{d}{T(d)})$
    with \WU{$\hat{\tau}_a(x):=\frac{\tau_a(T(d)\cdot x)}{\tau_b(T(d))}$}
    for \wu{all $a\in A$ and all $x\in [0,1].$} This holds since
    \WU{$\hat{\ga}=\Psi_{\tau_b(T(d))}\circ\Lambda_{T(d)}(\ga).$}
    
    \wusec{Consider the} auxiliary cost function vector
    $\sigma=(\sigma_a)_{a\in A}$
    with
    \begin{displaymath}
    \sigma_a(x):=\lambda_{a,b}\cdot x^{\beta}
    \quad\forall (a,x)\in A\times [0,1],
    \end{displaymath}
    and put $\ga':=(\sigma,\frac{d}{T(d)}).$
    Then $\rho(\ga')\equiv 1,$ as the cost functions of $\ga'$
    are monomials of the same degree $\beta$,
    see, \wusec{e.g.,} \citet{Wu2019} \wusec{and \citet{Roughgarden2000How}}. 
    
    \wusec{Both}
    $\ga'$ and $\hat{\ga}$
    belong to the same cost slice 
    $\Gamespace_{|\frac{d}{T(d)}}$ and 
    have the same total demand
    $T(\frac{d}{T(d)})\equiv 1.$
    Moreover, \WU{$\C^*(\ga')
    \ge \min_{a\in A}\lambda_{a,b}>0$}
    and $\sigma_{a'}(x)$
    is Lipschitz continuous on the compact interval
    $[0,1]$ with \wusec{Lipschitz} constant $\beta\cdot \max_{a\in A}\lambda_{a,b}$ for all $a'\in A.$
    
    Lemma~\ref{lemma:PoA_Lipschiz}d) then applies
    and yields that
    \begin{equation}\label{eq:Heavy-traffic-rate}
    \begin{split}
    |\rho(\ga)-&\rho(\ga')|=|\rho(\hat{\ga})-\rho(\ga')|\\
    &\le	2\cdot \frac{\rho(\ga')+\sqrt{ |A|\cdot T(\frac{d}{T(d)})\cdot \beta\cdot \max_{a\in A}\lambda_{a,b}}+2}{\C^*(\ga')}\cdot |A|\cdot T(\frac{d}{T(d)})\cdot \sqrt{\gnorm{\hat{\ga}-\ga'}}\\
    &\le 2\cdot \frac{1+\sqrt{ |A|\cdot \beta\cdot \max_{a\in A}\lambda_{a,b}}+2}{\min_{a\in A}\lambda_{a,b}}\cdot |A|\cdot \sqrt{\gnorm{\hat{\ga}-\ga'}}
    \end{split}
    \end{equation}
    when $\gnorm{\hat{\ga}-\ga'}\le \frac{\C^*(\ga')}{2\cdot |A|}.$ Here, we \wusec{have used} that
    \begin{align}
    \gnorm{\hat{\ga}-\ga'}&=
    \norm{\hat{\tau}_{|1}-
\sigma_{|1}}
    =\max_{a\in A,x\in [0,1]}|\hat{\tau}_a(x)
    -\sigma_a(x)|\notag\\
    &=\max_{a\in A,x\in [0,1]}\big|
    \frac{Q_a(T(d)\cdot x)}{Q_{b}(T(d))}\cdot x^{\beta}-\lambda_{a,b}\cdot x^{\beta}
    \big|=:w(T(d))\label{eq:RV_bound}\\
    &= \max_{a\in A,x\in [0,1]}\big|
    \frac{Q_a(T(d))}{Q_b(T(d))}
    \cdot \frac{Q_a(T(d)\cdot x)}{Q_{a}(T(d))}
    -\lambda_{a,b}
    \big|\cdot x^{\beta}\notag\\
    &\le \max_{a\in A,x\in [0,1]}\big|
    \frac{Q_a(T(d))}{Q_b(T(d))}-\lambda_{a,b}
    \big|
    \cdot \frac{Q_a(T(d)\cdot x)}{Q_{a}(T(d))}\cdot x^{\beta}+\max_{a\in A,x\in [0,1]}\big|
    \frac{Q_a(T(d)\cdot x)}{Q_{a}(T(d))}-1
    \big|
    \cdot \lambda_{a,b}\cdot x^{\beta},\notag
    \end{align}
    which tends to $0$ as $T(d)\to\infty.$
    Hence, $\rho(\ga)\to 1$ as $T(d)\to \infty,$
    since $\rho(\ga')\equiv 1.$
    
    Inequalities~\eqref{eq:Heavy-traffic-rate} \wusec{and} \eqref{eq:RV_bound} show that the convergence rate of $\rho(\ga)$ depends heavily on the properties
    of \wusec{the functions}
    $Q_a(x)$.
    As an example, we assume now that 
    these factors $Q_a(x)$ are of the form
    \begin{equation}\label{eq:SL_Template}
    \zeta_a\cdot 
    \ln^{\alpha} (x+1),\quad \zeta_a>0,\ 
    \alpha\ge  0.
    \end{equation}
    Then  we obtain by \eqref{eq:Heavy-traffic-rate} that
    \begin{displaymath}
    	\begin{split}
    	\gnorm{\ga'-\hat{\ga}}&=w(T(d))=\max_{a\in A,x\in [0,1]}\big|
    	\frac{Q_a(T(d)\cdot x)}{Q_{b}(T(d))}\cdot x^{\beta}-\lambda_{a,b}\cdot x^{\beta}
    	\big|\\
    	&=\max_{a\in A,x\in [0,1]}\lambda_{a,b}\cdot \Big(
    	x^{\beta}-\big(\frac{\ln (T(d)\cdot x+1)}{\ln (T(d)+1)}\big)^{\alpha}\cdot x^{\beta}
    	\Big)\\
    	&\le \max_{a\in A,x\in [0,1]}\lambda_{a,b}\cdot
    	\frac{\alpha}{\beta}\cdot 
    	\frac{T(d)\cdot x^{\beta+1}}{T(d)\cdot x+1}
    	\cdot \frac{\ln^{\alpha-1} (T(d)\cdot x+1)}{\ln^\alpha (T(d)+1)}\\
    	&\le \frac{\alpha}{\beta}\cdot 
    	\frac{T(d)}{T(d)+1}
    	\cdot \frac{1}{\ln (T(d)+1)}\cdot \max_{a\in A}\ \lambda_{a,b}
    	\end{split}
    \end{displaymath}
    and thus
    $\rho(\ga)= 1+O(\sqrt{\frac{1}{\ln (T(d)+1)}}).$
    \wu{This follows since} $\beta>0$ and the function
    \[
    x^{\beta}-\Big(\frac{\ln (T(d)\cdot x+1)}{\ln (T(d)+1)}\Big)^{\alpha}\cdot x^{\beta}, \quad x\in [0,1],
    \]
    reaches its maximum at a point $x\in [0,1]$ \wu{satisfying}
    the equation
    \[
    \ln^{\alpha}(T(d)+1)-\ln^{\alpha}(T(d)\cdot x+1)
    =\frac{\alpha}{\beta}\cdot \frac{T(d)\cdot x}{T(d)\cdot x+1}\cdot \ln^{\alpha-1} (T(d)\cdot x+1).
    \]
    
    We summarize this in Corollary~\ref{corollary:TailConvergence}.
    \begin{corollary}
    	\label{corollary:TailConvergence}
    	Consider
    	an arbitrary cost function vector $\tau=(\tau_a)_{a\in A}$ s.t.
    	all cost functions $\tau_a(\cdot)$ are regularly varying with the same regular
    	variation index $\beta>0$ and satisfy the condition in equation \eqref{eq:Salience}.
    	\begin{itemize}
    		\item[a)] The resulting cost slice $\Gamespace_{|\tau}$
    		behaves well for growing demands,
    		and
    		$\rho(\ga)=1+O(\sqrt{w(T(d))})$
    		with the upper bound $w(T(d))$ defined in \eqref{eq:RV_bound} for each game
    		$\ga=(\tau,d)\in\Gamespace_{|\tau}$ with a large
    		total demand $T(d).$
    		\item[b)] \wusec{For} cost functions $\tau_a(x)=\zeta_a\cdot 
    		x^{\beta}\cdot \ln^{\alpha} (x+1),$
    		\wusec{$a\in A,$}
    		\wusec{the PoA is} $\rho(\ga)=  1+O(\sqrt{\frac{1}{\ln (T(d)+1)}})$ for each game $\ga=(\tau,d)\in\Gamespace_{|\tau}$
    		with \wusec{a large} total demand $T(d).$ 
    		Here, $\alpha> 0$ and $\zeta_a>0$
    		are arbitrary constants independent of \wusec{the} demand vector $d$.
    	\end{itemize}
    \end{corollary}

    \wusec{Condition}~\eqref{eq:RV_bound} actually
    means that both the regularly varying cost functions
    $\tau_a(\cdot)$ and the slowly varying
    \wusec{functions} $Q_a(\cdot)$  are 
    \emph{mutually comparable},
    i.e., both $\lim_{T\to\infty} \frac{\tau_a(T)}{\tau_b(T)}$ and $\lim_{T\to\infty} \frac{Q_a(T)}{Q_b(T)}$ \wusec{exist} for \wu{any two} $a,b\in A$.
    These cost functions 
    in Corollary~\ref{corollary:TailConvergence}
    \wuu{can then be} thought of as a generalization
    of the BPR cost functions $q_a\cdot x^{\beta}+p_a$, \wusec{when}
    we substitute the positive constant factors $q_a$
    \wusec{by} mutually comparable and slowing varying \wusec{functions} $Q_a(\cdot)$. However, \wusec{this generalization comes at the price of a weaker
    convergence rate of the PoA.}

    \section{Summary and future work}
    \label{sec:Summary}
    
    \wusec{This} paper presents the first \WU{sensitivity} analysis for the PoA in non-atomic congestion games when both the demands and cost functions \wusec{may} change. To achieve this,
    we have introduced a topology and a metric on the class of games with the same combinatorial structure, which may also be of use for other research purposes. The PoA, the SO cost, and the WE cost turned out to be continuous maps w.r.t. that topology, see Theorem~\ref{thm:PoA_Continuity}.
    Their dependence on a small variation of the demands and/or Lipschitz continuous cost functions is thus small. 
    With the metric, we have quantified the variation of the PoA when demands and cost functions
    change simultaneously. 
    
    This \wusec{has led to}
    \wusec{an analysis of the} H{\"o}lder continuity of the PoA map
    $\rho(\cdot)$ on the game space.
    We \wu{have shown} first that the map $\rho(\cdot)$
    is not uniformly H{\"o}lder continuous on the whole
    game space, and \wusec{that} the H{\"o}lder neighborhood
    of each game is a proper subset of the game space, see 
    Theorem~\ref{thm:NotUniformlyHoelder}.
   So the PoA can \wu{in general} only be pointwise and locally H{\"o}lder 
   \wusec{continuous.}
    For each game $\ga=(\tau,d)$ with cost functions $\tau_a(\cdot)$
    that are Lipschitz continuous on the 
    compact interval $[0,T(d)],$ we \wu{have shown} in Theorem~\ref{thm:HalfHoelderContinuity} that 
    the PoA map $\rho(\cdot)$ is pointwise H{\"o}lder continuous
    at $\ga$ with H{\"o}lder exponent $\frac{1}{2},$
    i.e., there are constants 
    $\h_{\ga},\epsilon_{\ga}>0$ depending
    only on $\ga$
    such that $|\rho(\ga')-\rho(\ga)|<\h_{\ga}\cdot \sqrt{\gnorm{\ga'-\ga}}$
    for each game $\ga'$ with $\gnorm{\ga'-\ga}<\epsilon_{\ga}.$
    When the cost functions $\tau_a(\cdot)$ have
    \emph{\wusec{stronger} properties}, e.g., \wu{when they} \wusec{are constant} 
    or \wusec{have} strictly positive derivatives
    on the compact interval $[0,T(d)]$,
    we \wu{have shown} in Theorem~\ref{thm:Finer_PoA_Approx_Particular}a)--b) that the PoA map $\rho(\cdot)$
    then has  the \emph{\wusec{stronger} H{\"o}lder exponent} $1$ at the game $\ga$, which is
    of the same order as in the \wusec{recent} results by \citet{EnglertFraOlb2010}, \citet{TakallooKwon2020} and 
    \citet{Cominetti2019} for demand changes.
    
    Finally, we have applied our results to analyze the convergence behavior of the PoA when the total demand $T(d)$ tends to $0$ or $\infty.$ \wusec{We showed} that the PoA tends to $1$
    at a rate of $O(T(d))$ \wuu{for decreasing total demand $T(d)$}
    when the cost functions are strictly positive and
    Lipschitz continuous within a small neighborhood around
    the origin $0,$ and \wusec{identified} a \wusec{class} of non-polynomial regularly varying cost functions for which the PoA tends to $1$
    at a rate of $O(\sqrt{1/\ln (T(d)+1)})$ \wusec{for growing total demand.}
    These complement \wusec{recent results on the} convergence rate
    by \citet{Colini2017WINE,Colini2020OR} and \citet{Wu2019}.
    

    
    Theorem~\ref{thm:HalfHoelderContinuity} \wusec{yields}
    the H{\"o}lder exponent $\frac{1}{2}$
    for the PoA at games $\ga=(\tau,d)$ whose cost functions
    are Lipschitz continuous on the compact interval
    $[0,T(d)].$ We \wusec{conjecture} that this exponent should be $1,$
    but could only confirm this in Theorem~\ref{thm:Finer_PoA_Approx_Particular} for games $\ga=(\tau,d)$ with cost functions
    $\tau_a(\cdot)$ that are
    constant or
    have strictly positive derivatives on $[0,T(d)].$
    We have neither been able to confirm this, nor to provide a counterexample
    for games $\ga=(\tau,d)$ with 
    cost functions $\tau_a(\cdot)$ that are
Lipschitz continuous on $[0,T(d)].$ 

    \wuu{This is closely related to Open Question~\ref{OpenQuestion:TightnessOf1/2},
    	which concerns the tightness of the H{\"o}lder exponent $\frac{1}{2}$ 
    	for Lemma~\ref{lemma:PoA_Lipschiz}d).
    We guess that the exponent $\frac{1}{2}$ is not
tight for Lemma~\ref{lemma:PoA_Lipschiz}d). However, we cannot confirm this at present,
   and thus leave \wuu{it} as a topic for future research.}
 %
%
     
    \wuu{Open Question~\ref{open:Differentiability}
    is another interesting topic for future work, which may further develop
    the results obtained by \citet{Cominetti2019},
\citet{Patriksson2004,Patriksson2007}, \citet{Lu2017} and \citet{Klimm2021}
for the differentiability of the PoA and Wardrop equilibria in non-atomic congestion games.} 

	\section*{Acknowledgement}
	\wulast{We thank the two anonymous referees for their very deep and constructive 
	suggestions that have greatly helped to improve the paper.}
	
	 \wusec{Moreover,} the first author acknowledges support from the National Science Foundation of China with grant
	 No.~61906062, support from the Science Foundation of Anhui Science and Technology Department with grant
	 No.~1908085QF262, and support from the Talent Foundation of Hefei University with grant No.~1819RC29;
	 The first and second authors acknowledge support from the Science Foundation of the Anhui Education
	 Department with grant No.~KJ2019A0834.
	
	\bibliographystyle{plainnat}
	\bibliography{TO.bib}

\end{document}